\documentclass[10pt,journal,compsoc]{IEEEtran}
\usepackage{ifthen}
\usepackage{Rev}

\usepackage{amssymb}
\usepackage{pifont}
\usepackage{comment}
\usepackage{soul}
\usepackage{spverbatim}

\usepackage{float} 

\usepackage{graphicx}
\usepackage{dblfloatfix} 
\usepackage{epstopdf}
\usepackage{textcomp}
\usepackage{gensymb}
\usepackage{color}
\usepackage{amsmath,amsthm,amssymb}
\usepackage{bm}
\usepackage{etoolbox}
\usepackage{cite}
\usepackage{array}
\usepackage{booktabs}
\usepackage{multirow}

\usepackage{threeparttable} 
\usepackage{mathtools}
\usepackage{hyperref}
\PassOptionsToPackage{hyphens}{url}\usepackage{hyperref}
\usepackage{pifont}
\usepackage{breqn} 
\usepackage[framemethod=TikZ]{mdframed}

\usepackage[ruled]{algorithm}
\usepackage{algpseudocode}
\alglanguage{pseudocode}

\usepackage[font=footnotesize]{caption}
\usepackage{subcaption}

\DeclarePairedDelimiter\ceil{\lceil}{\rceil}
\DeclarePairedDelimiter\floor{\lfloor}{\rfloor}

\usepackage{url}
\urldef{\mailsa}\path|{alfred.hofmann, ursula.barth, ingrid.haas, frank.holzwarth,|
	\urldef{\mailsb}\path|anna.kramer, leonie.kunz, christine.reiss, nicole.sator,|
	\urldef{\mailsc}\path|erika.siebert-cole, peter.strasser, lncs}@springer.com|    
\newcommand{\keywords}[1]{\par\addvspace\baselineskip
	\noindent\keywordname\enspace\ignorespaces#1}
\usepackage[utf8]{inputenc}
\usepackage[english]{babel}

\usepackage{amsthm}

\theoremstyle{definition}
\newtheorem{definition}{Definition}

\theoremstyle{remark}

\definecolor{garrisonpink1}{rgb}{0.858, 0.188, 0.478}

\definecolor{R}{RGB}{0,150,0}
\definecolor{A}{RGB}{20,20,20}
\definecolor{T}{RGB}{0,0,150}

\usepackage{environ}
\NewEnviron{gather+}[1][1]{
  \begin{equation*}
  \scalebox{#1}{$\begin{gathered}\BODY\end{gathered}$}
  \end{equation*}
}

\newcommand{\aggregate}[2]{\underset{#2}{\operatornamewithlimits{#1\ }}}

\usepackage{tikz}
\newcommand{\circled}[2][]{
  \tikz[baseline=(char.base)]{%
    \node[anchor=text, shape=circle,draw, inner sep=0pt, minimum size=0.5em] (char){#1\strut};
    \node at (char.center) {\makebox[0pt][c]{#2}};}}
\robustify{\circled}

\usepackage{colortbl} 
\definecolor{LightCyan}{rgb}{0.7,1,1}

\usepackage{amsmath,booktabs}
\newcommand\mc[1]{\multicolumn{1}{c}{#1}} 

\newcommand{\f}[0]{${\rm {\bf f}}$~}
\newcommand{\ff}[0]{{\rm {\bf f}}}

\newcommand\upstrut{\rule{0pt}{10pt}}
\newcommand\downstrut{\rule[-5pt]{0pt}{5pt}}
\newcommand\mystrut{\upstrut\downstrut}

\usepackage{stackengine}
\newsavebox\mybox

\IEEEaftertitletext{\vspace{-3\baselineskip}}

\usepackage{cellspace} %
\renewcommand{\baselinestretch}{0.99}

\usepackage{setspace}

\usepackage{enumitem}
\setlist{itemsep=4pt, topsep=4pt}

\begin{document}
\bstctlcite{IEEEexample:BSTcontrol}

\title{NoisFre: Noise-Tolerant Memory Fingerprints from Commodity Devices for Security Functions}

\author{
Yansong Gao\IEEEauthorrefmark{1}, Yang Su\IEEEauthorrefmark{1}, Surya Nepal, Damith C.~Ranasinghe\IEEEauthorrefmark{2}

\IEEEcompsocitemizethanks{\IEEEcompsocthanksitem \IEEEauthorrefmark{1}Y. Gao and \IEEEauthorrefmark{1}Y. Su contributed equally to the study and are co-first authors in alphabetical order.}

\IEEEcompsocitemizethanks{\IEEEcompsocthanksitem Y.~Gao is with School of Computer Science and Engineering, Nanjing University of Science and Technology, China. This work was partially completed when he was with The University of Adelaide, Australia and CSIRO's Data61, Sydney, Australia. e-mail: yansong.gao@njust.edu.cn}

\IEEEcompsocitemizethanks{\IEEEcompsocthanksitem S.~Nepal is with Data61, CSIRO, Sydney, Australia. e-mail: surya.nepal@data61.csiro.au}

\IEEEcompsocitemizethanks{\IEEEcompsocthanksitem Y. Su and ~D.~C. Ranasinghe\IEEEauthorrefmark{2} (corresponding author) are with School of Computer Science, The University of Adelaide, Australia. e-mail: \{yang.su01; damith.ranasinghe\}@adelaide.edu.au}

\IEEEcompsocitemizethanks{\IEEEcompsocthanksitem We acknowledge the support from The Australian Research Council (DP140103448), National Natural Science Foundation of China (62002167), and Natural Science Foundation of JiangSu Province (BK20200461).}

}

\IEEEtitleabstractindextext{		
\begin{abstract}
Building hardware security primitives with on-device memory fingerprints is a compelling proposition given the ubiquity of memory in electronic devices, especially for low-end Internet of Things devices for which cryptographic modules are often unavailable. However, the use of fingerprints in security functions is challenged by the small, but unpredictable variations in fingerprint reproductions from the same device due to measurement noise. Our study formulates a novel and pragmatic approach to achieve highly reliable fingerprints from device memories. We investigate the transformation of raw fingerprints into a noise-tolerant space where the generation of fingerprints is intrinsically highly reliable. We derive formal performance bounds to support practitioners to easily adopt our methods for applications. Subsequently, we demonstrate the expressive power of our formalization by using it to investigate the practicability of extracting noise-tolerant fingerprints from commodity devices. Together with extensive simulations, we have employed 119 chips from five different manufacturers for extensive experimental validations. Our results, including an end-to-end implementation demonstration with a low-cost wearable Bluetooth inertial sensor capable of on-demand and runtime key generation, show that key generators with failure rates less than $10^-6$ can be efficiently obtained with noise-tolerant fingerprints with a single fingerprint snapshot to support ease-of-enrollment.

\end{abstract}

}

\maketitle

\section{Introduction}

Various schemes have investigated fingerprinting commercial-off-the-shelf (COTS) devices to build security applications: verifying the provenance of integrated circuits (IC) to guard against counterfeiting by fingerprinting IC packages~\cite{dhanuskodi2020counterfoil}; identifying unlawful 3D printed products by fingerprinting unique textures resulting from 3D printers~\cite{li2018printracker}; authenticating smartphones by fingerprinting the Photo-Response Non-Uniformity of a camera image sensor~\cite{ba2018abc}; and identifying commodity mobile devices by fingerprinting on-board sensors~\cite{bojinov2014mobile,zhang2019sensorid}.

Compared with fingerprinting methods for on-board sensors and other components like central processor units (CPUs)~\cite{bojinov2014mobile,son2018gyrosfinger,kim2018c,ba2018abc,willers2016mems,zhang2019sensorid,ba2019cim,quiring2019security,cheng2019demicpu,dhanuskodi2020counterfoil}, fingerprinting embedded memories---including static random access memory (SRAM)~\cite{guajardo2007fpga,holcomb2009power,van2015lightweight}, dynamic random access memory~\cite{schaller2018decay}, Flash memory~\cite{wang2012flash,guo2017ffd}, and electrically erasable programmable read-only memory (EEPROM)---pervasively embedded in COTS devices is a highly desirable proposition for provisioning security functions, especially in the absence of cryptographic modules. Fingerprinting embedded memory is attractive because: i) memory cells are intrinsic to computing platforms and available in large volumes to obtain many independent fingerprints or secret keys; ii)~memory biometrics provides a physical source of true randomness; iii) it removes the need for a protected non-volatile memory for secrets (root keys can be generated on-demand and ``\textit{forgotten}'' after usage); and iv) imparts no extra hardware costs to {\it existing COTS} devices such as medical devices, wireless sensors, credit cards, wearable devices, and a plethora of low-end Internet of Thing (IoT) devices, which are projected to grow to 75.44 billion worldwide by 2025~\cite{IoTNumberPrediction}.

\begin{figure}[!ht]
	\centering
	\includegraphics[trim=0 0 0 0,clip,width=.70\mylinewidth]{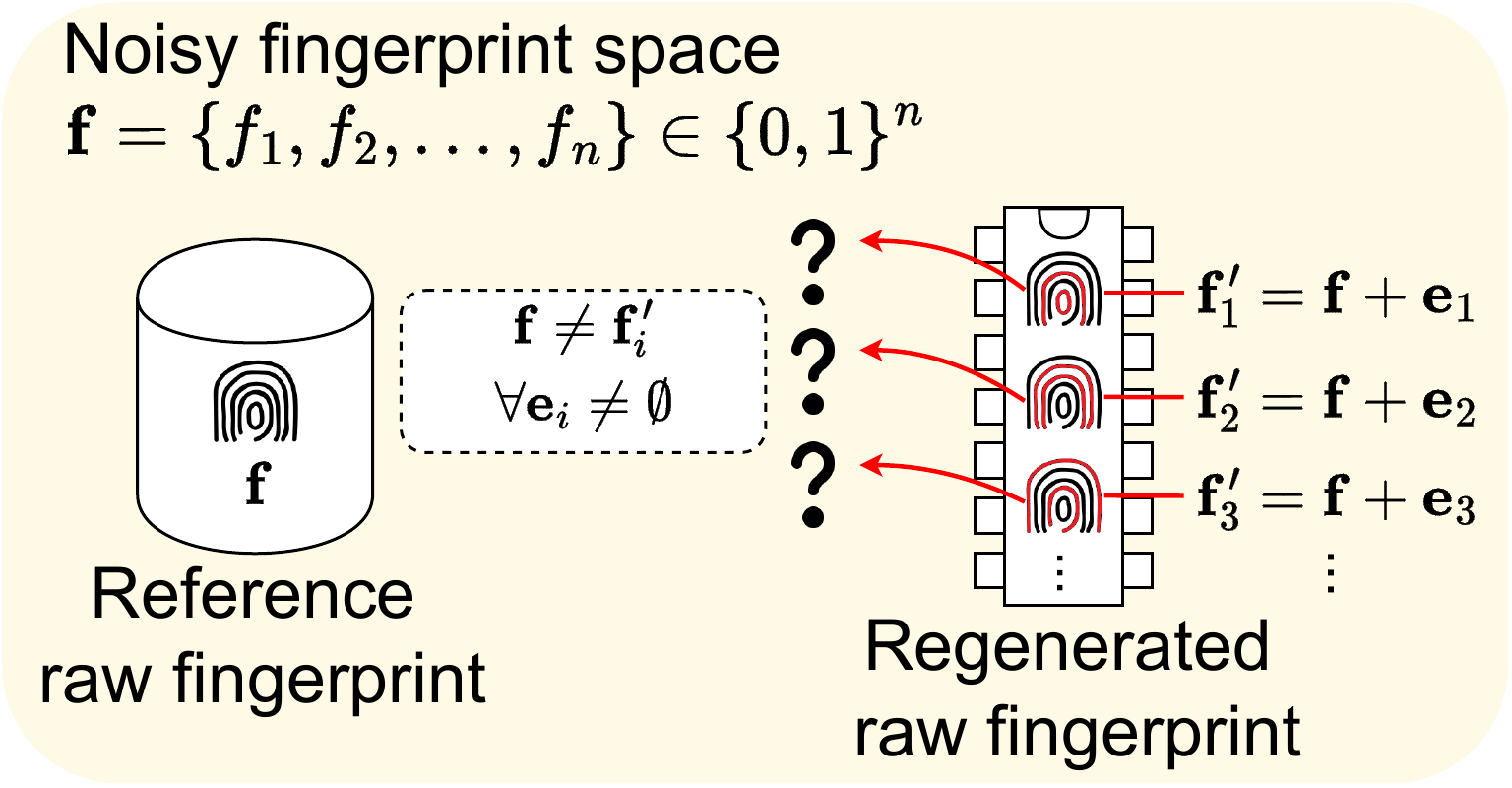}
	\caption{Conventional schemes extract \textit{raw} fingerprints from individual memory cells, such as from the random power-up state of each SRAM cell. However, raw fingerprints $\ff^\prime_i$ regenerated at different time instances from the same device can mismatch a reference raw fingerprint template $\ff$ due to native bit errors introduced by noise ${\bf e}_i$. The red curves in the {\it fingerprint symbol} depict errors resulting from noise.
	}
	\label{fig:ECClessF1a}
\end{figure}

\begin{figureRevsourceS}[t]
	\centering
	\includegraphics[trim=0 0 0 0,clip,width=0.85\mylinewidth]{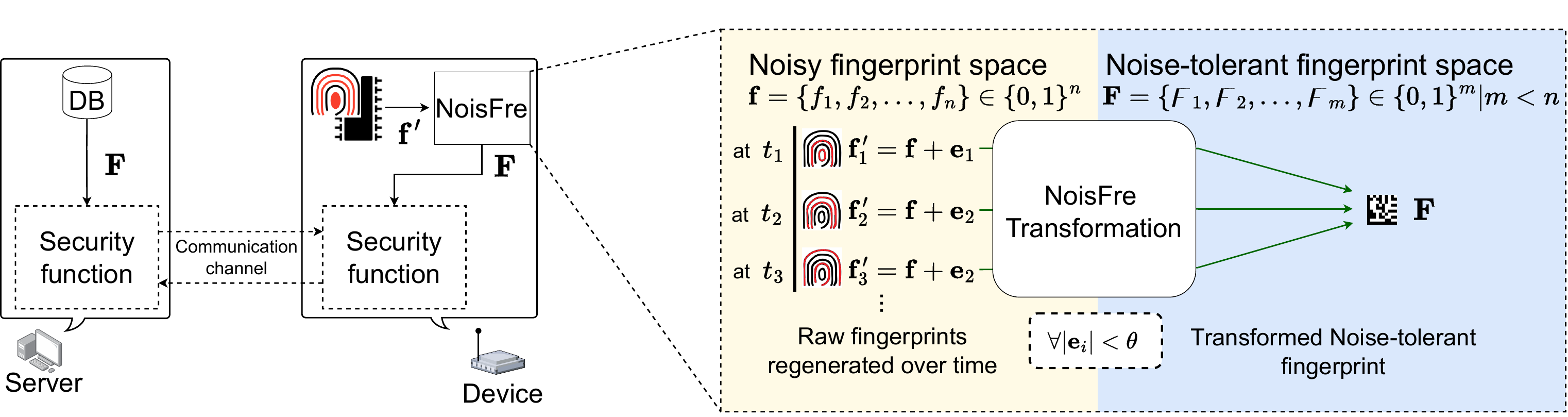}
	\caption{Illustrating the use of noise-tolerant memory fingerprints from commodity devices for security functions. We transform the raw fingerprint from an $n$-dimensional noisy fingerprint space to an $m$-dimensional space, we refer to as the noise-tolerant fingerprint space, where $m<n$. In the noise-tolerant space, as long as the noise ${\bf e}_i$ is less than a bound $\theta$, the regenerated and transformed fingerprint can be correctly projected to the reference transformed fingerprint template ${\bf F}$ securely enrolled and stored on the server. Red curves in the {\it fingerprint symbol} depict errors in the raw fingerprint upon regeneration at time instances $t_1$ to $t_3$. Now, the ${\bf F}$ obtained can serve as a root of trust or a root key for a security function.}
	\label{fig:ECClessF1b}
\end{figureRevsourceS}

\subsection{The Challenge} 
Whenever a fingerprint is generated from the same device, the digitized fingerprint should be exact for its use in security functions. However, fingerprints generated at different time instances are susceptible to unpredictable noise, such as thermal noise, supply voltage fluctuations, and device aging, and, consequently, differ in some bits. First, the positions of flipped raw bits vary. Second, the number of flipped raw bits varies from time to time. Thus, it is challenging to determine reliable raw bits, and existing memory fingerprinting schemes cannot naturally tolerate noise in the raw, noisy fingerprint space. Therefore, it is often infeasible to regenerate a fingerprint identical to a reference template that is securely stored (e.g., in a server) for directly building security functions between it and devices, as shown in Fig.~\ref{fig:ECClessF1a}.

\textit{Until now}, using approximate and noisy renditions of biometric fingerprint templates in security functions has been demonstrated to be possible with fuzzy extractor (FE) based methods~\cite{boyen2004reusable,dodis2004fuzzy}. The FE employs a generation function to transform ``fuzzy'' biometrics into private secrets together with helper data used in a subsequent reproduction function to reconcile errors and derive the exact private secret from an approximately close template of the original biometric~\cite{bosch2008efficient,maes2012pufky,delvaux2016efficient}. Employing an FE on a device leads to two fundamental problems: i)~the computation overhead introduced on a device by FE logic is high \cite{gao2018lightweight} and ii)~the associated helper data can be actively manipulated, in helper data manipulation (HDM) attacks, to weaken or even compromise the security of the derived fingerprint or private secret~\cite{delvaux2014helper,becker2017robust}. A generic countermeasure against HDM attacks remains an open challenge~\cite{becker2017robust}. 

\begin{center}
\textit{Hence, there remains a significant leap between the desire for re-purposing ubiquitously available memory for security functions and the practicability of exploiting memory fingerprints for security.}
\end{center}
While the notion of exploiting tiny hardware fabrication variations to generate memory fingerprints is not new, we challenge the traditional method of reliable fingerprint provisioning and pose the following research questions (RQs):
\vspace{-2mm}
\begin{mdframed}[backgroundcolor=black!10,rightline=false,leftline=false,topline=false,bottomline=false,roundcorner=2mm]
	\textbf{RQ1:} How can we extract \textbf{\textit{intrinsically reliable fingerprints}} from device memories? 
\end{mdframed}
\begin{mdframed}[backgroundcolor=black!10,rightline=false,leftline=false,topline=false,bottomline=false,roundcorner=2mm]
	\textbf{RQ2:}  If an approach does exist, is the method \textbf{\textit{pragmatic}} and \textbf{\textit{usable}} for fingerprinting memory resources on pervasive commodity computing devices?
\end{mdframed}
\vspace{-3mm}

\subsection{Our NoisFre Concept}
Current memory fingerprinting schemes extract a fingerprint bit from each memory cell. Fingerprinting under this scheme is susceptible to noise. To the best of our knowledge, for commodity memories, existing techniques fail to accurately capture the noise-tolerance degree of each raw bit to formally determine those extremely reliable bits for direct key usage without the problematic error reconciliation.

We recognize that device memories are a cost-free and abundant source of entropy. Attributing to the ever-decreasing fabrication costs, the size of memory pervasively embedded within devices has become increasingly large. Hundreds of kilobytes (KiB), even in low-end devices, are common (see the devices we tested in Table~\ref{tab:chipSpec}). Consequently, we envision that the entropy of extracted information may be sacrificed for improved reliability. Therefore, we propose the concept of transforming the raw fingerprint space of high information density into a lower-dimensional space with the attribute of being largely invariant to noise---bit flips---observed in the digitized raw fingerprint space or memory biometrics. We refer to this  noise-tolerant memory fingerprinting concept as \textbf{NoisFre}.

We illustrate our concept in Fig.~\ref{fig:ECClessF1b}. Building upon a raw memory biometric source that is {\it a noisy fingerprint space}, we propose extracting new fingerprints ${\bf F}$ {\it in the deliberately transformed noise-tolerant fingerprint space}, which can tolerate a desirable noise bound $\theta$. Here, as long as the noise ${\bf e}_i$ induced number of raw fingerprint bit errors is less than $\theta$, the regenerated and transformed raw fingerprints are guaranteed to be {\it projected} to its reference counterpart ${\bf F}$ enrolled at the server. More generally, the regenerated and transformed fingerprint ${\bf F}$ is insensitive to bit errors (resulting from noise) in the raw fingerprint space. Therefore, it can be directly employed---without error reconciliation---as a root key in a security function, despite the noisy renditions of the raw fingerprints at times $t_1$, $t_2$, and $t_3$. 

\vspace{1mm}
\noindent{\it Significantly, we recognize that the best strategy for fingerprint memory is not always directly from the raw noisy fingerprint space, such as directly treating the power-up state of an SRAM memory of {\it a cell} as a fingerprint bit, the foundation for {\it all current} memory fingerprinting schemes. We argue for exploiting the freely available, abundant entropy of memories. We do not focus on individual raw fingerprint bits but seek to find an invariant property of a group of raw bits to measure, so we can be less concerned of the complexity about the process generating those bits.} 

\subsection{Contributions and Results}
\begin{itemize}
    \item We {\it exploit the freely available and abundant} entropy from memories to propose a \textit{new concept}---NoisFre---for highly reliable fingerprinting of commodity device memories. The principle is based on transforming from a noisy raw fingerprint space to a lower dimensional, noise-tolerant fingerprint space capable of reconciling noise inherent across multiple measurements of the same raw fingerprint. To corroborate the proposed NoisFre concept, we have developed two specific transformation methods: i)~S-Norm and ii)~D-Norm (\textbf{RQ1}). 
    
    \item We formulate analytical models with the expressive power to support the design of security functions and evaluate the transformation methods. We express~i)~an upper bound for the unreliability of the transformed $\digamma$ bits with respect to the transform function parameters; and~ii)~the expected fingerprint extraction efficiency---the number of transformed $\digamma$ bits that can be extracted from a given memory size (\textbf{RQ2}).
    
    \item We conduct elaborate and extensive evaluations with a synthetic chip model to obtain the massive number of repeated fingerprint measurements necessary to validate our formalization of unreliability and extraction efficiency. Billions of repeated measurements were simulated using the synthetic chip model with bit-level modeling capable of capturing bit-error behavior in SRAM device memories. Our formal models are confirmed to be worst-case bounds in practice (\textbf{RQ2}).
    
    \item We extensively test: i)~110 SRAM memory devices from
    three different manufacturers to experimentally validate NoisFre performance. We focus on SRAM memory, as it is the most commonly embedded memory, especially for low-cost IoT devices. Further, we employ: ii)~seven Flash memories and iii)~two EEPROM memories for validating the \textit{generalizability} of NoisFre (\textbf{RQ2}).
    
    \item To demonstrate the expressive power of our formalization, we investigate the derivation of a root key---the foundation for realizing various security functions. We demonstrate a 128-bit root key with an extremely low key failure rate of less than $10^{-6}$ can be directly obtained by transformed fingerprints to obviate the need for costly noise reconciliation. Significantly, a fingerprint snapshot or single measurement is sufficient for enrolling a key, a process we follow in all our experiments (\textbf{RQ2}). 
    \item As a case study, we \textit{implement a NoisFre key generator} and a security function on a low-end wearable Bluetooth inertial sensor. We extract a root key directly from native SRAM fingerprints transformed into noise-tolerant $\digamma$ bits for use in a remote attestation primitive. By fundamentally obviating the state-of-the-art method necessary for reconciling noisy key bits, we demonstrate a significant overhead reduction (i.e., 54\% compared to reverse FE and 82\% compared to FE) and enhanced security. By utilizing the power isolation features, we also demonstrate the realization of {\it run-time and on-demand generation of robust SRAM fingerprints} ${\bf F}$ on this low-end device. A video demo is available at \href{https://youtu.be/O5NWZw-swpw}{\underline{https://youtu.be/O5NWZw-swpw}} (\textbf{RQ2}). 
    \item We release the 100 chip SRAM memory fingerprint dataset that we collected and open-source code artifacts to facilitate future research at \underline{https://github.com/AdelaideAuto-IDLab/NoisFre}.
\end{itemize}

\section{Background}\label{Sec:background}
We concisely describe the well-known methods of fingerprinting memories. Then we briefly introduce the commonly accepted (reverse) FE to reconstitute a ``fuzzy'' secret into cryptographic secrets for security functions.

\subsection{Fingerprinting Device Memories}\label{sec:SRAMBackground}
We consider the widely used, specifically in low-end devices, SRAM, Flash, and EEPROM memory fingerprinting. As for the SRAM memory, when SRAM is powered up, each cell exhibits a \textit{favored power-up state}; such an initial state varies from cell to cell and chip to chip. Therefore, each SRAM cell's power-up state is treated as a fingerprint bit. Fingerprinting SRAM is closely related to the SRAM physical unclonable functions~\cite{guajardo2007fpga,holcomb2009power}.

To extract fingerprints from Flash memory~\cite{wang2012flash,guo2017ffd}, all Flash cells on the same page are first erased to ``1''. Then partial programming is applied. As a result of tiny fabrication variations, some cells will remain in state ``1'' while others flip to ``0''. Which cell remains or flips is determined by the random and uncontrollable fabrication process variations. One can treat whether a cell flips as the fingerprint bit---a flip as logic ``0'', and otherwise ``1''. The partial programming period is pre-determined to ensure balanced ``0''/``1'' bits in practice. The same procedure is applicable to fingerprint EEPROM.

\subsection{Reliable Secrets with Fuzzy Extractors}\label{sec:rfe}

A widely accepted method to turn noisy hardware fingerprint bits (key material) into usable cryptographic keys is to use an FE~\cite{boyen2004reusable,dodis2004fuzzy}. In general, the FE consists of two procedures: i) a secure sketch and ii) an entropy extraction. The secure sketch reconciles errors in the regenerated bits. The entropy extraction (e.g., a cryptographic hash function) compresses the bits into a uniformly distributed cryptographic key with full bit entropy. 

\begin{figureRevsource}[!ht]
	\centering
    \includegraphics[trim=0 0 0 0,clip,width=\mylinewidth]{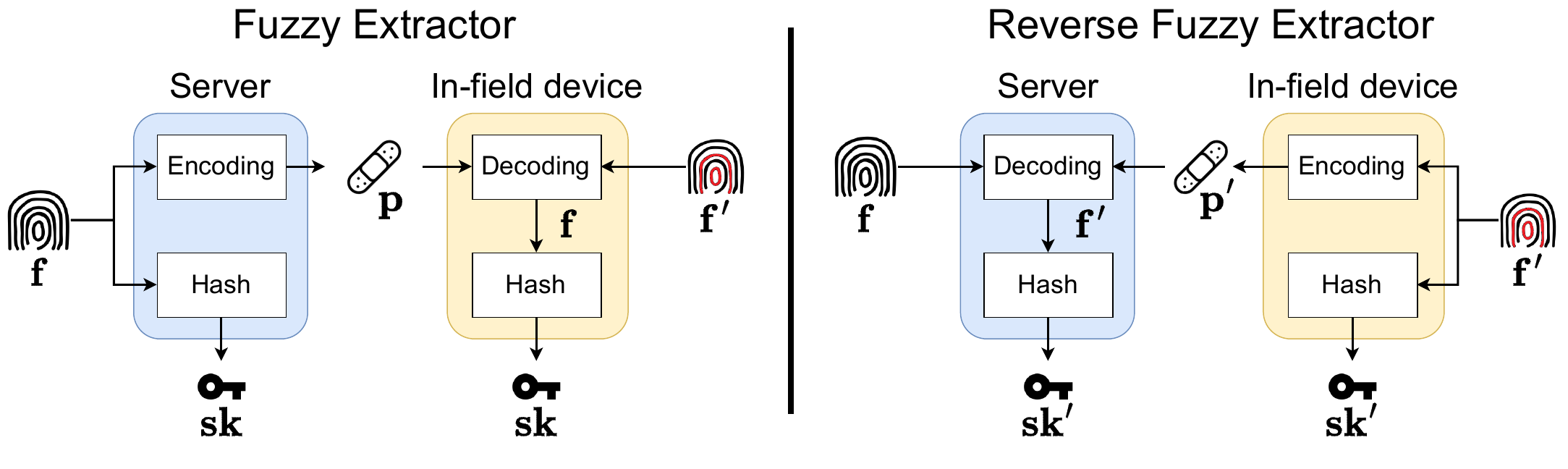}
	\caption{The fuzzy extractor (FE) and the lightweight state-of-the-art reverse fuzzy extractor (RFE)  for reconciling noisy fingerprint responses with the aid of helper data ${\bf p}$. In an RFE the encoding is embedded in a device and the computationally heavy decoding is offloaded to the server.}
	\label{fig:keyGen}
\end{figureRevsource}

\Revsource{rev:fuzzy_ext_description}{The secure sketch construction has a pair of operations, as shown in Fig.~\ref{fig:keyGen}: i)~encoding and ii)~decoding. Typically, in the FE setting, the encoding employing an error correction code (ECC) encoder is executed by the server during the fingerprint template enrollment phase to compute helper data ${\bf p}$ (redundant information). The decoding employing an ECC decoder is performed on the in-field device to recover a reliable fingerprint, ${\bf sk}$. By recognizing that the encoding function's computational burden is significantly higher than decoding, it is feasible to place the ECC encoder on the device-side while leaving the computationally complex ECC decoder to the resource-rich server; this method is termed reverse FE (RFE) or reusable FE~\cite{boyen2004reusable,van2012reverse,canetti2016reusable}. More specifically, the encoding function is implemented on the device-side to produce the associated helper data ${\bf p^{\prime}}$ as well as an ${\bf sk^{\prime}}$ based on on-device fingerprint evaluation ${\bf f^{\prime}}$, where ${\bf p^{\prime}}$ is now sent to the server to assist the reconstruction of ${\bf sk^{\prime}}$. Now, ${\bf p^{\prime}}$ and ${\bf sk^{\prime}}$ could vary every time as the on-device fingerprint ${\bf f^{\prime}}$ can differ during each reevaluation. In contrast to an FE setting, the server uses the enrolled ${\bf f}$ along with ${\bf p^{\prime}}$ for recovering the key ${\bf sk^{\prime}}$.}

\section{NoisFre Transformation}\label{Sec:method}

We provide the impetus for developing NoisFre fingerprinting and its key insights, followed by two specific and \textit{practical} NoisFre transformation methods.

\subsection{Our Pragmatic Approach} \label{sec:our_approach}

\begin{figureRevsource}[!ht]
	\centering
	\includegraphics[trim=0 0 0 0,clip,width=\mylinewidth]{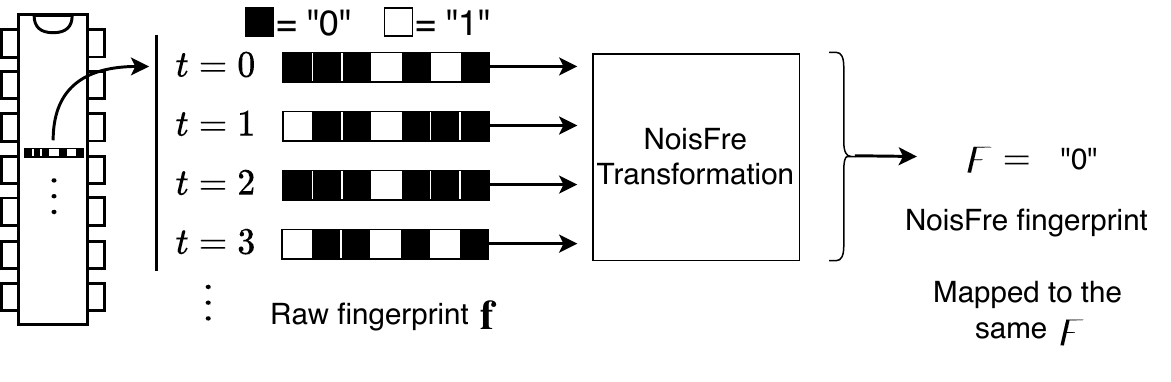}
	\caption{NoisFre transformation concept. A group of raw bit vectors exhibiting errors measured at different times can be transformed into the same NoisFre fingerprint bit $\digamma$ (e.g., ``0''), attributing to the \textit{invariance of the transform} patterns to:~i)~\textit{permutations} of raw bit vectors (e.g., at time $t=0$ and $t=1$) and ii)~vectors with different \textit{combinations} of raw bits (e.g., at times $t=2$ and $t=3$)}
	\label{fig:NoisFreeConcept}
\end{figureRevsource}

In contrast to extracting one fingerprint bit from each memory cell, we propose a many-to-one transformation possessing a \textit{property of invariance} to underlying raw bit patterns: more generally, invariant to the unpredictable, complex and dynamic raw fingerprint generating processes. \Revsource{rev:permutation}{Our desire is to project all of the fingerprint measurements conducted from the same block of memory at different time instances to the {\it exact} same transformed bit, $\digamma \in \{0,1\}^1$}. The concept is illustrated in Fig.~\ref{fig:NoisFreeConcept}. Note that:
\begin{itemize}
    \item Bit errors leading to permutations of raw bits---where the positions of ``1'' and ``0'' values change in a bit vector but the number of ``1''s and ``0''s do not, as seen at $t=0$ and $t=1$ in Fig.~\ref{fig:NoisFreeConcept}---are projected to the {\it exact} same transformed bit $\digamma \in \{0,1\}^1$.
    \item  Bit errors leading to vectors constituting different combinations of ``1''s and ``0''s (e.g., the fingerprint from the same block of memory at time $t=1$ with two ``1'' binary bits and that regenerated at time $t=2$ with only one ``1'' binary bit in Fig.~\ref{fig:NoisFreeConcept}) are projected to the {\it exact} same transformed bit $\digamma \in \{0,1\}^1$.
\end{itemize}

We observe that a transformed bit, $\digamma$, is able to mitigate the impact from multiple raw fingerprint bit errors manifesting as permutations or combinations of an $n$-bit raw fingerprint, $\bf f$. The concept we propose is surprisingly simple but efficient and practical because of the {\it important but inadvertent} reality of large memory volumes intrinsic to devices. From a practical consideration, our critical insight is that memory embedded within modern electronics is large and {\it provides abundant entropy} to be exploited without additional costs for security functions. This fact is the foundation for our NoisFre transformation method: \textit{trade-off entropy for reliability}.

This work proposes two specific NoisFre transformation methods: Single $\ell 1$-Norm (S-Norm) and Differential $\ell 1$-Norm (D-Norm). 

\subsection{Single \texorpdfstring{$\ell 1$}{TEXT}-Norm Transformation (S-Norm)}
\label{sec:noiseTolerantFingerprinting}
\Revsource{rev:L1_Norm_def}{The $\ell 1$-Norm of a vector is the distance of the vector from an all-zero vector---or the Hamming weight of a vector, as described in Definition~\ref{def:l1-norm}.}

\begin{definition}[\textbf{$\ell 1$-Norm}]\label{def:l1-norm}
Let \f be a binary vector length $n$ representing a noisy raw fingerprint where $f_j$ is the $j^{\rm th}$ bit in ${\bf f}$; then the $\ell 1$-Norm of ${\bf f}$ is defined as:
\vspace{-0.1cm}
\begin{equation}
    \|{\rm {\bf f}}\|_1 \triangleq \sum_{j = 1}^{n} f_j.
\end{equation}
\end{definition}

\begin{figureRevsource}[!ht]
	\centering
	\includegraphics[trim=0 0 0 0,clip,width=\mylinewidth]{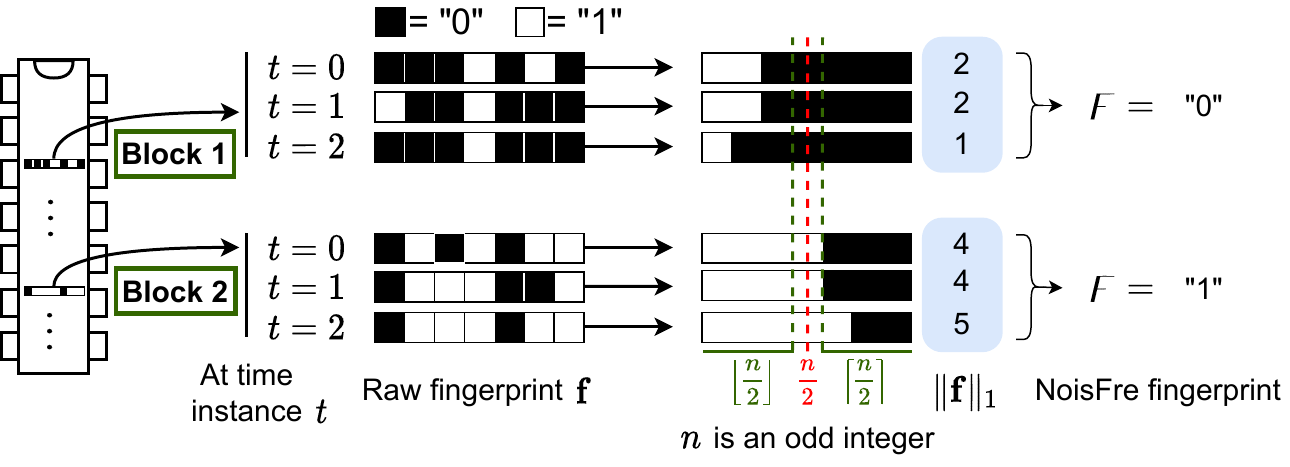}
	\caption{NoisFre transformation via S-Norm; the transformed bit $\digamma$ is extracted using the $\ell 1$-Norm of a group of raw bits. The group size is an odd integer number (e.g. 7 in this illustration) to ensure a balance between zeros and ones in the $\digamma$ bits. We illustrate the regenerated raw fingerprints from two blocks of memory---\textbf{Block 1} and \textbf{Block 2}---at times $t=0$, $1$, and $2$.
	}
	\label{fig:NoisFreeFull}
\end{figureRevsource}

Interestingly, an $\ell 1$-Norm of a vector is permutation invariant. Hence,  a new bit $\digamma$ can be obtained by applying an  $\ell 1$-Norm over a raw fingerprint vector $\bf f$ as $\|\ff\|_1$ and as described by the S-Norm below.

\begin{definition}[\textbf{S-Norm}]\label{def:s-norm} 
Let the $i$-th raw fingerprint bit vector of $n$ bits, where $n$ is an odd integer, be ${\rm {\bf f}}_i$. Then S-Norm transform is defined as:
\begin{equation}\label{eq:STrasform}
    \digamma = \left\{\begin{matrix}
 1, \Rev{~~\|{\rm {\bf f}}_i\|_1 \geq \ceil{\frac{n}{2}}}\\
 0, \Rev{~~\|{\rm {\bf f}}_i\|_1 \leq \floor{\frac{n}{2}} }
\end{matrix}\right.
\end{equation}
\end{definition}

\Revsource{rev:D_Norm_figure_description}{An illustrative example of S-Norm-based transformation is provided in Fig.~\ref{fig:NoisFreeFull}. When $\|\ff\|_1 > \frac{n}{2}$, where $n$ is an odd integer, the $\digamma$ bit is ``1'', and otherwise ``0''. This transform has the first desirable property of being insensitive to bit errors manifesting as permutations of a raw fingerprint ${\bf f}$. For example, despite the raw bit errors at time $t=1$ for the raw fingerprint from \textbf{Block~1} that lead to a permutation in respect to the bit vector referenced at time $t=0$, the corresponding $\ell 1$-Norm remains invariant; the memory \textbf{Block~1} is still projected to $\digamma=$ ``0''. The transform has the second desirable property of being insensitive to bit errors manifesting as combinations of an $n$ raw fingerprint bit vector. Notably, errors that lead to different combinations of raw binary ``1'' and ``0'' values can increment or decrement $\ell 1$-Norm but these can still be projected to the same transformed $\digamma$ bit. For example, see $\ell 1$-Norm values for memory \textbf{Block~1} at time $t=1$ compared to $t=2$ in Fig.~\ref{fig:NoisFreeFull}. Hence, S-Norm achieves the two qualities we desire from a transform expressed in Section~\ref{sec:our_approach}}. 

In Fig.~\ref{fig:NoisFreeFull}, we can expect the transformed $\digamma$ bit from \textbf{Block~2} with $\ell 1$-Norms at the decision boundary, defined in equation~\eqref{eq:STrasform}, to more likely be affected by error bits within the raw fingerprint, resulting in different combinations of an $n$-bit raw fingerprint. A resulting combination of $n$ bits with a single change in the number of raw binary ``1'' bits can lead to an $\ell 1$-Norm projection that crosses the decision boundary. We recognize the resulting $\digamma$ bits from such raw fingerprints to effectively display low reliability. Therefore, we propose winnowing raw fingerprints based on treating $\ell 1$-Norm as a reliability measure for $\digamma$ bits. For this purpose, we define the \textit{S-Norm-based Selection} method described in Definition~\ref{def:s-norm-sel} and generalize the approach using a noise tolerance parameter $\theta$ to provide an upper bound of tolerance on raw bit errors or the combinations of $n$-bit patterns; the transform will faithfully project to a specific $\digamma$ bit. Interestingly, the evaluation of $\ell 1$-Norm only require a \textit{single measurement}, while a larger $\theta$ can be chosen to facilitate higher noise tolerance. Therefore, we propose selecting based on the $\ell 1$-Norm of raw fingerprint vectors obtained from a single measurement defined as being at time $t=0$. Notably, this approach facilitates rapid characterization of raw fingerprints from device memories, as $\ell 1$-Norms can be acquired in a single measurement. All our experimental and theoretical analyses assume such a characterization.

\begin{definition}[\textbf{S-Norm-based Selection}]\label{def:s-norm-sel} 
\Revsource{rev:s-norm-sel}{Let the raw fingerprint in the $i$-th $n$-bit block extracted at time $t$ be ${\rm {\bf f}}_i^{t}$. Then for a chosen noise tolerance parameter $\theta \in \mathbb{N}^0$, an extracted raw fingerprint vector ${\rm {\bf f}}_i^{0}$ is selected at time $t=0$ if:}
\begin{align}\label{eqn:s-norm-selection}
        \|{\rm {\bf f}}_i^{0}\|_1 \leq \floor{\frac{n}{2}} - \theta~~~\text{or}~~~\|{\rm {\bf f}}_i^{0}\|_1 \geq \ceil{\frac{n}{2}} + \theta
\end{align}

\end{definition}

\begin{figure}[!ht]
	\centering
	\includegraphics[trim=0 0 0 0,clip,width=\mylinewidth]{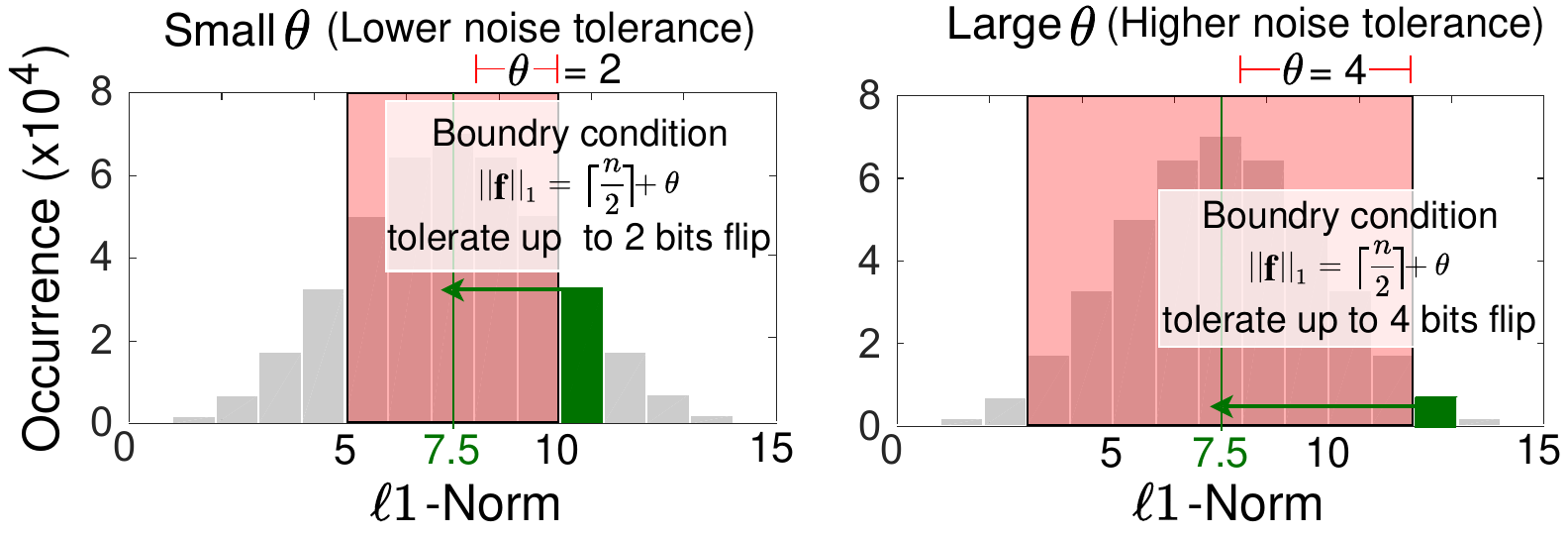}
	\caption{Illustrating the role of the noise tolerance parameter $\theta$ in S-Norm-based selection. The plots show the $\ell 1$-Norm distribution of raw noisy fingerprints. It approximates normal distribution. A larger $\theta$ ensures the transformed $\digamma$ bits can tolerate a higher degree of noise.} 
	\label{fig:reliabilityDist_HW}
\end{figure}

\Revsource{rev:equations_clear}{
To understand and demonstrate the significant role of the noise tolerance parameter $\theta$ in the mitigation of raw bit errors, we consider the distribution of $\|\ff^0\|_1$.} We used the experimental dataset obtained from Nordic Semiconductor chips detailed in Table~\ref{tab:chipSpec}. Fig.~\ref{fig:reliabilityDist_HW} plots the resulting distribution of enrolling measurements (at time $t=0$) for two cases of a small and a large $\theta$ for an $\bf f$ of $n=15$-bit. As expected, the distribution of $\|\ff^0\|_1$ approximates a bell curve. 

Consider the groups of $\ff^0$ raw fingerprints (green bar) at the boundary of the selection criteria in equation~\eqref{eqn:s-norm-selection}, where $\|\ff_i^0\|_1=\ceil{\frac{n}{2}} + \theta$ for the two cases of a small and large $\theta$. These groups represent those closest to the decision boundary,  $\|\ff_i^0\|_1=\ceil{\frac{n}{2}}$ (green line), consequently representing those most likely to lead to a bit error in a transformed bit $\digamma$ when raw bit errors changes the $\ell 1$-Norm of $\ff^0$. When $\theta$ is small ($\theta=2$), a change in more than two bits in a given pattern can lead to $\|\ff\|_1$ crossing the decision boundary $\frac{n}{2}$ in a subsequent raw fingerprint extraction, resulting in a $\digamma$ bit flip.  

In contrast, when $\theta$ is large, set to 4, the $\ell 1$-Norm of the selected fingerprint vectors or $\|\ff^0\|_1$ are further away from the decision boundary. Consequently, a change in more than four bits in a \f is needed to flip the corresponding $\digamma$ in a subsequent fingerprint evaluation. Such a probability is smaller than a change in more than two raw fingerprint bit flips for vectors selected with $\theta=2$. Therefore we can expect the $\digamma$ bits selected employing a larger $\theta$ to be significantly more reliable. Our study, thus far, leads to the following observations:

\begin{mdframed}[backgroundcolor=black!10,rightline=false,leftline=false,topline=false,bottomline=false,roundcorner=2mm]
	\noindent\textbf{Observation 1:} The S-Norm $\|\ff\|_1$ yields a representation analogous to the reliability of the new bit $\digamma$. 
	\vspace{1mm}
	
	\noindent\textbf{Observation 2:} The S-Norm transformed bits are invariant to permutations and  combinations of raw bit patterns. Further, $\theta$ provides a desirable lower bound on raw bit errors tolerated by the transform.
	\vspace{1mm}
	
	\noindent\textbf{Observation 3:} There is an expected trade-off evidence in Fig.~\ref{fig:reliabilityDist_HW}. While increasing $\theta$ increases the noise tolerance of the transform, it reduces the number of noise-tolerant fingerprint bits extractable from a given memory.
	
\end{mdframed}

\subsection{Differential \texorpdfstring{$\ell 1$}{TEXT}-Norm Transformation (D-Norm)}\label{sec:DNormConcept}
Considering \textit{Observation 3} and the distribution in Fig.~\ref{fig:reliabilityDist_HW}, we recognize that \textit{a distance measure capable of presenting a bimodal distribution could provide an intrinsic separation of groups of underlying raw fingerprint bits} with the potential to yield higher numbers of noise-tolerant bits. We hypothesize that a differential distance measure may afford such a desirable distribution and propose the D-Norm transform based on a differential distance measure.

\begin{definition}[\textbf{D-Norm}]\label{def:d-norm} 
Let the lowest and highest $\ell 1$-Norm of $m$ groups (each group is an $n$-bit vector) be $l$ and $h$, respectively, where: 
\begin{align}\label{eqn:lAndh1}
    h \triangleq \aggregate{arg~max}{\textbf{f}_i|i\in \{1,..,m\}} (\|\ff_i\|_1)\\
    l \triangleq \aggregate{arg~min}{\textbf{f}_i|i\in \{1,..,m\}} (\|\ff_i\|_1) \label{eqn:lAndh2}
\end{align}

\noindent Now, following the general definition in Section~\ref{sec:our_approach}, the D-Norm transform is defined as: 
\begin{equation}\label{eq:DTrasform}
    \digamma = \left\{\begin{matrix}
 1, ~~h - l\geq 0~~~\text{and}~~~[h] < [l]\\
 0, ~~l - h < 0~~~\text{and}~~~[h] > [l]
\end{matrix}\right.
\end{equation}
Here, we denote the spatial index $i$ (memory address, in practice) of the vector $\ff_i$ chosen for $h$ based on equation~\eqref{eqn:lAndh1} or $l$ based on equation~\eqref{eqn:lAndh2} using a square bracket, ``$[~]$''. 

\end{definition}
\begin{figureRevsource}[!ht]
	\centering
	\includegraphics[trim=0 0 0 0,clip,width=\mylinewidth]{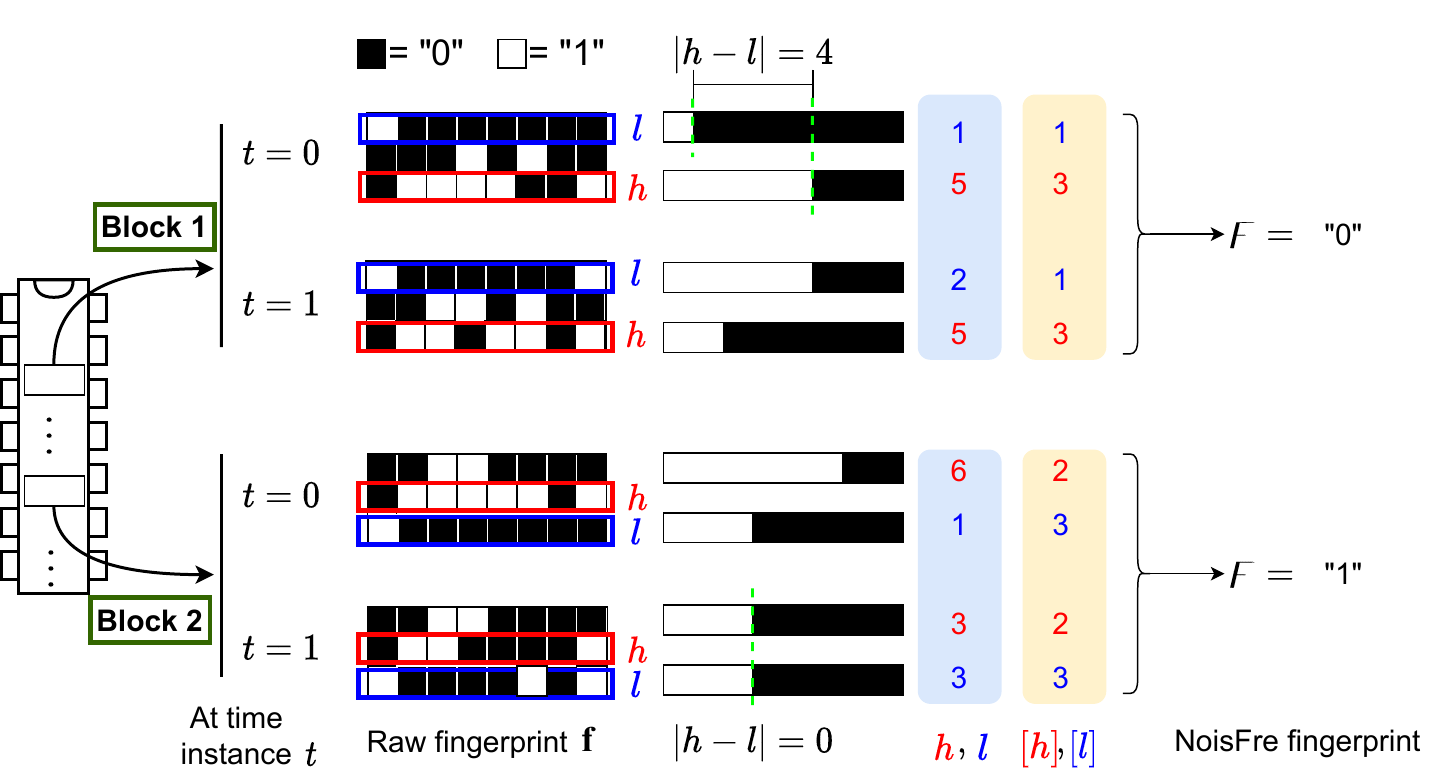}
	\caption{\Revsource{rev:D-Normfigure}{D-Norm-based NoisFre transform illustration, where $n=8$ and $m=3$. We show the results of reading out two blocks of memory (\textbf{Block 1} and \textbf{Block 2}). Here, each block ($n\times m$ bits) is formed by accessing three bytes from a byte-level addressable memory, and each block provides a new bit $\digamma$. Hence, the new bit $\digamma$ is a transformation from a block of raw bits with $m=3$ groups, where each group is an $n=8$~bit vector. The raw bit values are measured at two different time instances, $t=0$ and $t=1$, from each block to illustrate the manner in which the D-Norm transform is reliable against raw bit error.}}
	\label{fig:DNormFull}
\end{figureRevsource}

\Revsource{rev:D_norm_permutation}{The D-Norm-based transformation is illustrated in Fig.~\ref{fig:DNormFull}. In the illustration, the $\ell 1$-Norm of two blocks of $m=3$ groups of $n=8$ bit vectors are evaluated at time $t=0$.
In subsequent evaluations of the fingerprint at $t=1$:
\begin{itemize}
    \item In \textbf{Block 1}, we can observe the \textit{permutation invariance property}, similar to the S-Norm. For example, the highest $\ell 1$-Norm at $t=0$ and $t=1$ is $h=5$ for the third $8$-bit vector despite repeated generation of the raw bits not being exact.
    \item In \textbf{Block 2}, we further observe the difference of $h-l$ is $6-1=5$ at $t=0$ and shows an \textit{extreme case} of $3-3=0$ at $t=1$, where $\digamma$ bit of ``1'' remains invariant. Which reflects the \textit{combination invariance property}.
\end{itemize}
}

In both \textbf{Block 1} and \textbf{Block 2}, the fingerprint bit $\digamma$ remains robust to the raw fingerprint bit error patterns observed at different measurement times. However, a combination of $n$ bits with a single change in the number of raw binary ``1'' bits can lead a D-Norm projection at the proximity of the decision boundary in equation~\eqref{eq:DTrasform} to cross that boundary. Hence, the resulting $\digamma$ bits from such raw fingerprints effectively display low reliability. Therefore, similar to S-Norm, we propose winnowing raw fingerprints based on their $|h - l|$ projections. We describe the \textit{D-Norm-based Selection} method in Definition~\ref{def:d-norm-sel} and generalize the approach using a noise tolerance parameter $\theta$ to bound the combinations of bit patterns the transform needs to tolerate, using the differential $\ell 1$-Norm of the raw fingerprint vectors measured once (i.e., at $t=0$).

\begin{definition}[\textbf{D-Norm-based Selection}]\label{def:d-norm-sel} 
From a block of $m$ different $n$-bit raw noisy fingerprint vectors $\ff_{i}^{0}$ for $i \in \{1,\dots,m\}$ extracted at $t=0$, the block is selected for fingerprinting the device using D-Norm if $h$ and $l$ as defined in equation~\eqref{eqn:lAndh1} and equation~\eqref{eqn:lAndh2} satisfy: 
\begin{align}\label{eqn:d-norm-selection}
    |h - l|~\geq~\theta
\end{align}
\end{definition}

\begin{figure}[!hb]
	\centering
	\includegraphics[trim=0 0 0 0,clip,width=\mylinewidth]{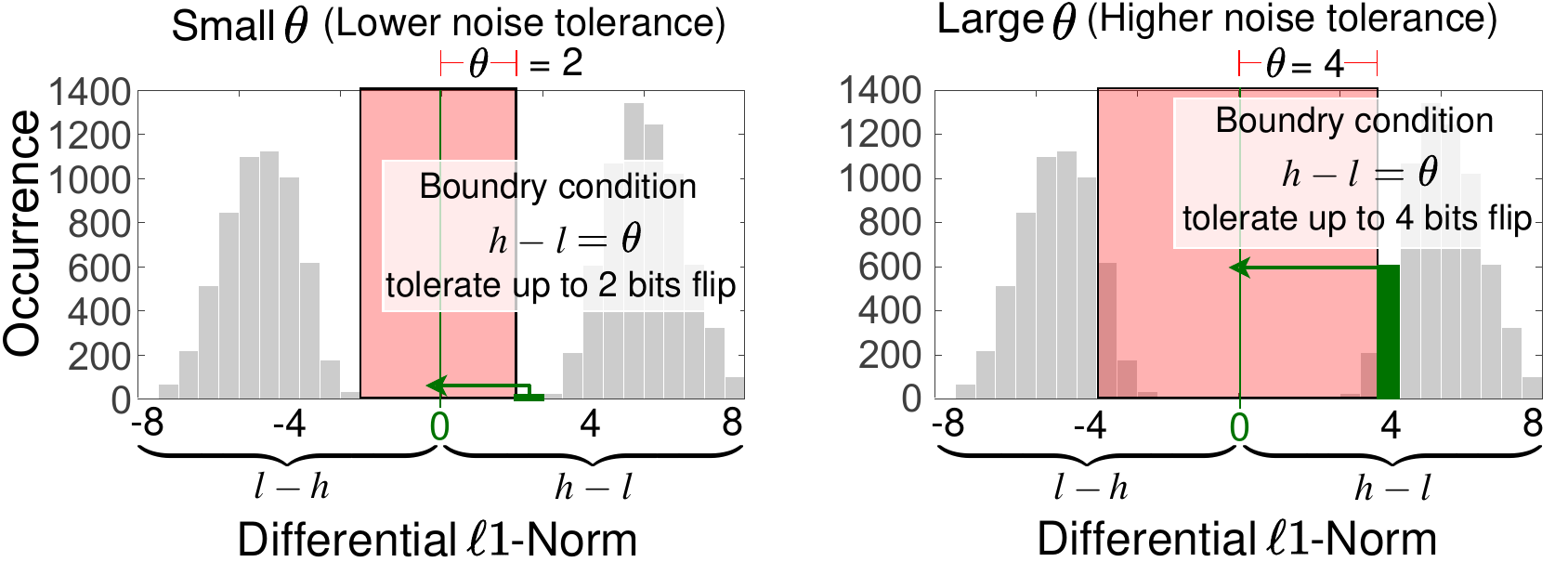}
	\caption{The role of the noise tolerance parameter $\theta$ in D-Norm-based selection. The plots depict the differential $\ell 1$-Norm distributions---the difference between the $h$ and $l$---of raw fingerprints. The distribution of differences leads to a bimodal distribution. Here, $n=16$ and $m=32$, where $n\times m$ raw bits are transformed into a 1-bit $\digamma$. If $[h]$ has a lower memory address than $[l]$, the D-Norm would be $h - l$ which is positive and plotted on the right half of the x-axis (otherwise, left). The reliability of $\digamma$ increases when the $\ell 1$-Norm difference between selected groups is away from 0.  {\it Significantly, the proportion of $\digamma$ that tolerates a chosen noise tolerance $\theta$ is greatly increased under D-Norm compared to the S-Norm}. For example, the uncovered (gray area) for D-Norm is much larger than that of S-Norm in  Fig.~\ref{fig:reliabilityDist_HW} for the same $\theta$ values.}
	\label{fig:reliabilityDist_DHW_Merge}
\end{figure}

To understand the significance of the D-Norm-based selection method and the role of the noise tolerance parameter $\theta$, we employ the Nordic Semiconductor chip fingerprint dataset used in S-Norm. The resulting distribution of enrolling measurements (at time $t=0$) for two cases of a small and a large $\theta$ for blocks of $n\times m$ raw fingerprint bits is shown in Fig.~\ref{fig:reliabilityDist_DHW_Merge}. Interestingly, the distribution of $|h-l|$ approximates a bimodal distribution; each mode represents those vectors mapping to $\digamma$ = ``1'' and ``0'', respectively.  \textit{Importantly, the two clear groupings of $n\times m$ bit blocks based on the D-Norm distance measure results in an intrinsic separation}. 

Now, consider the blocks of $h-l$ raw fingerprint bit vectors (green bar) in blocks at the boundary of the selection criteria, in equation~\eqref{eqn:d-norm-selection}, where $|h-l|~=~\theta$ for the two cases of a small and a large $\theta$. These blocks of bits represent those most likely to lead to a bit error in a transformed bit $\digamma$. When $\theta$ is small, e.g. $\theta=2$, two bit flips in the raw fingerprint  in a subsequent measurement is enough to push $|h-l|$ across the $h-l = 0$ decision boundary, defined in equation~\eqref{eq:DTrasform}, and result in a $\digamma$ bit flip. In contrast, when $\theta$ is large, e.g. $\theta = 4$, at least five raw fingerprint bit changes are required to flip the $\digamma$ bit in a subsequent evaluation. Therefore, we can expect the $\digamma$ bits selected upon a larger $\theta$ to be more reliable. 

The D-Norm method effectively sacrifices more of the available entropy ($n\times m$ raw bits are transformed into 1-bit $\digamma$) than S-Norm. However, the differential distance measure $h-l$ is bimodal and, thus, \textit{D-Norm is expected to yield a significantly higher number of noise-tolerant} $\digamma$ bits.

\section{Formalizing Performance Measures}
\label{sec:PerforMetrics}
We now formulate and derive analytical models to: i) provide an upper bound for the unreliability of noise-tolerant fingerprint bits and ii) evaluate the expected number of noise-tolerant fingerprint bits that can be extracted from each of the transform methods---the \textit{extraction efficiency}. We summarize the analytical formulations from our \textbf{detailed derivations differed to Appendix~\ref{app:derivation}} for interested readers. 
\subsection{Reliability}\label{sec:BER_F}
We employ the well-known measure of bit error rate (BER) to quantify the reliability of transformed fingerprint ${\bf F}$: 

\begin{equation}
{\rm BER}_{\bf F}=\textsf{FHD}({{\bf F}},{{\bf F}}^{\prime})
\end{equation} 

where ${\bf F}$ and ${{\bf F}}^{\prime}$ are two fingerprint measurements at distinct times from the same physical memory. The function \textsf{FHD}() is the fractional Hamming distance between the (binary) vectors  ${\bf F}$ and ${{\bf F}}^{\prime}$. Commonly, ${\bf F}$ is a reference fingerprint template measured at $t=0$, and ${{\bf F}}^{\prime}$ is the reevaluation under a potentially different device operating condition, such as temperature, and, therefore, is subject to noise. A lower ${\rm BER}_{\bf F}$ indicates higher tolerance to noise introduced from the raw fingerprints. 

\vspace{2mm}
\noindent{\bf S-Norm Reliability.~}The expected BER of noise-tolerant fingerprints for the S-Norm transformation ${\rm BER}_{\bf F}$ is formulated in equation~\eqref{eqn:HWBER}; we defer details of the derivation to {\bf Appendix~\ref{app:SnormBER}}:

\begin{dmath}
\label{eqn:HWBER}
    {\rm BER}_{{\bf F}} = \sum_{i=0}^{\floor{\frac{n}{2}}-\theta} \big( (1 - \textsf{binocdf}(\theta + i,\ceil{\frac{n}{2}}+\theta,{\rm BER}_{\textbf{f}})) \times \textsf{binopdf}(i,\floor{\frac{n}{2}}-\theta,{\rm BER}_{\textbf{f}}) \big) 
\end{dmath}

Here, \textsf{binopdf} and \textsf{binocdf} are density and cumulative density functions of a binomial distribution, respectively.  The ${\rm BER}_{\bf F}$ is a function of $\theta$, $n$ and the BER of the raw noisy fingerprint bits, ${\rm BER}_{\textbf{f}}$. If we select the worst-case ${\rm BER}_{\textbf{f}}$ of any memory chip, equation~\eqref{eqn:HWBER} provides a worst-case (upper-bound) assessment of ${\rm BER}_{\bf F}$.

\vspace{2mm}
\noindent{\bf D-Norm Reliability.~}A D-Norm transform employs a block of $m$ groups---each group with $n$ raw fingerprint bits---to be transformed into a 1-bit $\digamma$. The ${\rm BER}_{\bf F}$ of D-Norm is expressed in equation~\eqref{eqn:DHW_failure_rate}; we defer the derivation of the formula to {\bf Appendix~\ref{app:DnormBER}}: 

\begin{dmath}
\label{eqn:DHW_failure_rate}
    {\rm BER}_{{\bf F}} = \sum_{i=0}^{n-\theta} \big( (1 - \textsf{binocdf}(\theta + i-1,n+\theta,{\rm BER}_{\textbf{f}})) \times \textsf{binopdf}(i,n-\theta,{\rm BER}_{\textbf{f}}) \big) 
\end{dmath}

We can observe the reliability of transformed bits from the D-Norm to be related to  $\theta$ (the selection criterion), $n$ and BER$_{\textbf{f}}$. Again, equation~\eqref{eqn:DHW_failure_rate} provides an upper-bound estimation when the worst-case BER$_{\textbf{f}}$ is assumed. Notably, the ${\rm BER}_{{\bf F}}$ is independent of the number of groups $m$ within the block.

\begin{table*}
	\centering 
	\caption{Memory Datasets.}
			\resizebox{\textwidth}{!}{
	\begin{tabular}{c|| c || c || c || c || c || c || c || c || c || c  } %
		\toprule 
		\hline 
				
		\begin{tabular}{@{}c@{}}Manufacturer \\ Model $^1$ \end{tabular}& Abbr & \begin{tabular}{@{}c@{}} Tech \\ Node \end{tabular} & \begin{tabular}{@{}c@{}} Memory \\ Type \end{tabular} & \begin{tabular}{@{}c@{}} Memory \\ Size \end{tabular} & Quantity &  \begin{tabular}{@{}c@{}} Repeat \\ Times \end{tabular} & \begin{tabular}{@{}c@{}} Operating \\ Range$^2$ \end{tabular} & \begin{tabular}{@{}c@{}} Enrolling \\ Condition \end{tabular} & \begin{tabular}{@{}c@{}} Worst \\ Condition \end{tabular} & Worst ${\rm BER}_{\textbf{f}}$  \\ 
		\hline
		
		 \begin{tabular}{@{}c@{}} Nordic \textbf{(ours)} \\ nRF52832 \end{tabular} & NORDIC & 55~nm & SRAM & 64 KiB & 12 + 88$^6$ & 100 & $-15$-$80\celsius$ & $25\celsius$ & $80\celsius$ & 6.09\%  \\	\hline	
		 
		\begin{tabular}{@{}c@{}} ISSI \cite{guo2017scare,rahman2017systematic} \\ IS61WV25616BLL  \end{tabular} & ISSI & 110~nm & SRAM & 256 KiB & 4 & 30 & $25$-$80\celsius$ & $25\celsius$ & $80\celsius$ & 8.29\%  \\  \hline

		 \begin{tabular}{@{}c@{}} IDT \cite{guo2017scare,rahman2017systematic} \\ IDT71V416S  \end{tabular} & IDT & 130~nm & SRAM & 512 KiB & 6 & 50 & $25$-$80\celsius$ & $25\celsius$ & $80\celsius$ & 5.42\%  \\\hline
		 
		 \begin{tabular}{@{}c@{}} Winbond \cite{guo2017ffd} \\ W29N02GV  \end{tabular} & Flash & 46~nm & FLASH & \begin{tabular}{@{}c@{}} 69,696 Bytes/\\256 MiB$^3$  \end{tabular}  & 7 & 99 & \begin{tabular}{@{}c@{}} 0-100,000 \\ P/E Cycles  \end{tabular}
		   & \begin{tabular}{@{}c@{}} 0th P/E \\ Cycle  \end{tabular} & \begin{tabular}{@{}c@{}} after 100,000th  \\ P/E Cycles$^4$ \end{tabular} & 16.26\% \\
		   \hline

		 \begin{tabular}{@{}c@{}} Microchip \textbf{(ours)} \\ 24LC256  \end{tabular} & EEPROM & 350~nm & EEPROM & \begin{tabular}{@{}c@{}} 2 KiB/\\32 KiB$^5$ \end{tabular}  & 2 & 100 & $14$-$80\celsius$ & $14\celsius$ & $80\celsius$ & 16.37\%  \\ \hline
		 
		\bottomrule
	\end{tabular}
			}
	\label{tab:chipSpec}
		\begin{tablenotes}[flushleft]
		\footnotesize
		\item{$^1$The NORDIC and EEPROM datasets we collected will be released, remaining public datasets are from \href{https://www.trust-hub.org/data}{\underline{https://www.trust-hub.org/data}}.}
		
		\item{$^2$Notably these public datasets focus on room temperature and high-temperature evaluations. Other operating corners are incomplete.}
		
		\item $^3$The tested Flash memory size in the public dataset is 69,696 bytes, while the total memory size is 256~MiB.
		
		\item $^4$Experimental studies demonstrate that the ${\rm BER}_{\textbf{f}}$ of Flash memory is mainly affected by the programming/erase (P/E) cycles, equivalent to wear-out or aging, but negligibly affected by voltage and temperature~\cite{guo2017ffd}. The maximum endurance is 100,000 according to the datasheet. 
		
		\item $^5$The EEPROM chip has 32~KiB capacity, while the first 2~KiB memory is evaluated here.
	   
	   \item $^6$There are 12 chips with three corner measurements $\{-15,25,80\}\celsius$ and 88 chips with a single $25\celsius$ corner measurement.
	    \end{tablenotes}
\end{table*}

\subsection{Extraction Efficiency} \label{sec:ext_eff}
We define extraction efficiency $\eta$ as the number of obtainable transformed bits, $\digamma$,  \textit{subject to a given noise-tolerance} $\theta$, from the total number of available memory bits expressed in KiB. 

\vspace{2mm}
\noindent{\bf S-Norm Extraction Efficiency.~}The extraction efficiency of S-Norm can be expressed as below; the detailed derivation is deferred to \textbf{Appendix~\ref{app:SnormEf}}:

\begin{dmath}
    \label{eqn:HW_SR}
    {\eta}_{\rm SNorm} = \frac{1}{n} \times \big(1 - \textsf{binocdf}(\floor{\frac{n}{2}}+\theta,n,0.5) + \textsf{binocdf}(\ceil{\frac{n}{2}} - \theta - 1,n,0.5)\big)\times (1024 \times 8)
\end{dmath}

Here, the term $1-\textsf{binocdf}(\floor{\frac{n}{2}}+\theta,n,0.5)$ expresses the case when the $\ell 1 $-Norm of an $n$-bit \textbf{f} is larger than the selection threshold $\floor{\frac{n}{2}}+\theta$, assuming that the probability of each bit being ``1''/``0'' is 50\%. While the term $\textsf{binocdf}(\ceil{\frac{n}{2}} - \theta - 1,n,0.5)$ formulates the alternative case when the $\ell 1 $-Norm of a $n$-bit \textbf{f} is less than or equal to $\ceil{\frac{n}{2}} - \theta - 1$. Both cases comprise vectors that satisfy the selection criterion in equation~\eqref{eqn:s-norm-selection}. We can see that the overall extraction efficiency should be the sum of the above two cases divided by $n$---recall that $n$ raw bits transform into a 1-bit $\digamma$. The $1024 \times 8$ term expresses the extraction efficiency as bit/KiB---number of selected reliable bits $ \digamma$ out of 1 KiB memory.

\vspace{2mm}
\noindent{\bf D-Norm Extraction Efficiency.~}A D-Norm transform obtains a 1-bit $\digamma$ from a block of $m$, $n$-bit raw fingerprint vectors. We define the probability that a given block will meet the selection criterion ($|~h - l ~|\ge \theta$) in equation~\eqref{eqn:d-norm-selection} as $P_{\rm DNorm}^{\rm select}$ (recall that we refer to the lowest $\ell 1$-Norm as $l$, and the highest $\ell 1$-Norm as $h$, out of all $m$ groups within a block). The direct derivation of $P_{\rm DNorm}^{\rm select}$ is non-trivial. Instead, we use a different but equivalent problem and defer the details to \textbf{Appendix~\ref{app:DnormEf}}. We formulate the extraction efficiency of D-Norm as:

\begin{equation}\label{eqn:DHW_SR}
    {\eta}_{\rm DNorm} = \frac{1}{n \times m} \times P_{\rm DNorm}^{\rm select} \times (1024 \times 8)
\end{equation}

Here, the term of $\frac{1}{n\times m}$ expresses $n\times m$ raw bits producing a single $\digamma$ bit, while the $1024\times 8$ constant facilities express the result in terms of bits/KiB of memory. Given the complexity of formulating ${\eta}_{\rm DNorm}$, the fitness of the formalized expression is further validated through running extensive numerical experiments (defined in Section~\ref{Sec:validation}), with the results detailed in  Fig.~\ref{fig:DHWn32} in {\bf Appendix}.

\subsection{Summary}
Our formulation of \textit{reliability} allows a security practitioner to evaluate, for a given transform, suitable transform parameters (e.g., the number of bits $n$ to employ in a raw fingerprint vector \f and noise tolerance parameter $\theta$) for extracting new fingerprint ${\bf F}$. The extracted ${\bf F}$ will have an expected worst-case error bound given by ${\rm BER}_{\bf F}$. Then, the $\eta$ yields the total number of such noise-tolerant bits ${\rm BER}_{\bf F}$ that can be extracted from a given memory size. 

\section{Experimental Validations}\label{Sec:validation}
For comprehensively evaluating NoisFre we used 119 commodity chips consisting of three memory types pervasive in COTS devices, especially in low-end IoT devices and extensive simulation based experiments with billions of bit generations to overcome the practical hurdle of demonstrating extremely low BER and key failure rates with physical chips. In the following:
\begin{itemize}
    \item We validate our analytical models for \textit{reliability} and \textit{extraction efficiency}.
    \item We we assess the performance of the noise-tolerant fingerprints by evaluating the uniqueness and uniformity of ${\bf F}$. 
\end{itemize}

\subsection{Evaluation Approaches}\label{sec:evaApproach}
We consider three evaluation approaches described below.

\vspace{1mm}
\noindent {\bf Predictions (Analytical model).} In this evaluation, we use the analytical models formalized in Section~\ref{sec:PerforMetrics} to predict extraction efficiency and the BER$_{\bf F}$ of the transformed fingerprints. 

\vspace{1mm}
\noindent {\bf Simulations (Synthetic chip model).} To evaluate the reliability of the transformed fingerprint, a massive number of repeated measurements and the management of the data for analysis are required. For example, if we want to validate whether a 128-bit NoisFre enabled key can achieve a failure rate of $10^{-6}$ as done in Section~\ref{sec:key-gen-eval}, the ${\rm BER}_{\bf F}$ needs to be no more than $7.81 \times 10^{-9}$. To test this with physical measurements, approximately $10^{8}$ repeated measurements are required from the same chip instance. Such a measurement process would take more than nine years and generates roughly 6 Terabyte (TB) of data---this is merely for one 64-KiB chip. Such a massive testing regime is impractical, as detailed in \textbf{Measurement (Physical chips)}. Instead, we employed a synthetic memory chip model (detailed in \textbf{Appendix~\ref{sec:apx_Random_chip}}). The model follows the \textit{physical unclonable function (PUF) response model} summarized in~\cite{maes2013accurate} and assumes each bit to have a binomial probability of a bit flip across repeated measurements based on employing a \textit{worst-case} BER  measured from a physical SRAM chip (see~\autoref{tab:chipSpec}) as the binomial probability parameter value for ${\bf p}$. Using the synthetic chip model, for instance, 100 million ($10^8$) times of simulations can be completed in approximately  53~hours or 2.2 days using a laptop equipped with quad-core Intel Core i7-10510U CPU and 16 Gigabyte (GiB) RAM. The synthetic chip, models bit errors and when applied with the worst-case BER is sufficient for evaluating reliability and extraction efficiency. Therefore, we employ the data from the simulated measurements to determine $\eta$ and the ${\rm BER}_{\bf F}$.

\vspace{1mm}
\noindent {\bf Measurements (Physical chips).~}\Revsource{rev:processvariation}{Performing massive testing on physical chips is impractical. For example, obtaining 100 repeated measurements from an nRF52832 physical chip (in the NORDIC dataset) takes four minutes and 45 seconds, and this generates 6.25 Megabyte (MiB) of data (the SRAM memory size of the single chip is 64~KiB). Then, we can estimate that 100 million ($10^8$) repeated measurements for a \textit{single} physical chip under a \textit{single} operating corner will take 3,298.6 days (or nine years) and generate 5.96~TB of data. Therefore, we confirm extraction efficiency and the transformed fingerprints' BER  validated using the synthetic chip model with a limited number of \textit{repeated} physical chip measurements. However, we dedicate the physical chip measurements across a large batch of 100 chips to evaluate the quality of the transformed bits because properties such as fingerprint uniformity and uniqueness are affected by fabrication variations not incorporated in the synthetic chip model used for simulations. The datasets we used are described in \textit{Physical chips--Fingerprint Datasets} below.}

\vspace{1mm}
\noindent\textbf{Physical chips--Fingerprint Datasets.~}\label{sec:mem-dataset}
Specifications of: i) three SRAM; ii) one Flash memory; and iii) one EEPROM datasets are summarized in Table~\ref{tab:chipSpec} and described in detail in Appendix.~\ref{app:memory_dataset}. Each dataset is obtained from chips from a \textit{different manufacturer}. Further, the datasets describe multiple repeated measurements of raw fingerprint bits under {\it each} operating condition---see the operating range in Table~\ref{tab:chipSpec}.

We use the NORDIC dataset for extensive validations, considering the fact that it is collected with the broadest operating range and highest number of repeated measurements (100 repeated measurements). In addition, we use the remaining datasets to corroborate the generality of our approaches. When we evaluate the BER of transformed bits, ${\rm BER}_{\bf F}$, we report the average from repeated evaluations. Notably, the enrolled reference template is only based on the first \textit{single measurement}.

\subsection{Validating Extraction Efficiency and Bit Error Rate}

We employ simulations with the synthetic chip model to conduct the necessary massive number of repeated measurements to assess the reliability of transformed fingerprint ${\bf F}$. To generate the results, for each parameter combination (i.e., $n$, $\theta$ in S-Norm and $m$, $n$, and $\theta$ in D-Norm) of a NoisFre transform in Fig.~\ref{fig:BER_S} for S-Norm and Fig.~\ref{fig:BER_D} for D-Norm, we simulated \textit{one~million} repeated measurements using a synthetic chip with a memory capacity of up to 16 MiB\footnote{We start with a memory size of 64~KiB, the SRAM capacity of the NORDIC chip, but we double the size when a 128-bit ${\bf F}$ cannot be obtained. Thus, for each parameter setting, at least 128 bits are ensured to be produced.} 

\subsubsection{Using Simulations}
\label{Sec:UsingSimulations}

\begin{figureRevsource}[!ht]
	\centering
	\includegraphics[trim=0 0 0 0,clip,width=\mylinewidth]{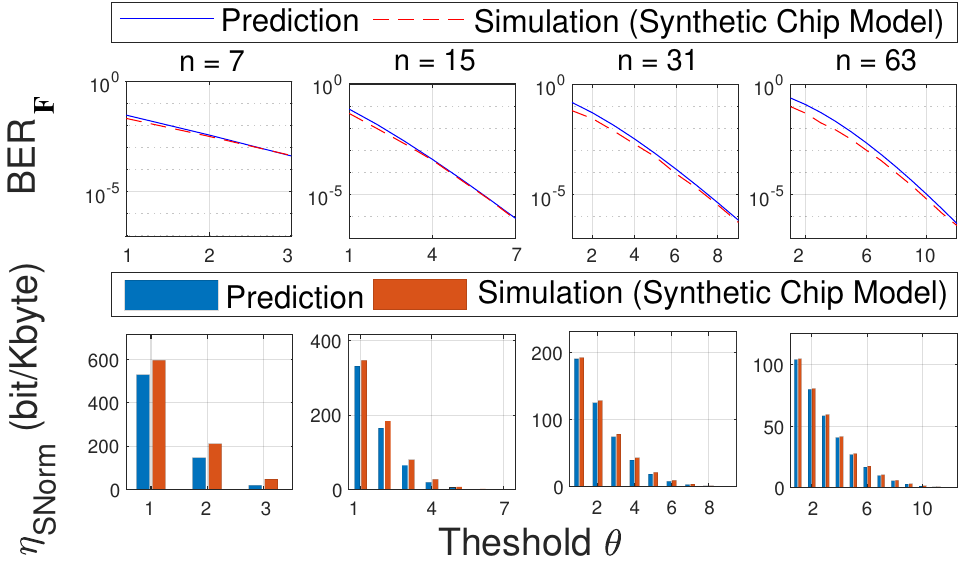}
	\caption{S-Norm ${\rm BER}_{\bf F}$ and extraction efficiency $\eta_{\rm SNorm}$ validation using the synthetic chip  model constructed based on the NORDIC chip dataset. The evaluation is conducted for different $n$ and noise-tolerance parameter $\theta$, where $\theta$ ranges from $1$ to $n/2$. Here, $n$ raw bits are transformed into one $\digamma$ bit. 
}
	\label{fig:BER_S}
\end{figureRevsource}

\noindent\Revsource{rev:S_model_test1}{\noindent\textbf{S-Norm.~}The evaluation results from S-Norm are detailed in Fig.~\ref{fig:BER_S} under various $n$ and $\theta$ settings. Based on Fig.~\ref{fig:BER_S}, we can confirm that the formalization of BER$_{{\bf F}}$ in equation~\eqref{eqn:HWBER} provides a conservative estimation of the selected $\digamma$ bits. The results for extraction efficiency are in good agreement with equation~\eqref{eqn:HW_SR} used to predict the number of $\digamma$ bits that can be expected from a given chip.}
\vspace{2mm}

\Revsource{rev:D_model_test1}{\noindent\noindent\textbf{D-Norm.~}The validation results of D-Norm are shown in Fig.~\ref{fig:BER_D}. As expected, the ${\rm BER}_{\bf F}$ plotted in Fig.~\ref{fig:BER_D} reduces substantially as the $\theta$ is increased. Again, we can confirm that the formalized ${\rm BER}_{\bf F}$ in equation~\eqref{eqn:DHW_failure_rate} is a conservative estimate because it is always shown to be higher than the synthetic chip model results. Further, the extraction efficiency derived in equation~\eqref{eqn:DHW_SR} provides an accurate prediction of the number of bits of ${\bf F}$ that can be expected from a given chip under various D-Norm settings ($n$, $m$, and $\theta$).}

\begin{figureRevsource}[!ht]
	\centering
	\includegraphics[trim=0 0 0 0,clip,width=\mylinewidth]{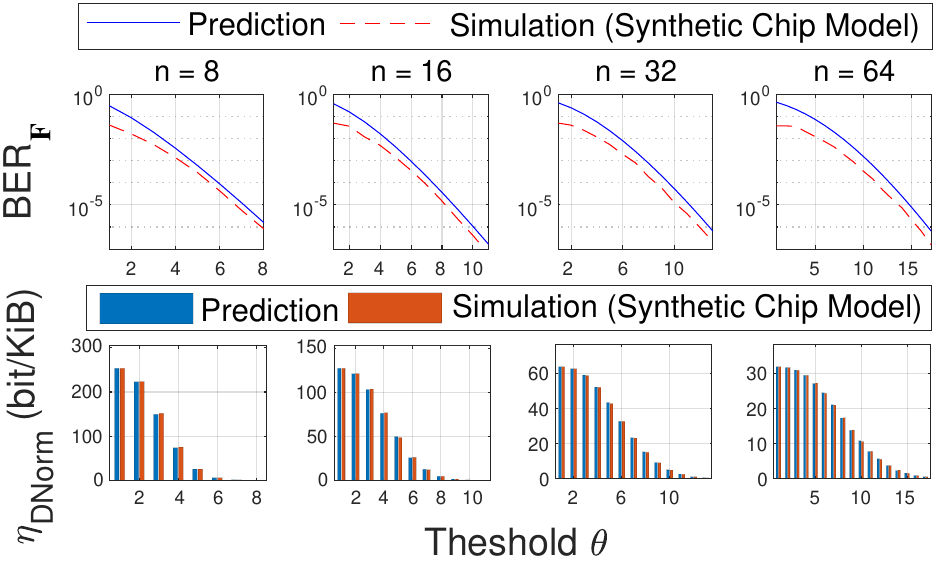}
	\caption{D-Norm ${\rm BER}_{\bf F}$ and extraction efficiency $\eta_{\rm DNorm}$ validation on the synthetic chip model constructed based on the NORDIC chip dataset. \Revsource{rev:DNorm_transformed}{The evaluation is conducted under different $n$ and noise-tolerance parameter $\theta$ where $\theta$ ranges from $1$ to $n$ and $m=4$. Here $n\times m$ raw bits are \Rev{transformed} into one $\digamma$ bit.}}
	\label{fig:BER_D}
\end{figureRevsource}

\begin{figure*}[tp]
	\centering
	\includegraphics[trim=0 0 0 0,clip,width=\textwidth]{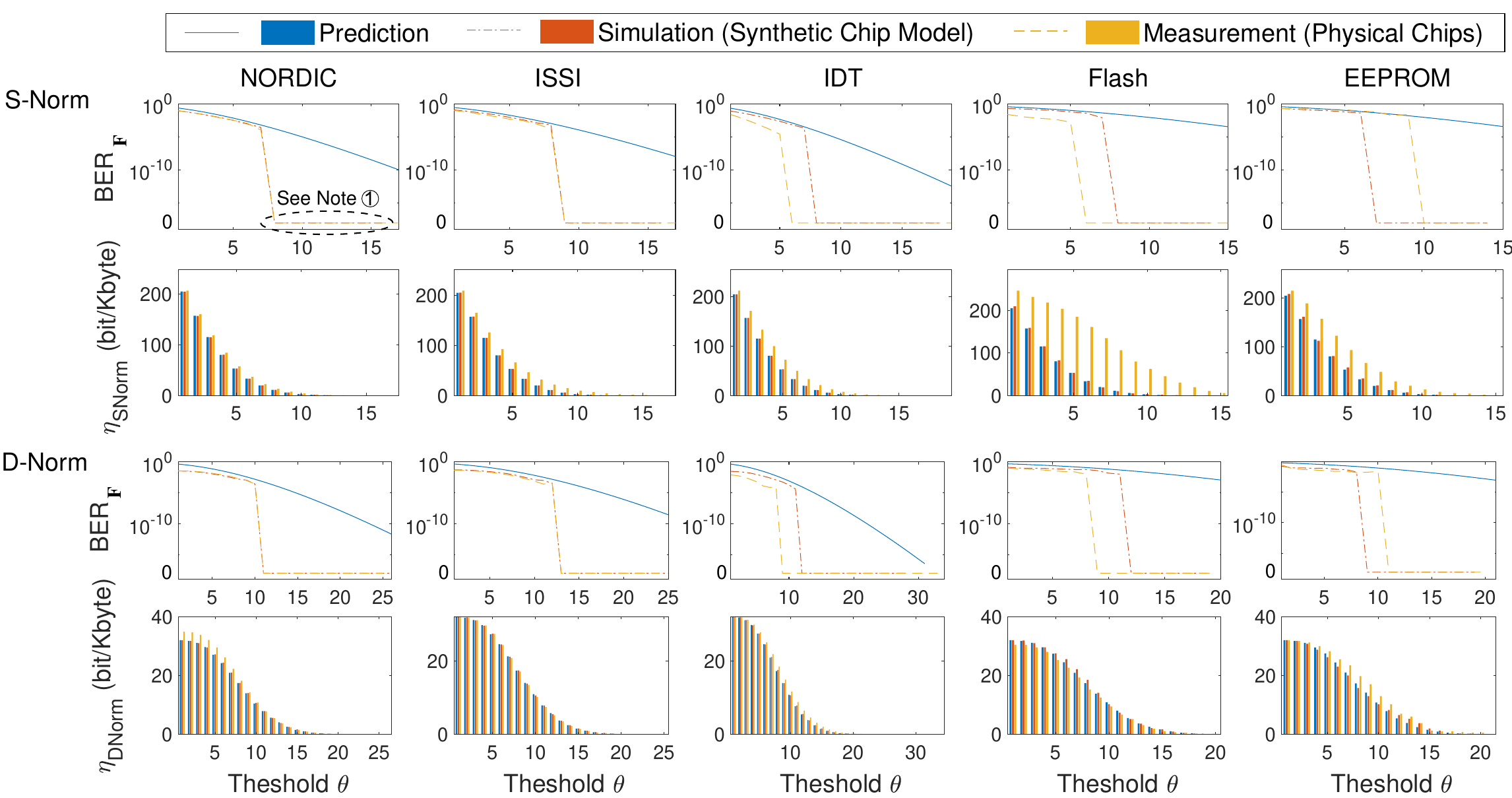}
	\caption{S-Norm and D-Norm validation on the datasets including three types of SRAM memories from three different manufacturers (NORDIC, ISSI, and IDT), one type of Flash memory, and one type of EEPROM memory with S-Norm setting ($n=63$) and D-Norm settings ($n=64$, $m=4$). Note \circled{1}: the number of repeated measurements is finite (see \autoref{tab:chipSpec}) and inadequate to demonstrate any errors when the expected ${\rm BER}_{{\bf F}}$ is considerably less than $1/(\text{number of repeated measurements})$.}
	\label{fig:BER_SpD_gener}
\end{figure*}

\subsubsection{Using Measurements} \label{sec:generalize_on_phy_chips}
Three SRAM datasets (our NORDIC, public ISSI and IDT datasets); one Flash dataset; and our EEPROM dataset are used to validate the generality of the NoisFre approach based on \textit{physical chip measurements}. The results of S-Norm and D-Norm validated on these five datasets are detailed in Fig.~\ref{fig:BER_SpD_gener}. In contrast to our simulation-based study in Section~\ref{Sec:UsingSimulations}, here we use a synthetic chip of identical data capacity and worst-case BER$_{f}$ matching the physical chip under investigation and simulate 100 repeated measurements in sympathy with the physical measurement regime.\footnote{One hundred measurements are due to the impracticality of conducting the necessary number of repeated measurements with physical chips, as detailed in Section~\ref{sec:evaApproach}} Overall, we can observe from the plots in Fig.~\ref{fig:BER_SpD_gener} that simulations with the synthetic chip model agree well with the measurements for both S-Norm and D-Norm. 

\textit{Based on our comprehensive experimental validations on SRAM memories from three different manufacturers, Flash memories, and EEPROM memories, we can now conclude that our formalized model of the unreliability, ${\rm BER}_{{\bf F}}$, and extraction efficiency, $\eta$, in Section~\ref{sec:PerforMetrics} are indeed reliable measures. Most importantly, the formalized models serve as bounds for ${\rm BER}_{{\bf F}}$ and $\eta$ in practice; the measurement results and synthetic chip model results are the same or better than those predicted by the analytical models.}

\subsection{Evaluating Uniformity and Uniqueness} \label{sec:quality_quality_of_F_bits}
\Revsource{rev:threemetrics}{In addition to the two crucial performance measures we formulated, reliability and extraction efficiency, we further consider measures that evaluate other qualities of the transformed bits in terms of \textit{uniqueness} and \textit{uniformity} (see~\cite{maes2012physically} for a definition of these measures). In the following, our evaluations are based on the measurements obtained from the 100 physical chips in the \textit{augmented NORDIC dataset}.}

\begin{figureRevsource}[!ht]
	\centering
	\includegraphics[width=\mylinewidth]{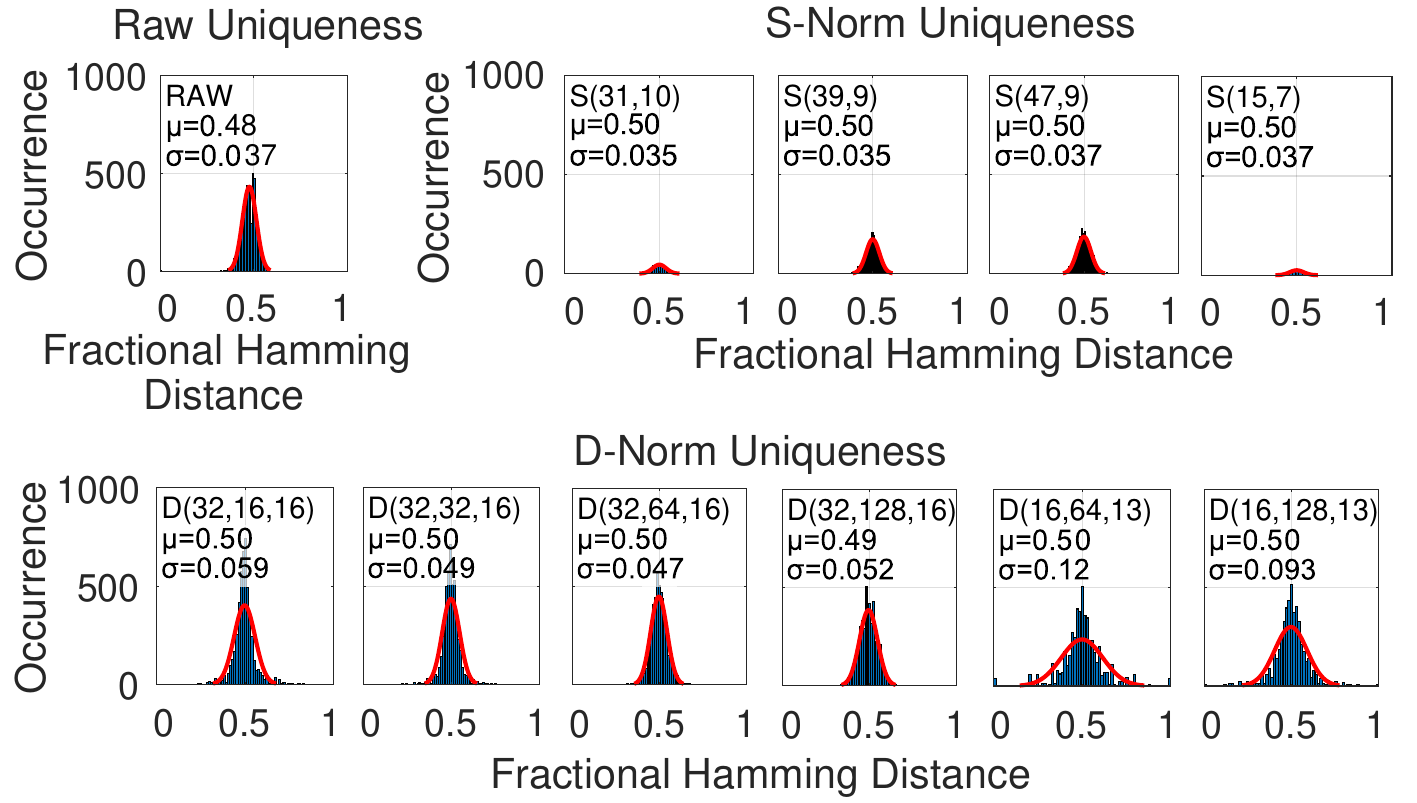}
	\caption{Uniqueness evaluation using physical nRF52832 chips. The plots summarize the mean ($\mu$) and standard deviation ($\sigma$) across a subset of S-Norm parameters, S(n, $\theta$), and D-Norm parameters, D(n, m, $\theta$) applied to our large dataset of 100 chips.}
	\label{fig:Uniqueness}
\end{figureRevsource}

\vspace{2mm}
\noindent\textbf{Uniqueness Evaluation.~}\Revsource{rev:uniqueness}{Essentially, uniqueness measures how different the fingerprints are between devices. However, the formal  definition of uniqueness based on fractional Hamming distance~\footnote{\footnotesize Fractional Hamming distance (FHD) is a distance measure  between two vectors of equal length, defined as the number of positions in the two vectors with different values, normalized by the vector length.}}~\cite{maes2012physically} cannot be directly applied for ${\bf F}$ fingerprints because the transformed  bits are generated from different physical memory blocks from chip to chip, and the number of such bits obtained could also vary from chip to chip. To account for this, we propose evaluating the uniqueness of S-Norm and D-Norm transformed fingerprints based on the following approach:
\begin{enumerate}
    \item Given a set of transformation parameters (such as $n, m$, and $\theta$ for D-Norm), we extract all the $\digamma$ bits for each of the $N$ (i.e., 100) devices.
    \item Given a pair of devices out of $\binom{N}{2}$, we identify the device which produces the ${\bf F}$ string with the smaller number of $\digamma$ bits within this pair, and truncate the longer bit string to the same length. Then, we calculate  the fractional inter-chip Hamming distance for this pair.
    \item We repeat the process in Step 2) for all the $\binom{N}{2}$ pairs to obtain the uniqueness measurement over the 100-chip dataset.
\end{enumerate}
The uniqueness of raw fingerprints, S-Norm fingerprints, and D-Norm fingerprints is illustrated in Fig.~\ref{fig:Uniqueness}. The mean uniqueness of the raw fingerprints is 0.48, with a standard derivation of 0.037. For S-Norm, the mean uniqueness achieves the ideal value of 0.50 under all tested settings, and the largest standard derivation is 0.037 under the setting of ($n=15$ and $\theta=7$) and ($n=47$ and $\theta=9$). The mean uniqueness of the D-Norm fingerprints also exhibits the ideal value of 0.50, except for settings ($n=32$, $m=128$,  $\theta=16$), but the mean uniqueness of 0.49 is still nearly the ideal value. In general, as the number of extracted fingerprints in a tested sample decreases, we also observe an increase in standard deviation; this is expected because of the resulting small sample size for statistical analysis.

\vspace{2mm}
\noindent\textbf{Uniformity (Bias) Evaluation.~}\Revsource{rev:uniformity}{Uniformity measures the balance between zeros and ones in a fingerprint vector. The uniformity distribution of raw fingerprints, S-Norm fingerprints and D-Norm fingerprints is illustrated in Fig.~\ref{fig:Uniformity}. The uniformity of the raw fingerprint is very close to the ideal value of 0.5, with a very small standard deviation. The uniformity of S-Norm and D-Norm methods---across the various parameter settings---is close to the ideal value, albeit with a slight bias toward ``1''. Notably, such slight biases are acceptable for key derivation and can be simply compensated by using a few more fingerprint bits when deriving a key~\cite{maes2016secure}.}

\begin{figureRevsource}[!ht]
	\centering
	\includegraphics[width=\mylinewidth]{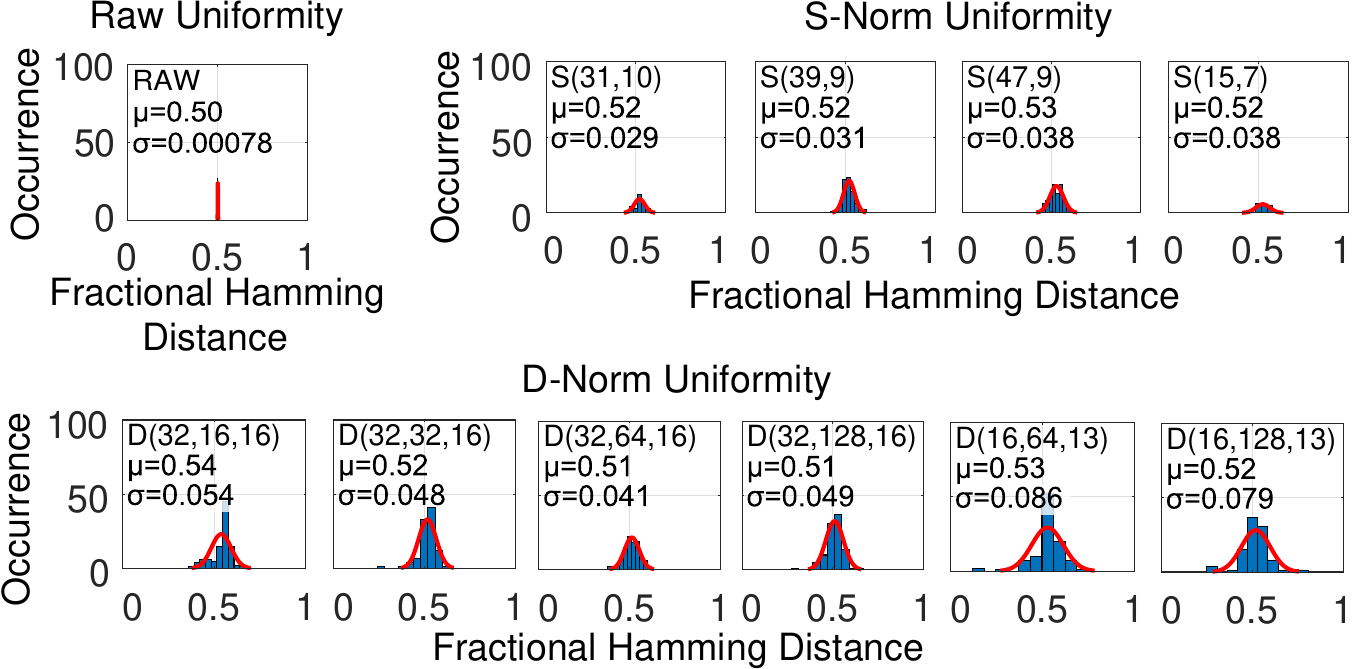}
	\caption{Uniformity evaluation using physical nRF52832 chips. The plots show the mean ($\mu$) and standard deviation ($\sigma$) for uniformity across a subset of S-Norm parameters, S(n, $\theta$), and D-Norm parameters, D(n, m, $\theta$) applied to our large dataset of 100 chips.}
	\label{fig:Uniformity}
\end{figureRevsource}

\section{Deriving Cryptographic Keys for Security Functions}\label{Sec:keyDerivation}
We demonstrate the \textit{expressive power of our formalization} by investigating the derivation of root keys from commodity memory chips facilitated by our analytical models. The dynamic and direct generation of cryptographic keys from memory fingerprint transformations into noise-tolerant bits is a basis for building security functions because: i) memory biometrics is a true source of randomness and ii) it removes the need for a protected non-volatile memory---keys can be generated on-demand and ``\textit{forgotten}" after usage. 

\Revsource{rev:keyDerivation}{In the following sections, we elaborate on a method for employing the new fingerprint ${\bf F}$ obtained from the NoisFre transformation to realize a cryptographic key generator (Section~\ref{sec:secure_key_generator}) and evaluate the practical realization of such a key generator (Section~\ref{sec:key-gen-eval}); we defer the security analysis of the key generation process to Section~\ref{sec:security_analysis}.}

\subsection{A Method for Realizing a NoisFre Key Generator} \label{sec:secure_key_generator}

  \Revsource{rev:secure_key_generator}{A typical memory fingerprint-based key generation method involves two steps: i)~a one-time secure key enrollment on the server-side and ii)~on-demand secure key regeneration on the device-side~\cite{maes2012pufky,herder2014physical,gao2018lightweight,su2019secucode}. 
  Positions of transformed bits $ \digamma$ should be provisioned during the key enrollment phase and provided during the key regeneration phase. We refer to these positions using a \textbf{mask}. Recall that we have referred to those raw bits that produce a 1-bit $\digamma$ as a block. For the S-Norm, one block has $n$ raw bits, while one block has $n\times m$ raw bits in the D-Norm; for both methods, $n$ raw bits form one $\ell 1$-Norm. In the discussion that follows, we consider key generation under two practical settings: 
\begin{itemize}
    \item Devices with write once read many (WORM) memory for storage of the $\textbf{mask}$ defined to select the memory regions to be used in the NoisFre transform prior to deployment. 
    \item Devices without WORM memory where the $\textbf{mask}$ has to be transmitted, for example, through a wireless communication channel. 
\end{itemize}}

\subsubsection{On-Server Secure Key Enrollment} \label{sec:secure_key_enrollment}
\Revsource{rev:secure_key_enrollment1}{First, we describe the one-time secure key enrollment process, depicted in Fig.~\ref{fig:NoisFre_enrollment}. This process is performed in a secure environment by the server. }
\renewcommand{\mylinewidth}{0.83\linewidth} 
\begin{figureRevsource}[!ht]
	\centering
	\includegraphics[width=\mylinewidth]{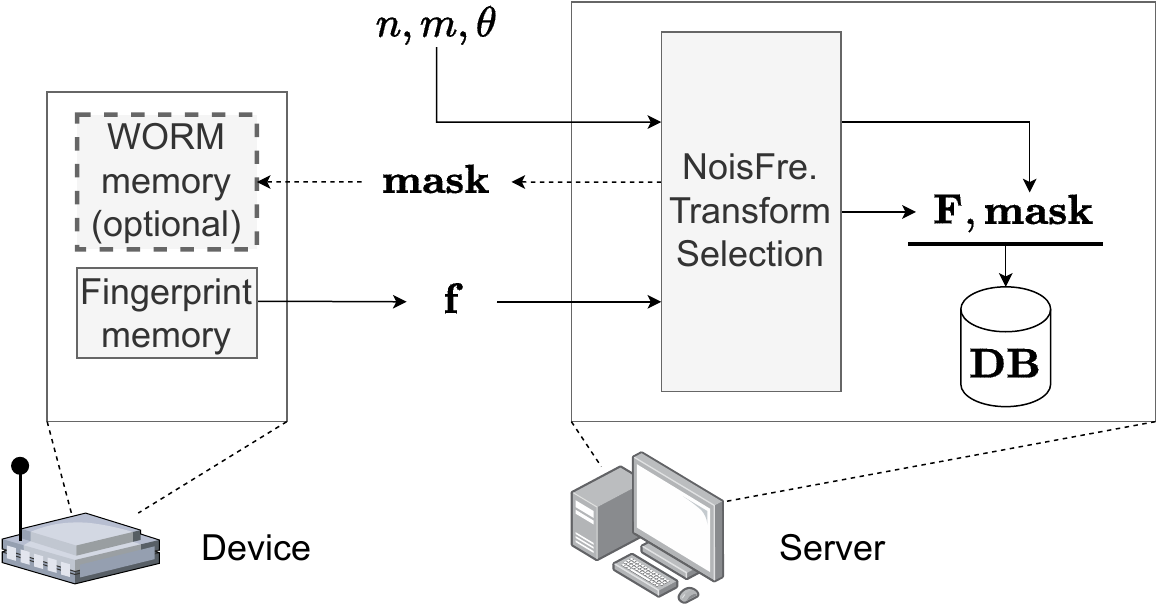}
	\caption{NoisFre secure key enrollment process. The raw fingerprints are extracted and used in the NoisFre Transform Selection process (as defined in equations \eqref{eq:DTrasform} and \eqref{eqn:d-norm-selection} for D-Norm) to determine the selected memory addresses (\textbf{mask})
	for subsequent use in security functions. Although we have only focused on the generation of a single mask, several masks may be defined to allow the server to subsequently generate different secret keys on-demand.}
	\label{fig:NoisFre_enrollment}
\end{figureRevsource}
\renewcommand{\mylinewidth}{\linewidth}

\Revsource{rev:secure_key_enrollment2}{\noindent\textit{Protocol.} The one-off on-server secure key enrollment protocol with NoisFre is as follows:
\begin{enumerate}
    \item Fingerprint memory is a memory region from which the raw device memory fingerprint $\textbf{f}$ is extracted.
    \item The raw fingerprint $\textbf{f}$ is processed by the server. The \textsf{NoisFre Transform Selection} process determines a noise-tolerant fingerprint vector ${\bf F}$ and the corresponding $\textbf{mask}$ based on the parameters $n$, $m$, and $\theta$ determined by a security practitioner. Notably, a practitioner can employ the analytical expressions derived in Section~\ref{sec:PerforMetrics} to determine the appropriate parameter values.
    \item Both ${{\bf F}}$ and the $\textbf{mask}$ are stored in the server's secure database ($\textbf{DB}$, indexed by, for example, the device identification number [$\textbf{id}$], although not explicitly shown here for simplicity).
    \item \textit{Optionally}, the $\textbf{mask}$ can be stored inside the device's WORM memory.
\end{enumerate}}

\subsubsection{Dynamic On-Device Secure Key Generation} \label{sec:secure_key_generation}
\Revsource{rev:secure_key_generation1}{Now, we consider the realization of on-device secure key generation with a device memory fingerprint biometric. We illustrate the key generation method in Fig.~\ref{fig:NoisFre_generation}.}

\vspace{3mm}
\noindent\textit{Protocol.} The dynamic on-device secure key generation protocol with NoisFre is as follows:
\begin{enumerate}
    \item If the device implements WORM memory to store the $\textbf{mask}$, as in Fig.~\ref{fig:NoisFre_generation}~(a), the server fetches device-specific information from the $\textbf{DB}$, such as the enrolled ${\bf F}$.
    \item If the device does not implement WORM memory, the server fetches device-specific information from the $\textbf{DB}$, such as the enrolled ${\bf F}$ and $\textbf{mask}$, as in Fig.~\ref{fig:NoisFre_generation}~(b). The $\textbf{mask}$ is transferred from the server to the device over a (non-secure) wireless communication channel. To ensure the integrity of the $\textbf{mask}$, a message authentication code (MAC) tag is computed by the server as $\textbf{tag} \leftarrow \textsf{MAC}_{{\bf F}}(\textbf{mask})$ and appended to the $\textbf{mask}$.
    \item The device dynamically generates a new noisy raw fingerprint $\textbf{f}'$ from the fingerprint memory.
    \item The device computes ${\bf F} \leftarrow \textsf{NoisFre.Transform}(\textbf{f}', \allowbreak \textbf{mask})$, where $\textsf{NoisFre.Transform}()$ is a function defined by, for example, the D-Norm transform in equation~\eqref{eq:DTrasform}.
    \item If the device does not implement a WORM memory, then the \textbf{mask} is sent by the server, as shown in Fig.~\ref{fig:NoisFre_generation}~(b); the device computes $\textbf{tag}' \leftarrow \textsf{MAC}_{{\bf F}}(\textbf{mask})$. To check the integrity of the mask, the $\textbf{tag}'$ is compared to the $\textbf{tag}$ supplied by the server. If the two values match, output the NoisFre fingerprint ${\bf F}$; otherwise, output $\bot$.
    \item Now both the server and the device share the same highly reliable ${\bf F}$ to be used as a shared secret in a security function.
\end{enumerate}

\renewcommand{\mylinewidth}{0.90\linewidth} 
\begin{figureRevsource}[!ht]
	\centering
	\includegraphics[width=\mylinewidth]{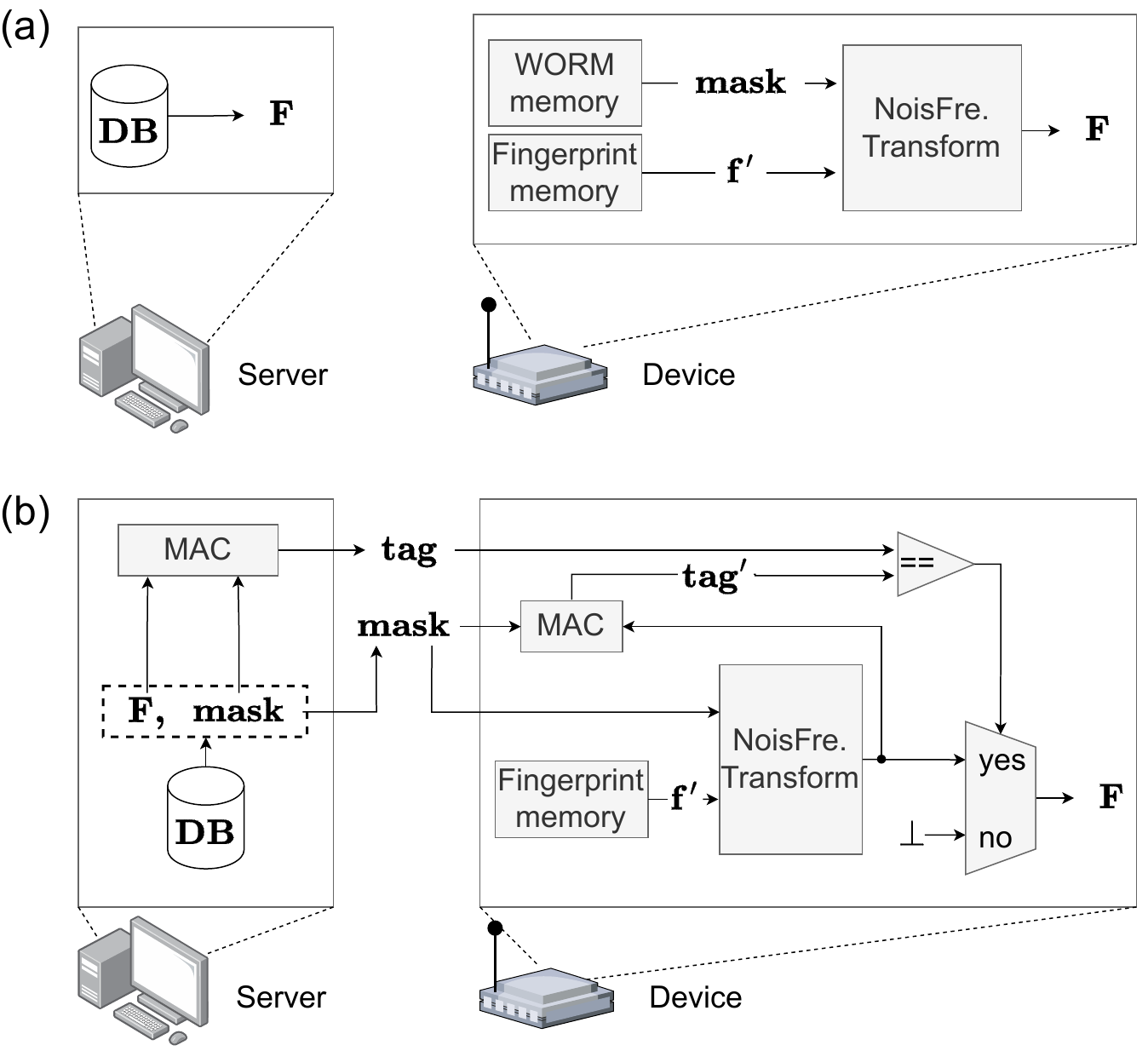}
	\caption{On-device NoisFre key generation: (a) the $\textbf{mask}$ is stored in a device's WORM memory; (b) the $\textbf{mask}$ is supplied by the server over the wireless communication channel, if there is no WORM memory available on-device, for example. Although the illustration shows the production of one ${\bf F}$-based key using one $\textbf{mask}$, it is possible to enroll and generate several keys if desired.}
	\label{fig:NoisFre_generation}
\end{figureRevsource}
\renewcommand{\mylinewidth}{\linewidth}

\subsection{Evaluations}\label{sec:key-gen-eval}
We begin our systematic evaluation of cryptographic key generation with the following question and employ the formal models and the physical chip measurements for our evaluations.

\begin{mdframed}[backgroundcolor=black!10,rightline=false,leftline=false,topline=false,bottomline=false,roundcorner=2mm]
	What is the reliability of a $k$-bit NoisFre fingerprint ${\bf F}$? 
\end{mdframed}

Transformed fingerprint $ {\bf F}$ can be directly utilized as a cryptographic key because they are invariant to a desirably high number of noise-induced bit error patterns---these  $\digamma$ bits exhibit a high noise tolerance. The overall failure rate $P_{\bf F}^{\rm fail}$ of a $k$-bit noise-tolerant key ${\bf F}$ can be expressed as: 

\begin{equation}
    \label{eqn:key_failure_rate}
    P_{\bf F}^{\rm fail} = 1-(1-{\rm BER}_{\bf F})^k
\end{equation}

Recall that the formalized ${\rm BER}_{\bf F}$ in Section~\ref{sec:BER_F} is conservative. Therefore, the $P_{\bf F}^{\rm fail}$ in equation~\eqref{eqn:key_failure_rate} will also yield a conservative estimation. We expect a key failure rate in practice to be lower than our prediction here. This hypothesis is validated with an extensive simulation-based on a large simulated chip with up to \textit{one billion} repeated noise-tolerant key bit extraction, as illustrated in Fig.~\ref{fig:PKF_validation}.

\begin{figureRevsource}[!ht]
	\centering
	\includegraphics[width=\mylinewidth]{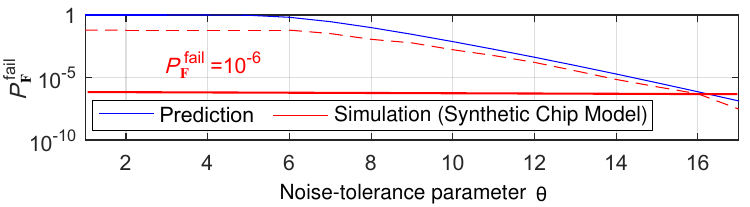}
	\caption{Validation of equation~\eqref{eqn:key_failure_rate}. The simulated chips are based on the worst-case ${\rm BER}_{\bf f}=6.09\%$ from the NORDIC chip set. The parameters selected are $n = 32$, $m = 16$ 
	and varied D-norm parameter $\theta$ from $1,\dots,17$. We conducted 10 \textit{million} re-evaluations of a 128-bit noise-tolerant fingerprint for each value of $\theta=1,\dots,16$ and one \textit{billion} evaluations for $\theta = 17$. Our results corroborates equation~\eqref{eqn:DHW_failure_rate} and equation~\eqref{eqn:key_failure_rate} as an upper bound on the failure rate of a NoisFre fingerprint employed as a cryptographic key. An even lower $P_{\bf F}^{\rm fail}$ is achievable if a larger $\theta$ is used. We halted our investigation at $\theta$=17 as it answers the question we investigated.}
	\label{fig:PKF_validation}
\end{figureRevsource}

\Revsource{rev:P_key_fail_target}{Next, considering a practitioner's desire for a $P_{\bf F}^{\rm fail}<10^{-6}$ performance target}\footnote{Notably, there is nothing fundamentally preventing us from aiming for a lower key failure rate. We can see from Fig.~\ref{fig:PKF_validation} that a larger $\theta$ will achieve a lower failure probability.} for typical industrial applications, as highlighted in~\cite{maes2015secure} and recent studies~\cite{maes2013physically,herder2014physical,herder2016trapdoor,becker2017robust,hiller2016cherry,wang2018efficient}, we investigate the following question. 

\begin{mdframed}[backgroundcolor=black!10,rightline=false,leftline=false,topline=false,bottomline=false,roundcorner=2mm]
	What is the most efficient transformation method presenting the highest extraction efficiency while ensuring sufficient reliability for ${\bf F}$ to be {\it direct} use as a 128-bit cryptographic key with a failure rate lower than $10^{-6}$ under \textit{worst-case} raw fingerprint BER$_{\rm f}$?
\end{mdframed}
\vspace{-2mm}

We employ NORDIC SRAM-based synthetic data
to facilitate the massive number of evaluations necessary to address the question. The evaluation process is described below: 

\begin{figure}[!ht]
	\centering
	\includegraphics[width=\mylinewidth]{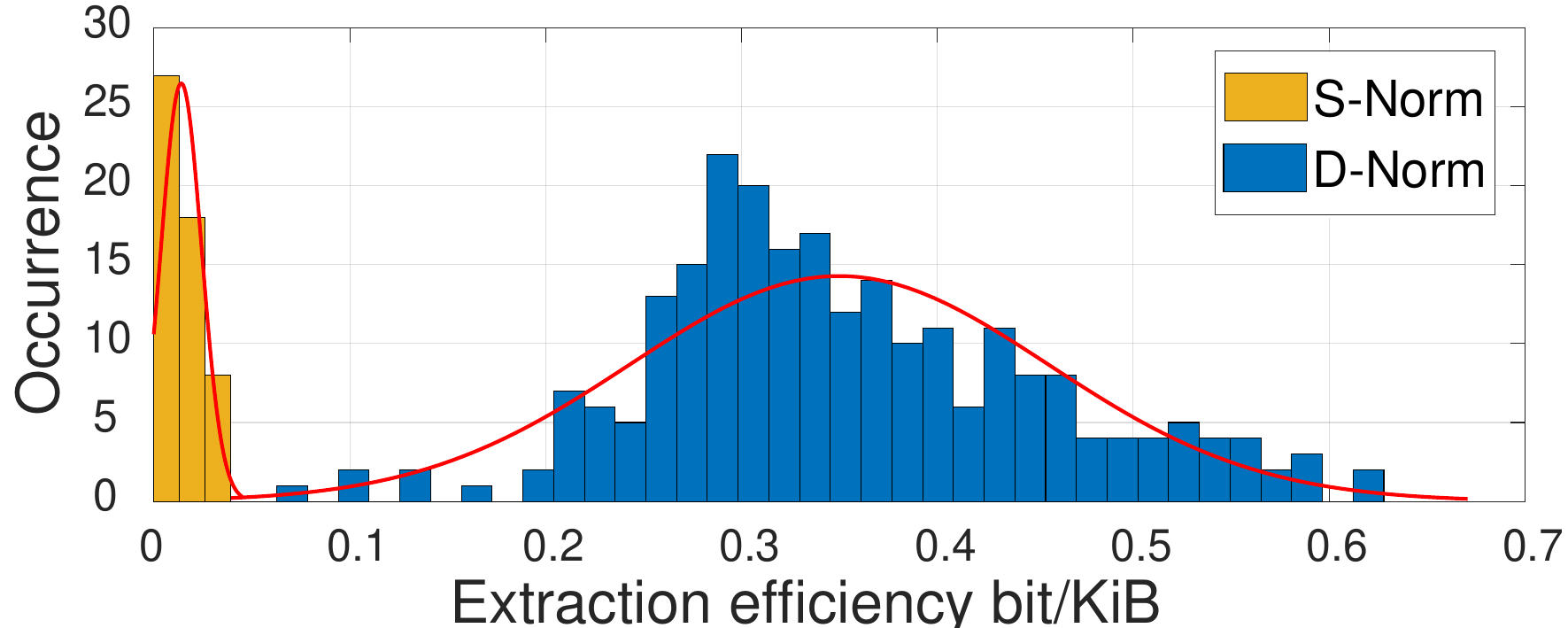}
	\caption{Extraction efficiency comparison between D-Norm and S-Norm.
	For S-Norm and D-Norm extractions, we have evaluated 16,384 and 8,388,480 parameters combinations, respectively. The $\theta$ of D-Norm may take any value in [$1,n$], while in S-Norm, the $\theta$ is restricted within [$1,n/2$]. Meanwhile, D-Norm extraction employs the additional parameter, $m$. Therefore, the possible combinations of parameters for D-Norm are magnitudes larger than that of S-Norm.}
	\label{fig:ExtEff_DvsS}
\end{figure}

\begin{enumerate}
    \item We determine the ${\rm BER}_{{\bf F}}$ corresponding to a 128-bit key with a failure rate of $10^{-6}$ using  equation~\eqref{eqn:key_failure_rate}. The resulting ${\rm BER}_{{\bf F}}$ is $7.81 \times 10^{-9}$.
    \item For each $n \in \{1,2,3,...,256\}$, evaluate the minimum $\theta$ for the required ${\rm BER}_{\bf F}$ (equation~\eqref{eqn:HWBER} for S-Norm and equation~\eqref{eqn:DHW_failure_rate} for D-Norm) to ensure ${\rm BER}_{{\bf F}} < 7.81 \times 10^{-9}$. In these equations, we employ the mean of the worst-case BER$_{\rm f} = 6.09\%$ of NORDIC dataset to compute an upper bound for  ${\rm BER}_{{\bf F}}$. 
    \item For S-Norm, the extraction efficiency $\eta$ is calculated with equation~\eqref{eqn:HW_SR} using the $n$ and $\theta$ determined in the previous step. 
    \item D-Norm requires us to further determine the $m$ value that can provide the highest $\eta$. As observed in Fig.~\ref{fig:DHWn32} in the Appendix, $\eta$ changes smoothly with respect to $m$. To reduce the search-time overhead, we applied a grid-based search technique: i)~evenly select $j$ sample points from the entire domain of $m \in \{1, 256\}$. ii)~calculate the $\eta$ for each $m=1,2,3,..,j$; iii)~find the $m$ values corresponding to the highest and the second highest $\eta$; iv)~refine the search domain to be between the two points found in step iii); and v)~repeat from i) to iv) to locate the $m$ that gives the highest $\eta$. 
\end{enumerate}

The results from our investigation are depicted in Fig.~\ref{fig:ExtEff_DvsS}; here, we plot the occurrences of extraction efficiency as a function of $\eta$ from all the combinations of S-Norm parameters ($n$ and $\theta$) and  D-Norm parameters( $n$, $\theta$ and $m$). We can conclude that the D-Norm always affords significantly higher extraction efficiencies conditioned on the 128-bit $P_{\bf F}^{\rm fail}<10^{-6}$ constraint. {\it Therefore, in the following discussion, we focus on the D-Norm.}

Given: i)~different sizes of memories embedded within various COTS electronics and ii)~${\rm BER}_{\textbf{f}}$ characteristics of noisy fingerprints from different memory technologies: 

\begin{mdframed}[backgroundcolor=black!10,rightline=false,leftline=false,topline=false,bottomline=false,roundcorner=2mm]
	What is the \textit{lowest key failure rate} $P_{\bf F}^{\rm fail}$ achievable for a 128-bit key ${\bf F}$ from \textit{each} memory technology and manufacturer considered in our study? 
\end{mdframed}

This scenario resembles a practical application setting where the computing platform or micro-controller unit, for example, needs to be selected based on meeting security and application requirements. We can assume that inherent (worst-case) ${\rm BER}_{\textbf{f}}$ of raw fingerprints are known (i.e., published measurement studies on memory technologies). Thus, we test our suite of memory technologies using the following approach:
\begin{enumerate}
    \item For each memory dataset listed in Table~\ref{tab:PkeySingleChip}, we conduct an exhaustive parameter search using possible combinations of D-Norm parameters ($n,m \in [1, 128]$, and $\theta \in [1,n]$) using our analytical models. This step identifies the ($n$, $m$, $\theta$) combination exhibiting the lowest $P_{\bf F}^{\rm fail}$ while still providing least 128-bit ${\bf F}$. 
    \item We employ the formulated equation~\eqref{eqn:DHW_failure_rate} to obtain the ${\rm BER}_{\bf F}$ of the extracted $ {\bf F}$ using the identified $m$, $n$, $\theta$ and the mean of ${\rm BER}_{\textbf{f}}$ characterized across the chips in a given memory type dataset. 
    \item We use ${\rm BER}_{\bf F}$ substituted into equation~\eqref{eqn:key_failure_rate} to determine the best $P_{\bf F}^{\rm fail}$ of the selected and transformed  ${\bf F}$ with at least 128 bits.
\end{enumerate}

\begin{table}[ht]
	\centering 
	\caption{The lowest key failure rate $P_{\bf F}^{\rm fail}$ achievable for obtaining a 128-bit key ${\bf F}$ for \textit{each} investigated memory dataset using D-Norm. Here Mem. size is the abbreviation for Memory size.  Worst case BER$_{\rm f}$ is the mean of the value calculated across the chips in a given dataset. Notably, as described in Section~\ref{sec:key-gen-eval} and illustrated in Fig.~\ref{fig:PKF_validation}, equation \eqref{eqn:key_failure_rate} provides a conservative upper bound, the actual key failure rates will be much lower in practice.} 
	\label{tab:PkeySingleChip} 
	\resizebox{\mylinewidth}{!}{
	\begin{tabular}{ c| c |c | c | c | c | c } %
		\toprule 
		\toprule 
				
		 \begin{tabular}{@{}c@{}} Dataset \\ (Type)  \end{tabular} & \begin{tabular}{@{}c@{}} worst case \\ ${\rm BER}_{\textbf{f}}$ (mean) \end{tabular}& \begin{tabular}{@{}c@{}} Mem. \\ size \end{tabular} & n & \begin{tabular}{@{}c@{}} m \end{tabular} &  $\theta$ & \begin{tabular}{@{}c@{}}  $P_{\bf F}^{\rm fail}$\\ equation~\eqref{eqn:key_failure_rate} \end{tabular} \\ 
		\midrule
	    \hline
	    
	    \begin{tabular}{@{}c@{}} NORDIC \\ (SRAM)  \end{tabular} 
	       &\multirow{1}*{6.09\%} 
	        &  \multirow{1}*{64 KiB} 
	            &  \multirow{1}*{29}
	                & \multirow{1}*{65}
                        & 13 & $\mystrut 4.04\times 10^{-5}$ \\ \hline

	     \begin{tabular}{@{}c@{}} ISSI \\ (SRAM)  \end{tabular}
	       &\multirow{1}*{8.29\%} 	    
	        &  \multirow{1}*{256 KiB}
	            &  \multirow{1}*{50}
	                & \multirow{1}*{128}
                        & 19    & $\mystrut3.56\times 10^{-5}$ \\ \hline
        
	     \begin{tabular}{@{}c@{}} IDT \\ (SRAM)  \end{tabular} 
	       &\multirow{1}*{5.42\%} 	    
	        &  \multirow{1}*{512 KiB}
	            &  \multirow{1}*{83}
	                & \multirow{1}*{128}
                        & 25    & $\mystrut5.29\times 10^{-9}$ \\ \hline
        
	    \begin{tabular}{@{}c@{}} Winbond \\ (Flash)  \end{tabular}
	       &\multirow{1}*{16.26\%} 	    
	        &  \multirow{1}*{256 MiB $^1$}
	            &  \multirow{1}*{120}
	                & \multirow{1}*{128}
                        & 41    & $\mystrut2.52 \times 10^{-4}$ \\ \hline

        \begin{tabular}{@{}c@{}} Microchip \\ (EEPROM)  \end{tabular} 
	       &\multirow{1}*{16.37\%}         
	        &  \multirow{1}*{32 KiB $^1$}
	            &  \multirow{1}*{14}
	                & \multirow{1}*{61}
                        & 9    & $\mystrut4.01\times 10^{-1}$ \\ \hline

		\bottomrule
	\end{tabular}
    }
	 \begin{tablenotes}[flushleft]
      \footnotesize
      \vspace{1mm}
      
      \item $^1$Recall that the tested size of Flash and EEPROM memory are 69~KiB and 2~KiB. When calculating the number of selected noise-tolerant bits, the memory sizes are scaled up by assuming the entire 256~MiB Flash memory and 32~KiB EEPROM memory are available for fingerprinting.

    \end{tablenotes}
\end{table}

Results are summarized in Table~\ref{tab:PkeySingleChip}. Taking the expected BER$_{\rm f}$ across the smallest SRAM dataset, the lowest $P_{\bf F}^{\rm fail}$ expected from a chip with SRAM capacity of 64~KiB is in the magnitude of $10^{-5}$. Notably, $P_{\bf F}^{\rm fail}$ reported in Table~\ref{tab:PkeySingleChip} is conservatively estimated from formulations. In practice, $P_{\bf F}^{\rm fail}$ is expected to be much better. Importantly, with more abundant and freely available on-chip SRAM, represented in the IDT dataset, a remarkably low key failure rate of $5.29\times 10^{-9}$ is achievable.
 
As expected, the higher worst-case BER$_{\rm f}$ of the EEPROM and Flash datasets  implies that the techniques in NoisFre are not able to select a 128-bit $\bf F$ with a satisfactory $P_{\bf F}^{\rm fail}$. However, the Flash memory tested benefits from a high memory capacity (256~MiB compared to 32~KiB for EEPROM) and  we can achieve orders-of-magnitude better $P_{\bf F}^{\rm fail}$ than EEPROM. 

\Revsource{rev:SRAM_prevalent}{In summary, for SRAM---the most prevalent memory type in IoT devices---a 128-bit key with a key failure rate less than $10^{-6}$ can be efficiently obtained given \Rev{an adequate SRAM} memory capacity.} However, for memory types exhibiting severely high ${\rm BER}_{\textbf{f}}$, for example, EEPROM and Flash, the method itself is insufficient to gain a satisfactory $P_{\bf F}^{\rm fail}$. Although, NoisFre does significantly reduce the key failure rate given the higher capacity of Flash memory for selecting bits. Notably, with such high BER$_{\rm f}$ memory characteristics, even the state-of-the-art, efficient method of RFE-based key generators are unlikely to deliver a computationally tractable solution on resource limited devices. We discuss this limitation further in Section~\ref{sec:limitations}.

\section{Security Function Implementation for Comparison}\label{Sec:CaseStudy}
Here, we describe a case study implementing a NoisFre-based key generator followed by performance and implementation overhead comparisons against the lightweight, state-of-the-art (R)FE-based method.
\begin{figure}[!ht]
	\centering
	\includegraphics[width=\mylinewidth]{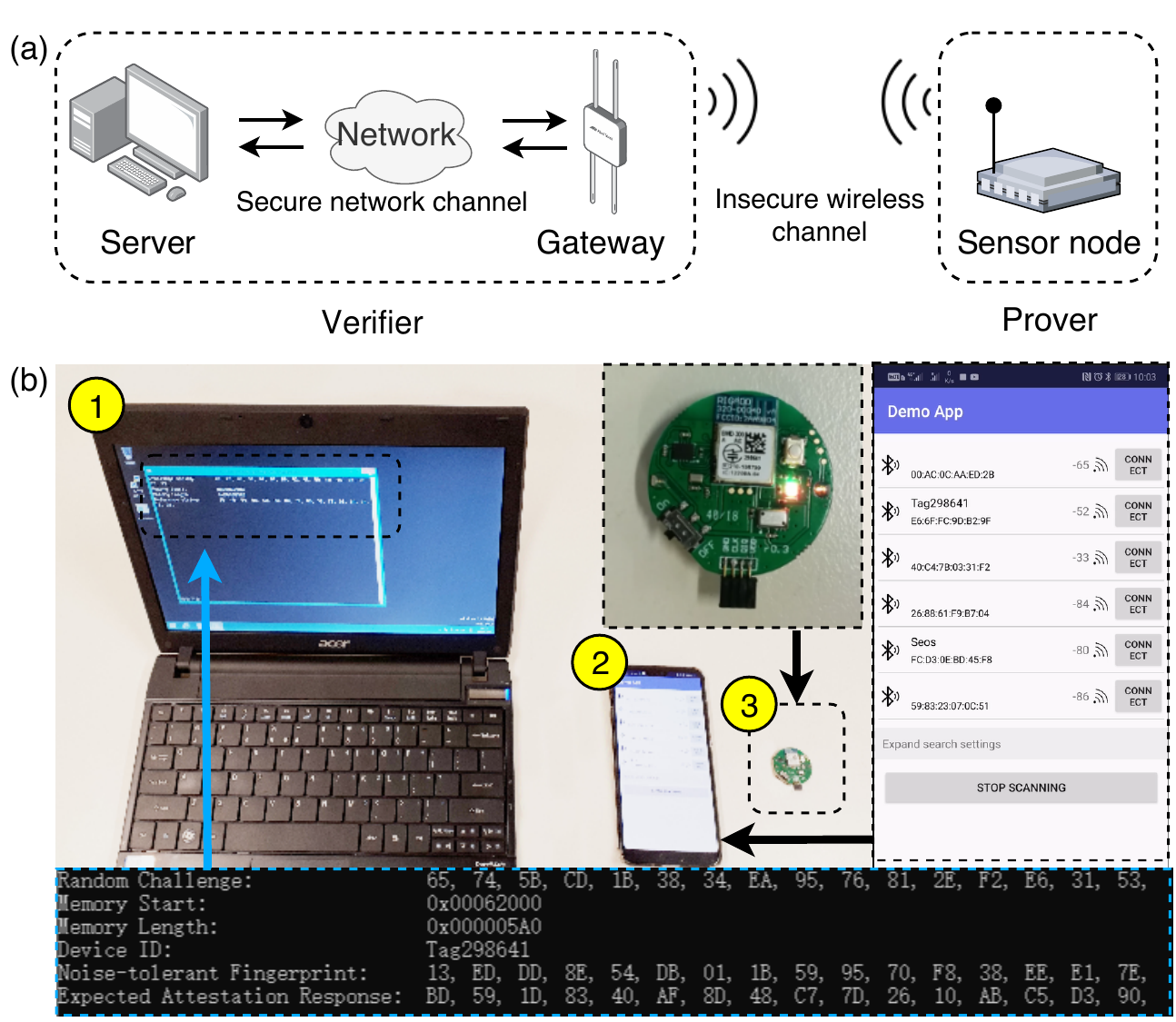}
	\caption{(a) System overview. (b) Experiment setup: Verifier consists of \circled{1} a laptop as the cloud server and \circled{2} a smartphone as the gateway; device is \circled{3} a commercial widely used nRF52832 Bluetooth-LE sensor. See the demo video for more details \href{https://youtu.be/O5NWZw-swpw}{\underline{https://youtu.be/O5NWZw-swpw}}.}
	\label{fig:Overview}
\end{figure}

\subsection{An Overview}
\begin{figure}[!ht]
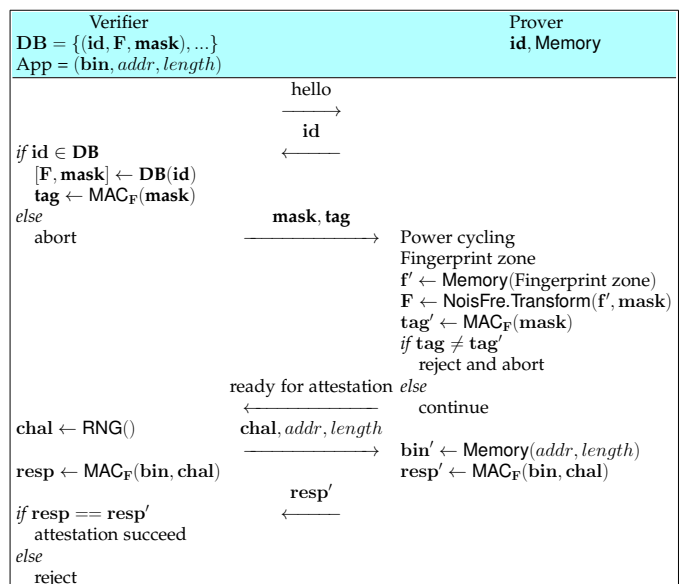

\centering
\resizebox{\mylinewidth}{!}{
    \fbox{$
        \Huge
        \begin{array}{lcl}
        \rowcolor{LightCyan}
        \mc{\Huge \text{Verifier}} & & \mc{\Huge \text{Prover}} \\
        \rowcolor{LightCyan}{\bf DB} =  \{({\bf id},{\bf F}, \textbf{mask}),...\} & & \qquad \qquad \qquad \textbf{id},\textsf{Memory}\\
        \rowcolor{LightCyan}\text{App = }({\bf bin},{addr},{length}) & & \\
        
        \cmidrule{1-3}
        & \text{hello} & \\
        & \xrightarrow{~~~~~~~~~~} &\\
        & {\bf id} &\\
        \textit{if } {\bf id} \in \textbf{DB} & \xleftarrow{~~~~~~~~~~} & \\
        \quad [{\bf F}, \textbf{mask}] \leftarrow \textbf{DB}(\textbf{id})\\
        \quad \textbf{tag} \leftarrow \textsf{MAC}_{{\bf F}}(\textbf{mask})& & \\
        \textit{else}& \textbf{mask}, \textbf{tag} & \\
        \quad \text{abort}& \xrightarrow{~~~~~~~~~~~~~~~~~~~~~~~~~} & \text{Power cycling }\\
        & & \text{Fingerprint zone}\\
        & & {\bf f}' \leftarrow \textsf{Memory}(\text{Fingerprint zone})\\
        & & {\bf F} \leftarrow \textsf{NoisFre.Transform}({\bf f}', {\bf mask})\\
        & & {\bf tag}' \leftarrow \textsf{MAC}_{{\bf F}}({\bf mask})\\
        & &\textit{if}~{\bf tag} \neq {\bf tag}'\\
        &  & \quad \text{reject and abort}\\
        & \text{ready for attestation} &\textit{else} \\
        & \xleftarrow{~~~~~~~~~~~~~~~~~~~~~~~~~} & \quad \text{continue} \\
        {\bf chal} \leftarrow \textsf{RNG}() & {\bf chal}, {addr}, {length} & \\
        & \xrightarrow{~~~~~~~~~~~~~~~~~~~~~~~~~} & {\bf bin}' \leftarrow \textsf{Memory}({addr},{length})\\
        {\bf resp} \leftarrow \textsf{MAC}_{{\bf F}}({\bf bin},{\bf chal}) &  & {\bf resp}' \leftarrow \textsf{MAC}_{{\bf F}}({\bf bin},{\bf chal})\\
        & {\bf resp}' & \\
        \textit{if } {\bf resp} == {\bf resp}' & \xleftarrow{~~~~~~~~~~} & \\
        \quad \text{attestation succeed}\\
        \textit{else}\\
        \quad \text{reject}
\end{array}
    $}
}
\caption{Remote attestation protocol, with mask transmitted from the Verifier, In case the WORM is not supported on the Prover device.}
\label{fig:Protocol_transmit}
\end{figure}

The entities, a Verifier and a Prover, involved in this case study are illustrated in Fig.~\ref{fig:Overview} (a). The Verifier consists of a server and a wireless network gateway (smartphone). The Prover refers to a wireless sensor node (Bluetooth sensor). In this setup, the server functions as a coordinator, holds the enrolled Prover's information in the database, and issues commands to instruct the Prover to perform remote attestation. The gateway bridges the communication between the server and the Prover. The traffic between the server and the gateway is assumed to be secure by applying standard security protection mechanisms. The Prover, communicating wirelessly, is deployed in an (insecure) environment. Details of the corresponding attestation protocol are provided in Fig.~\ref{fig:Protocol_transmit}. Our case study aims to:
\begin{itemize}
    \item Implement a lightweight remote attestation routine suitable for a Prover with a constrained resource by following~\cite{dinu2019sia}.
    \item Experimentally demonstrate that \textit{SRAM fingerprints can be accessed on-demand} and at \textit{run-time} by exploiting the low-cost micro controller unit (MCU)'s memory power control features---SRAM regions are arranged in blocks can be individually powered \textit{on} or \textit{off}.
\end{itemize}

\begin{figure}[!ht]
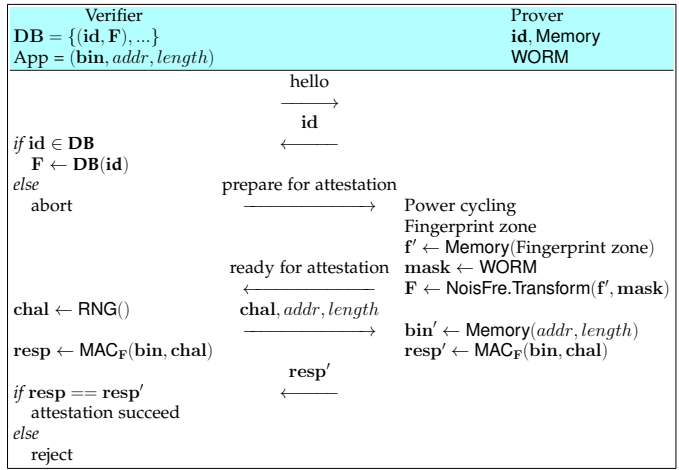

\centering
\resizebox{\mylinewidth}{!}{
    \fbox{$
        \Huge
        \begin{array}{lcl}
        \rowcolor{LightCyan}
        \mc{\Huge \text{Verifier}} & & \mc{\Huge \text{Prover}} \\
        \rowcolor{LightCyan}{\bf DB} =  \{({\bf id},{\bf F}),...\} & & \qquad \qquad \qquad \textbf{id},\textsf{Memory}\\
        \rowcolor{LightCyan}\text{App = }({\bf bin},{addr},{length}) & & \qquad \qquad \qquad \textsf{WORM}\\
        
        \cmidrule{1-3}
        & \text{hello} & \\
        & \xrightarrow{~~~~~~~~~~} &\\
        & {\bf id} &\\
        \textit{if } {\bf id} \in \textbf{DB} & \xleftarrow{~~~~~~~~~~} & \\
        \quad {\bf F} \leftarrow \textbf{DB}(\textbf{id})\\
        \textit{else}& \text{prepare for attestation} & \\
        \quad \text{abort}& \xrightarrow{~~~~~~~~~~~~~~~~~~~~~~~~~} & \text{Power cycling }\\
        & & \text{Fingerprint zone}\\
        & & {\bf f}' \leftarrow \textsf{Memory}(\text{Fingerprint zone})\\
        & \text{ready for attestation} & {\bf mask} \leftarrow \textsf{WORM}\\
        & \xleftarrow{~~~~~~~~~~~~~~~~~~~~~~~~~} & {\bf F} \leftarrow \textsf{NoisFre.Transform}({\bf f}', {\bf mask})\\
        {\bf chal} \leftarrow \textsf{RNG}() & {\bf chal}, {addr}, {length} & \\
        & \xrightarrow{~~~~~~~~~~~~~~~~~~~~~~~~~} & {\bf bin}' \leftarrow \textsf{Memory}({addr},{length})\\
        {\bf resp} \leftarrow \textsf{MAC}_{{\bf F}}({\bf bin},{\bf chal}) &  & {\bf resp}' \leftarrow \textsf{MAC}_{{\bf F}}({\bf bin},{\bf chal})\\
        & {\bf resp}' & \\
        \textit{if } {\bf resp} == {\bf resp}' & \xleftarrow{~~~~~~~~~~} & \\
        \quad \text{attestation succeed}\\
        \textit{else}\\
        \quad \text{reject}
\end{array}
    $}
}
\caption{Remote attestation protocol, with mask stored in Prover's WORM. In the demo, we implement this version.}
\label{fig:Protocol_WORM}
\end{figure}

\vspace{2mm}
\noindent{\it Remote Attestation Mechanism.} An overview of the remote attestation mechanism based on a NoisFre key generator is illustrated in: i)~Fig.~\ref{fig:Protocol_transmit}, where we assume the Prover has no WORM memory available for storing a $\textbf{mask}$ and that it has to be transmitted over the wireless communication channel (the worst-case setting in terms of implementation overhead) and ii)~Fig.~\ref{fig:Protocol_WORM}, where we assume the Prover has  available WORM memory. We assume the Prover has already undergone the enrollment phase we described in Section~\ref{sec:secure_key_enrollment}. The enrollment is conducted by the Verifier in the current setting.

A remote attestation can be requested anytime. First, the Verifier scans for visible Provers by sending a ``hello" message. Once there is a Prover in the horizon responding with its unique identifier ${\bf id}$, the Verifier fetches the Prover's information (e.g., ${\bf F}$ and $\textbf{mask}$) from the secure database $\textbf{DB}$ by using the ${\bf id}$ as an index. Second, if the received ${\bf id}$ matches one of that stored in the Verifier's $\textbf{DB}$, the Verifier instructs the Prover to perform attestation---by sending the $\textbf{mask}$ and MAC  $\textbf{tag}$ for Provers with no WORM memory, as in Fig.~\ref{fig:Protocol_transmit}. In this context, the Prover performs a power cycling of memory banks {\it solely} corresponding to the fingerprint zone\footnote{Each memory bank can be individually powered off by exploiting particular power control registers, thus enabling run-time SRAM fingerprinting.} and dynamically generates ${{\bf F}}_{i}$ following the steps described in Section~\ref{sec:secure_key_generation}. After confirming a ready acknowledgment from the Prover, the Verifier randomly generates a challenge (a nonce) ${\bf chal}$, and sends it to the Prover along with the address ${addr}$ and the ${length}$ of the target application program (App) code ${\bf bin}$ in the Prover's memory. The Prover's response ${\bf resp}'$ is generated using MAC computed with the noise-tolerant fingerprint ${{\bf F}}$. The Verifier compares the received response ${\bf resp}$ with a locally calculated reference response ${\bf resp}'$. The remote attestation is accepted if ${\bf resp}$ and $ {\bf resp}'$ match and rejected otherwise. 

If the Prover device implements WORM memory for storing a $\textbf{mask}$, the protocol can be simplified as shown in Fig.~\ref{fig:Protocol_WORM}; in our end-to-end demo implementation, we consider this simpler case, and describe the implementation details in Fig.~\ref{fig:dataFlow} in Appendix.~\ref{app:remAtt}.

\subsection{Overhead Comparisons}\label{sec:overheadCompare}

\noindent{\bf Implementation Details.} We provide an overview of the system implemented in Fig.~\ref{fig:Overview} (b) and defer details to~\textbf{Appendix~\ref{app:remAtt}}. Further, we refer the reader to our open-source code release\footnote{See \underline{https://github.com/AdelaideAuto-IDLab/NoisFre}} for detailed descriptions of our implementation, including dynamic and run-time key generation from SRAM fingerprints. A video demonstration of the end-to-end implementation is available at \href{https://youtu.be/O5NWZw-swpw}{\underline{https://youtu.be/O5NWZw-swpw}}.

We implemented a D-Norm-based key generator on an nRF52832 chip with the smallest on-chip SRAM capacity and ${\rm BER}_{\bf f}$ of  4.93\% tested under $-15$ to $80\celsius$ operating range. We used $n$=32, $m$=48, $\theta$=13 for D-Norm parameters determined by equation~\eqref{eqn:DHW_failure_rate}, \eqref{eqn:DHW_SR}, and \eqref{eqn:key_failure_rate} to be able to extract a 128-bit NoisFre key capable of a key failure rate below $9.15 \times 10^{-6}$. 

For comparisons, we implemented the (R)FE-based key generators summarized in~\autoref{tab:FE_overhead} to achieve a key failure rate to closely match $10^{-6}$. As discussed in Section \ref{sec:rfe}, in an FE, the device executes the computationally-heavy decoding function, while in an RFE, the device executes the more lightweight encoding function. In our end-to-end demonstration, to achieve a comparable failure rate to that of the D-Norm-based NoisFre key generator, the (R)FE implementation needs 13 parallel blocks of ($n=63$, $k=10$, $t=13$) BCH code\footnote{BCH code is a class of cyclic error-correcting codes, named after its inventors Bose, Chaudhuri, and Hocquenghem, constructed using polynomials over Galois field} to provide a similar key failure rate. 

\vspace{2mm}
\noindent{\bf Implementation Overhead.}\label{Sec:EndtoEnd}
The implementation overhead evaluates the usage of two system resources: random-access memory (for run-time data) and clock cycles (for code executions). Overall, in terms of obtaining a 128-bit reliable key with a key failure rate of $9.15\times10^{-6}$, the implementation of the D-Norm-based NoisFre method with parameters ($n$=32, $m$=48, $\theta$=13) takes 51,044 clock cycles\footnote{This was tested with nRF52832 SoC, via J-link EDU V10.1 debugger, with nRF5 SDK Ver. 15.3.0, Keil uvision 5.25.2.0 and ARM CC compiler Ver. 5.06 Update 6. Optimization setting = -O3.}. If the $\textbf{mask}$ is provided by the server and transmitted over a wireless channel, an additional 45,622 clock cycles are required for mask integrity checks. 

In contrast, the FE-based and the lightweight state-of-the-art RFE-based method introduces significantly higher overheads to achieve a 128-bit reliable key with a slightly inferior key failure rate of $2.45\times10^{-5}$. Specifically, as evaluated and shown in Fig.~\ref{fig:Implementation_overhead}, the on-device FE decoding and RFE encoding functions consume 285,311 and 109,850 clock cycles, respectively. Both methods need an additional 60,755 clock cycles for helper data integrity checks. In comparison with the state-of-the-art FE and RFE, for meeting a comparable key failure rate, NoisFre reduces clock overhead by 72\% and 43\%, respectively, if the mask or helper data is transmitted over the wireless channel requiring helper data integrity checks. However, if the mask or helper data for all of the method are stored on a device's WORM memory, clock cycles required in comparison to NoisFre reduces by 82\% (compared to FE) and 54\% (compared to RFE).

\begin{figureRevsource}[!ht]
	\centering
	\includegraphics[width=\mylinewidth]{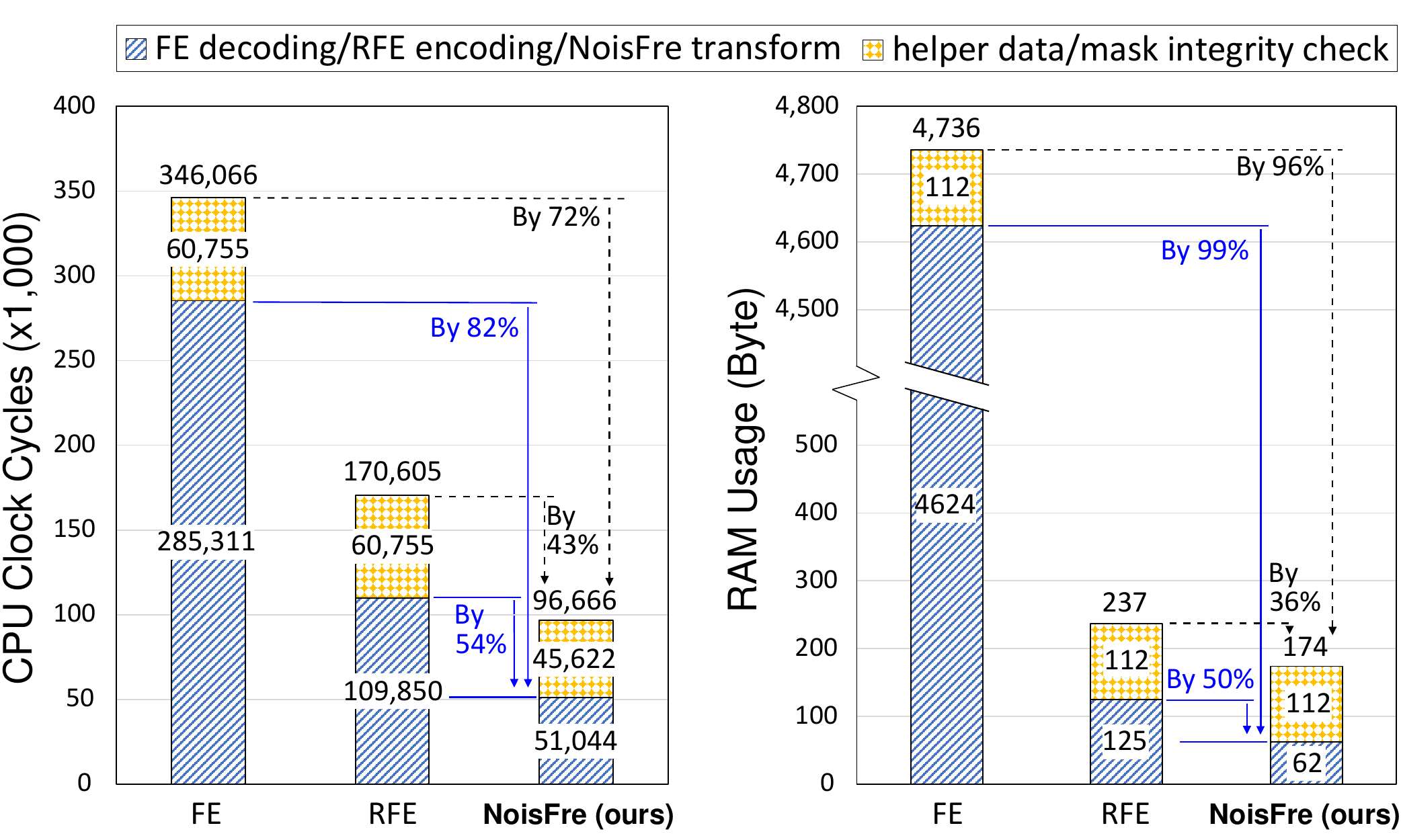}
	\caption{Comparison of implementation overhead of the proposed NoisFre key derivation against traditional (R)FE-based key derivation. The integrity checks are necessary if the helper data or the mask is transmitted over a wireless channel.}
	\label{fig:Implementation_overhead}
\end{figureRevsource}

\begin{table*}[!pb]
\caption{Implementation overhead of R(FE) employing BCH codes. 
}
\label{tab:FE_overhead}
\centering
\resizebox{\textwidth}{!}{%
\footnotesize
\begin{tabular}{c|c|c|c|c|c|ccc}
\toprule \hline
\multirow{3}{*}{\begin{tabular}[c]{@{}Sc@{}}Fingerprint source\\
(${\rm BER}_{\bf f}$)\end{tabular}} & \multirow{3}{*}{BCH(n,k,t)} & \multirow{3}{*}{\begin{tabular}[c]{@{}c@{}}Block number\end{tabular}} & \multirow{3}{*}{Key failure rate} & \multirow{3}{*}{Key size} & \multirow{3}{*}{Helper data size} & \multicolumn{3}{c}{Clock cycles} \\ \cline{7-9}

& &  &  &  &  & \begin{tabular}[c]{@{}Sc@{}}Fuzzy Extractor \\
decoding\end{tabular} & \begin{tabular}[c]{@{}Sc@{}}Reverse Fuzzy\\ Extractor encoding\end{tabular} & \begin{tabular}[c]{@{}Sc@{}}Helper data \\ integrity check\end{tabular} \\ \midrule

 NORDIC (4.93\%) & (63,10,13) & 13 & $2.45 \times 10^{-5}$ &  130 &  689 &  285,311 &  109,850 &  60,755 \\

 IDT (5.42\%) & (127,15,27) & 9 &  $1.69 \times 10^{-9}$ &  135 &  1008 &  967,599 &  188,487 &  84,013 \\
 \hline \bottomrule
\end{tabular}%
}
\end{table*}

It is worth ephasizing that we have compared NoisFre with an RFE capable of deriving a key with a failure rate of $P_{\bf F}^{\rm fail} \approx 10^{-6}$. However, as we show in Table~\ref{tab:PkeySingleChip}, if an IDT chip is used in the implementation, we can obtain a key with a significantly lower key failure rate by exploiting the  \textit{free}, abundant memory; now, $P_{\bf F}^{\rm fail}$ can be $\approx5\times10^{-9}$. Attempting to achieve such a small $P_{\bf F}^{\rm fail}$ using an (R)FE will lead to significantly higher overheads. The (R)FE-based key provisioning method introduces increasing execution overheads if a lower key failure rate is desired as illustrated in \autoref{tab:FE_overhead}. For example, if an IDT SRAM chip is used as the fingerprinting barometric source instead of the NORDIC chips' internal SRAM, a 128-bit key with failure rate of $1.69 \times 10^{-9}$ requires 188,487 (3.69 times larger) clock cycles with the REF-based method or 967,599 (18.95 times larger) clock cycles with the FE-based method, compared with 51,044 clock cycles for our NoisFre-based method. \textit{Hence, in contrast to (R)FE methods, the on-device computational overhead of the proposed NoisFre key generator remains constant, regardless of the desired key reliability and only depends on the size of the key to be derived}.

\section{Discussion}\label{Sec:discussion}
\vspace{2mm}
\subsection{Generality of NoisFre}\label{sec:generality}
\Revsource{rev:simple_extraction}{Although our work focused predominantly on  SRAM, considering its {\it ubiquity in low-end IoT devices \Rev{and the simplistic nature of fingerprint extraction}}, the NoisFre fingerprinting methods presented are applicable for other memories, including Flash and EEPROM memories validated in our study.} In principle, it can be applied to other hardware fingerprinting methods~\cite{pappu2002physical,ruhrmair2010applications}, given an abundant raw digital fingerprint bit space.

\subsection{Provisioning Fingerprints at Run-time}
Flash and EEPROM memory fingerprints can be accessed during run-time. However, for SRAM fingerprinting, the most common method is to utilize its initialization pattern at power-up as a fingerprint, although there are other means~\cite{xu2015reliable}; for example using data retention voltage~\cite{xu2015reliable} or intentionally putting SRAM cells under a meta-stable state. Those methods usually require customized peripheral circuitry, which tends to be unavailable in COTS devices. Thus, SRAM fingerprinting generally requires power cycling to read the start-up values. As a matter of fact, some low-end microcontrollers allow direct control over the powering of individual SRAM banks~\cite{semiconductor2016nrf52832} (e.g., the low-end nRF52832 studied in this work). Consequently, by leveraging such a feature, SRAM fingerprint-based root keys can be requested during {\it run-time}.

\subsection{Security Analysis} \label{sec:security_analysis}
\Revsource{rev:security_analysis}{We have looked at the problem of achieving a pragmatic, on-device key derivation method using noisy memory fingerprints. NoisFre fundamentally obviates the need for computationally intensive on-device ECC logic for the task. It is thus immune to HDM attacks~\cite{delvaux2014helper,becker2017robust} that strategically tamper the helper data \textit{associated with the ECC} to weaken or compromise the key extracted using the state-of-the-art (R)FE methods. The vulnerability is induced by the usage of {\it ECCs} (see Section~\ref{sec:rfe}). Various ECCs are examined and shown to be vulnerable to HDM attacks~\cite{becker2017robust}. A generic countermeasure against HDM attacks appears to be an open challenge. The NoisFre scheme has sought to remove the necessity for helper data associated with key generation in an RFE and, thus, avoid the HDM attacks that exploit helper data. In the following, we consider the security of our proposed key derivation method in the context of prior methods based on the state-of-the-art (R)FE methods.}

\subsubsection{Threat Model}\label{sec:Threat_model}
\Revsource{rev:Threat_model}{Memory fingerprint-based key provisioning studies rarely explicitly define a threat model~\cite{kusters2019secret,maes2016secure,hiller2016cherry} and operate under the assumption that the key material (i.e., memory fingerprint) cannot be directly accessed. However, studies focusing on incorporating key derivation methods to provide a security function, such as authentication or remote attestation~\cite{aysu2015end,feng2017secure,kohnhauser2016secure,qureshi2021puf}, follow a threat model. Therefore, we follow the threat model reasoned therein, along with the assumption that the key material cannot be directly accessed.} 

Specifically, we consider that an adversary cannot access the raw fingerprint and temporary data stored in RAM or internal chip registers during key derivation. The attacker can tamper with public information used to assist the key derivation. Notably, in prior work, such information would be the ECC associated helper data in a (R)FE-based reliable key derivation method~\cite{becker2017robust}---in our key derivation approach, we assume the \textit{mask} is public information. The mask is sent to a device over a communication channel together with a method for assessing the integrity of the mask or is stored in WORM.

\subsubsection{Mask Manipulation Attack}\label{sec:mask_manip_att}

\Revsource{rev:mask_manip_att}{In use cases where the mask is sent to a device over a  communication channel, it is possible for an attacker to manipulate the mask. Therefore, we consider \textit{mask manipulation attacks}.} 

In the context of a NoisFre-based key generator, a MAC \textbf{tag} is produced over the mask using the derived ${\bf F}$ to ensure the integrity of the mask---more specifically, $\textbf{tag} \leftarrow  $\textsf{MAC}$_{{\bf F}}$(\textbf{mask}), with ${\bf F}$ being the reliable secret key, as illustrated in Fig.~\ref{fig:NoisFre_generation} (b). The $\textbf{mask}$ and MAC \textbf{tag} can be publicly stored off-chip and/or stored on-chip. Subsequently, the MAC \textbf{tag} can be regenerated to validate the integrity of a mask stored on-device or transmitted to the device prior to the use of the key derived on-demand, as illustrated in Fig.~\ref{fig:NoisFre_generation} (b). Now, the probability of making a modification without being detected is $\frac{1}{2^k}$ with $k$ the length of the derived key. It will be $\frac{1}{2^{128}}$ for a typical 128-bit key.

Although we adopted a simple mechanism in this study to ensure mask integrity, other mechanisms have been proposed to ensure the integrity of helper data in the context of state-of-the-art (R)FE methods~\cite{delvaux2014helper}. Thus, we can also employ these existing methods to ensure the integrity of the mask for NoisFre key derivation method.

\subsubsection{Brute-force Attack}\label{sec:brute_force_att} 
\Revsource{rev:brute_force_att}{For completeness, we also assess the attack complexity of a brute-force attack on a NoisFre-based key derivation method. The attacker may utilize a brute-force attack to determine the derived key. However, this is extremely challenging when the key is appropriately sized. For a brute-force attack, the probability of finding the correct derived key is $\frac{1}{2^k}$, which is computationally infeasible given a reliable key with a typical length of $k=128$ bits.
}

\subsubsection{Aging Attack}\label{sec:aging_att} 
\Revsource{rev:aging_att}{The data stored in a SRAM cell can gradually affect its start-up state. This is called data-dependent aging~\cite{maes2014countering}. Given that the key derivation is based on a physical primitive, we also consider \textit{aging attacks} that may attempt to exploit the small changes in behavior of memory cells that occur as a result of aging the underlying electronic components.}

In use cases where a write access protected (e.g., using a memory protection unit [MPU]) memory cannot be allocated for generating fingerprints and where the memory space used for fingerprints must be shared with user application code, an attacker may utilize malicious code on the device to continuously write specific memory patterns to the SRAM used for device fingerprinting. Such an attempt can accelerate aging and can potentially degrade the reliability of a NoisFre key generation method.

In use cases where a dedicated memory cannot be allocated for generating fingerprints, several simple mitigation strategies already exist. First, the aging effect is data dependent. The user can employ an anti-aging method, such as writing reverse data patterns to mitigate the aging effect validated as an efficient approach to counter aging~\cite{maes2014countering}. Second, the SRAM unreliability induced by aging, even over six years, is small---only 2\%~\cite{maes2014countering}. Hence, a simple anti-aging method for NoisFre is to allow the server to intentionally assume a higher worst-case ${\rm BER}_{\bf f}$ during the enrollment phase to count for or tolerate the aging effect by trading off a slight increase in SRAM volume required to retain the same NoisFre key reliability. If the available memory volume is constrained, a further low-cost anti-aging measure is for the server to adopt the trial-and-error method reported in~\cite{gao2018treverse}  to recover the least reliable transformed $\digamma$ bits, because the server can ascertain the bit-specific reliability of each $\digamma$ bit. Notably, in this approach, all the computation overhead is offloaded to the server without imparting any overhead to the device.

\subsection{Limitations and Future Work} \label{sec:limitations}
\Revsource{rev:limitation}{Our study is not without limitations. As shown in Fig.~\ref{fig:ExtEff_DvsS}, the highest extraction efficiency (i.e., the number of fingerprint bits with a BER$_{\bf F}< 7.81\times 10^{-9}$ that can be extracted from a unit-sized memory block) that NoisFre can achieve is 0.62 bits per KiB. Hence, extracting a usable (e.g., 128-bit) secure key from a highly resource-limited device with a mere 2~KiB memory space (i.e., the SRAM size of the passively powered computational radio frequency identification (CRFID) device studied in~\cite{su2019secucode}) with NoisFre is not immediately possible.}

The investigation of potential methods for improving the performance, in particular enhancing the ${\rm BER}_{\bf F}$ and/or extraction efficiency $\eta$, is left out of scope for our current study focused on developing NoisFre, formalizations, and extensive evaluations. As a potential direction for future work, it will be interesting to consider approaches, for example, to extract more bits from a given memory. Although our formulation for reliability is applicable for such a method, the analytical formulation of extraction efficiency for such new methods will likely require considerable effort to develop. Importantly, the complexity of the task will provide an interesting direction for future work. 
Therefore, we leave the investigation of potential means for improving the NoisFre performance, in particular enhancing the ${\rm BER}_{\bf F}$ and/or extraction efficiency $\eta$, as potential directions for future research.

\section{Related Work}\label{Sec:relatedwork}
Besides memories, various on-board sensors, such as cameras, accelerometers, gyroscopes,  magnetometers, and other components, such as CPU magnetic radiations, are utilized to provide fingerprints~\cite{bojinov2014mobile,son2018gyrosfinger,kim2018c,ba2018abc,willers2016mems,zhang2019sensorid,ba2019cim,quiring2019security,cheng2019demicpu}. Other recent works also explore commodity scanners to fingerprint 3D objects to track them~\cite{li2018printracker} and exploit the package variations as fingerprints for anti-counterfeiting~\cite{dhanuskodi2020counterfoil}. However, to obtain hardware fingerprints, those fingerprint extractions are relatively complicated in comparison with memory, especially SRAM, enabled fingerprints. 

Notably, hardware fingerprinting is closely related to the notion of PUFs~\cite{chang2017retrospective,herder2014physical,ruhrmair2013pufs,rahman2015aging,rahman2014aro}. Commodity memory fingerprinting, such as SRAM PUF and Flash PUF, is not new. However, mounting them on low-end IoT devices to derive a usable key for security functions relies on post-error correction to reconcile bits errors, which is cumbersome in terms of both overhead and security in practice. Our simple yet efficient NoisFre memory-fingerprinting approach addresses this gap.

\Revsource{rev:aging_RO_PUF}{We exploit the idea of a differential measurement in the formulation of the D-Norm method based on the base distance ($\ell 1$-Norm) to improve the \textit{extraction efficiency} (number of $\digamma$ bits with a desired noise tolerance) from a given memory. Interestingly, in PUF studies, formulating methods to exploit a differential gap or comparison has been utilized by \textit{extrinsic PUFs}--implemented with additional hardware---such as ring oscillator PUF (RO-PUF)~\cite{rahman2015aging,rahman2014aro,suh2007physical,hesselbarth2018large} and arbiter PUF (APUF)~\cite{gassend2002silicon} to obtain responses with improved reliability. The concept has subsequently been applied in~\cite{rahman2015aging,rahman2014aro} to enhance reliability and address aging of electronic components in RO-PUFs facilitated by the ease with which RO frequency differences are already measured and can be directly used. In our \textit{intrinsic memory} studies, we exploit a base distance, $\ell 1$-Norm, to generate a differential measurement to build the D-Norm transform for memory PUFs \textit{intrinsic} to COTS devices. As discussed in Section~\ref{sec:DNormConcept}, we recognized that the differential formulation can yield significantly more reliable bits compared to S-Norm employing (simply, the $\ell 1$-Norm base distance). Thus, in our work, we combine these two mathematical concepts (base distance with a differential measure) together to extract more noise-tolerant bits from a memory PUF---a method that can be used with intrinsic memories widely exist in COTS devices.}

\section{Conclusion}\label{Sec:conclusion}
By exploiting ubiquitously embedded memory within commodity computing devices, the proposed NoisFre approach constructively extracts transformed memory fingerprints that were embodied with a high tolerance to noise affecting the generation of fingerprints. With a simple, single, one-off fingerprint enrollment measurement, NoisFre is able to judiciously identify highly reliable transformed fingerprints serving as hardware root key or root of trust to directly support various security functions for a wide range of COTS electronic devices. Besides formalization of two specific S-Norm and D-Norm fingerprint transformation methods and extensive empirical validations on SRAM, Flash, and EEPROM memories using 119 physical chips in total, we have conducted a case study with an end-to-end implementation of a remote attestation security service employing NoisFre fingerprints to significantly reduce the overhead in comparison with the state-of-the-art RFE method for constructing reliable fingerprints for a key generator. We also demonstrate how SRAM fingerprints can be generated  at {\rm run-time} by utilizing individual memory-bank power control features on MCUs. Overall, NoisFre is a {\it simple but practical} method,  especially for existing low-end commodity electronic devices.

\Urlmuskip=0mu plus 1mu\relax
\bibliographystyle{IEEEtran}
\bibliography{IEEEfull}

\vspace{-1.0cm}
\begin{IEEEbiography}[{\includegraphics[width=.8in,clip,keepaspectratio]{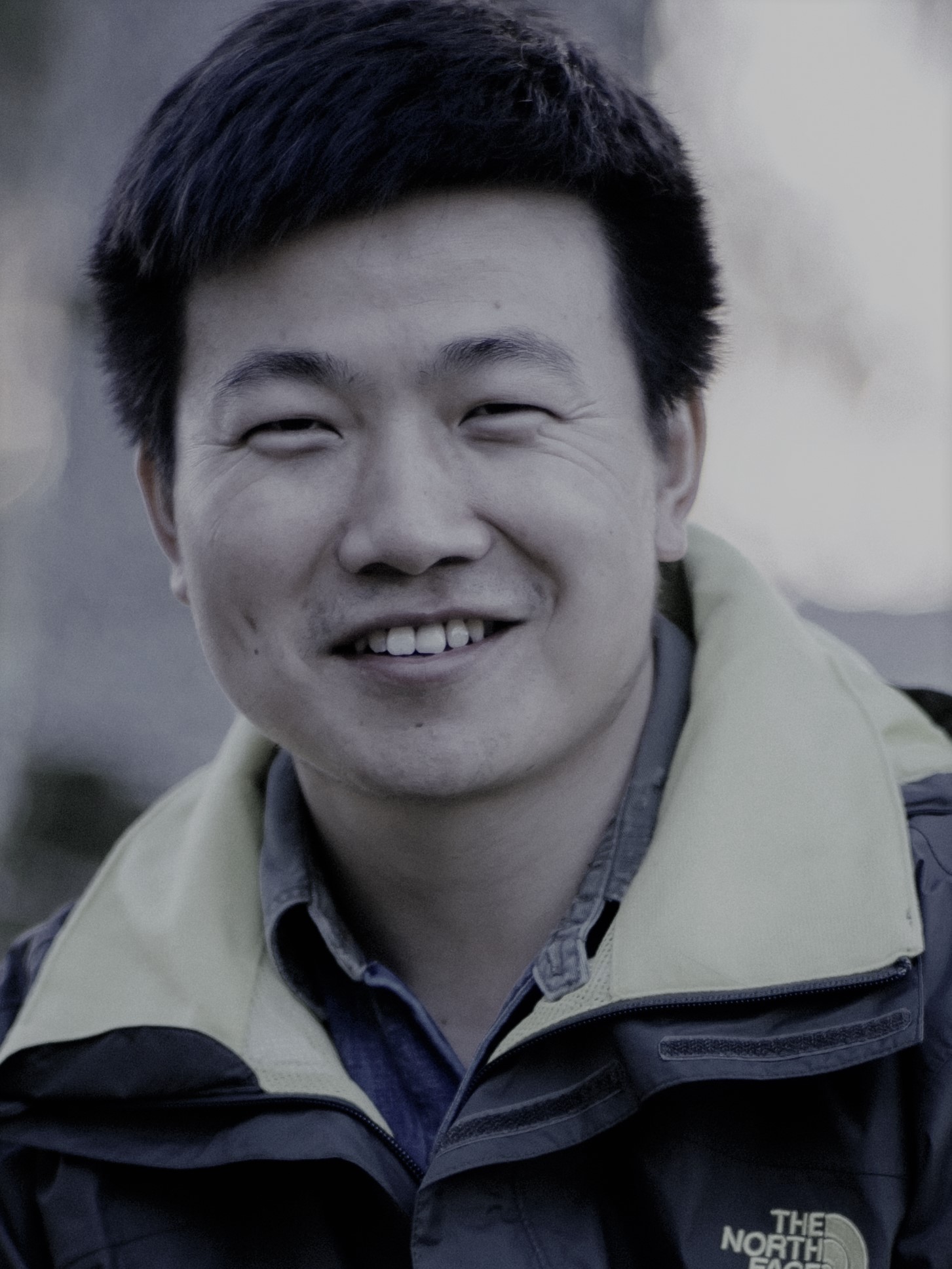}}]{Yansong Gao} received the Ph.D degree from the University of Adelaide, Australia, in 2017. He was a Postdoc Research Fellow with Data61, CSIRO, Australia. He is now with Nanjing University of Science and Technology, China, as an Associate professor at School of Computer Science and Engineering.
His current research interests are hardware security, system security, and AI security and privacy.
\end{IEEEbiography}

\vspace{-1.5cm}
\begin{IEEEbiography}[{\includegraphics[width=.8in,clip,keepaspectratio]{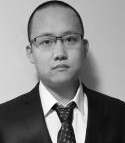}}]{Yang Su}
(S'14) Yang Su received the B.Eng. degree with first class Honours in Electrical and Electronic Engineering from The University of Adelaide, Australia in 2015. He worked as a Research Associate at The University of Adelaide from 2015-2016 and he is currently pursuing the Ph.D. degree. His research interests include hardware security, physical cryptography, embedded systems and system security. 
\end{IEEEbiography}

\vspace{-1.5cm}
\begin{IEEEbiography}[{\includegraphics[width=.8in,clip,keepaspectratio]{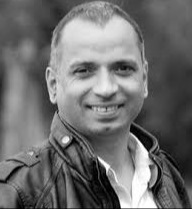}}]{Surya Nepal} received the Ph.D. degree from RMIT University, Australia. He joined CSIRO in 2000. He is currently a Principal Research Scientist with Data61, CSIRO. He is the leader of the Distributed System Security Group, Data61, CSIRO. His main research interest is in the development and implementation of technologies in the area of distributed systems including Web services, cloud computing, IoT, and big data, with a specific focus on security, privacy, and trust. He is currently one of the associate editors of the {\sc IEEE Transactions on Services Computing} and {\sc IEEE Transactions on Dependable and Secure Computing}.
\end{IEEEbiography}

\vspace{-1.5cm}
\begin{IEEEbiography}[{\includegraphics[width=0.8in,height=1.07in,clip]{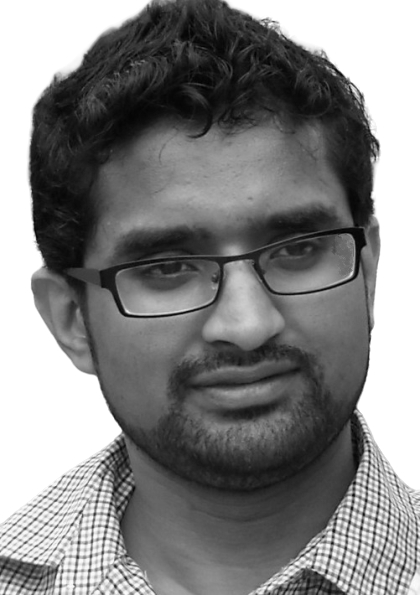}}]{Damith C. Ranasinghe} received the Ph.D. degree in Electrical and Electronic Engineering from The University of Adelaide, Australia. In the past, he was a Visiting Scholar with the Massachusetts Institute of Technology, Cambridge, MA, USA, and a Post-Doctoral Research Fellow with the University of Cambridge, Cambridge, U.K. Currently, he is an Associate Professor at The University of Adelaide. His research interests include embedded systems, system security, autonomous systems, and deep learning.
\end{IEEEbiography}

\clearpage
\appendices
    \setcounter{page}{1}
    \renewcommand{\thepage}{A \arabic{page}}%
\section{Derivation of Performance Metric Models}\label{app:derivation}

In this section, we detail the derivation of equations in Section~\ref{sec:PerforMetrics}, where the  following ideal chip model is adopted.

\subsection{Synthetic Chip Model} \label{sec:apx_Random_chip}
Due to the size limitation of physical chips and difficulty of taking a massive number of repeated measurements from physical chips, we adopt a synthetic chip model to evaluate our analytic predictions. The synthetic chip model used is from~\cite{maes2013accurate} and is built under the following settings:
\begin{enumerate}
    \item Each bit has a 50\% chance being logic ``1'' or ``0'' during the enrollment phase. Each initial bit value is randomly generated during chip initialization.
    \item Each bit has an equal probability, ${\rm BER}_{\textbf{f}}$, of being flipped during a regeneration.
    \item The values of bits are independent and identically distributed (iid); hence, we assume no  spatial or temporal correlations.
\end{enumerate}

\subsection{Unreliability Formalization of S-Norm Transformation}\label{app:SnormBER}
As described in Section~\ref{sec:noiseTolerantFingerprinting}, all possible cases of noise-tolerant S-Norm transformed bit $\digamma$ are shown below. 
\vspace{-1mm}
\[
        \digamma \leftarrow \mathcal{T}_{\rm SNorm}({\bf f}^{n\times 1},\theta) = 
    \begin{cases}
        1, & \|\ff\|_1 \geq \ceil{\frac{n}{2}} + \theta\\
        0, & \|\ff\|_1 \leq \floor{\frac{n}{2}} - \theta\\
        \bot, & {\rm Otherwise}
    \end{cases}
\]

\begin{figure}[!ht]
	\centering
	\includegraphics[width=1\mylinewidth]{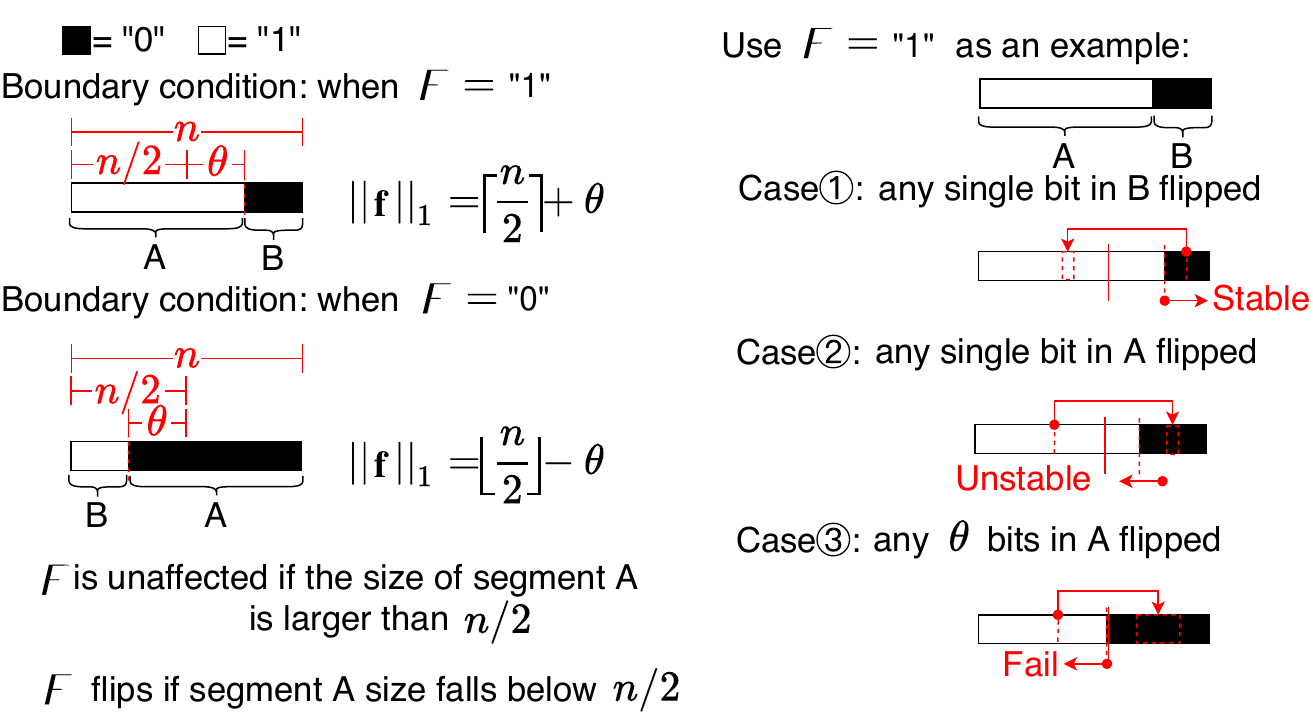}
	\caption{S-Norm-based NoisFre. Two boundary conditions are illustrated: when $\digamma$ = ``1'' where $\|\ff\|_1$=$\ceil{\frac{n}{2}}+\theta$, and when $\digamma$ = ``0'' where $\|\ff\|_1$=$\floor{\frac{n}{2}}-\theta$. Here we use $\digamma = $``1'' as an example to demonstrate the influence of flipped raw bits on their transformed 1-bit $\digamma$. The generated $\ell 1$-Norm is partitioned into two segments: A and B. Consider three representative cases:  \circled{1} any single raw bit flip in B will enhance the reliability of the transformed bit $\digamma$; otherwise  \circled{2} any single raw bits flip in segment A will deteriorate reliability of the $\digamma$; \circled{3} the $\digamma$ will fail/flip if there are $\theta$ or more raw bits flipped in segment A.}
	\label{fig:HW_failure_rate}
\end{figure}

The formalization is visualized in Fig.~\ref{fig:HW_failure_rate}. Recall that a $\digamma$ bit can be transformed from $n$ raw bits and the $\ell 1$-Norm of the $\digamma$ is between $[0,n]$. To assess the worst-case ${\rm BER}_{{\bf F}}$, we consider the condition where the selected word's $\ell 1$-Norm is exactly equal to $\ceil{\frac{n}{2}}+\theta$, as shown in boundary condition $\digamma=$``1'' in Fig.~\ref{fig:HW_failure_rate}. Here, $\theta$ is a threshold to select highly reliable $ \digamma$ bits.  

Each raw bit is with ${\rm BER}_{\textbf{f}}$ probability to be flipped under reevaluation. Using boundary condition $\digamma=$``1'' as an example, on the one hand, \circled{1}, if there are raw bits of ``0'' (marked as segment B) flipping, it will increase the tolerance of the number of {\it raw bits of ``1''} that allows being flipped (in segment A) {\it without influencing $\digamma$ bit}. In contrast, \circled{2} flipping raw bits of ``1'' (marked as segment A) will potentially result in an error to the $\digamma$ bit. Further, \circled{3}, supposing that raw bits of ``0'' (marked as segment B) remain unchanged, if more than $\theta$ raw bits of ``1'' flip, the $\digamma$ will exhibit an error---flipping from ``1'' to ``0''. To be precise, the transformed $\digamma$ bit {\it will not exhibit} error unless more than $\theta+i$ raw bits of ``1'' flipping.

Overall, bit flipping within raw bits of ``0'' (marked as segment B) increases the reliability of extracted ${\bf F}$. In contrast, bit flipping within raw bits of ``1'' (marked as segment A) decreases the reliability of extracted ${\bf F}$. 
The boundary condition $\digamma=$ ``0'' is logically equivalent to the case $\digamma=$ ``1'' but, only inverts $\digamma$'s ``0''/``1'' value rather than its ${\rm BER}_{\bf F}$.

\begin{figure*}[!ht]
	\centering
	\includegraphics[width=\mylinewidth]{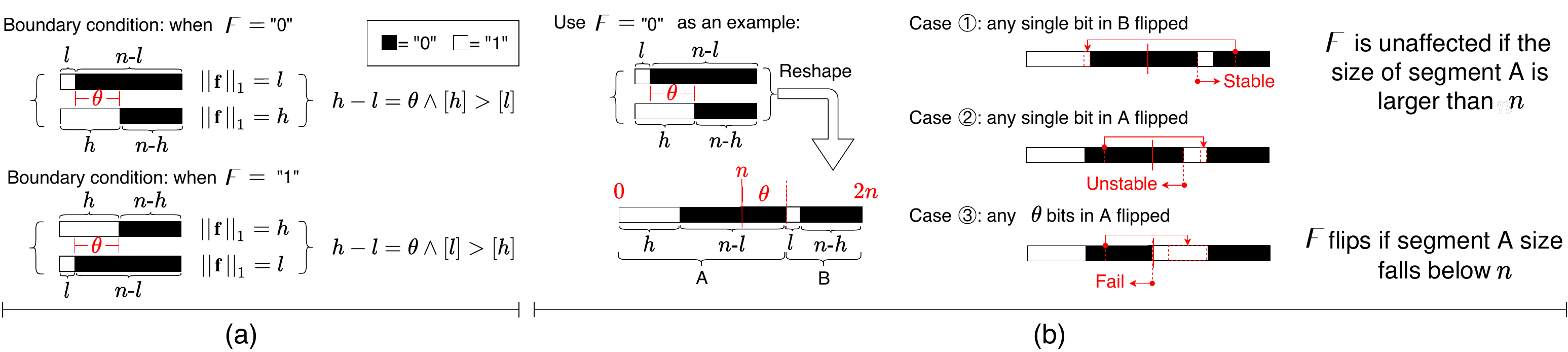}
	\caption{D-Norm-based NoisFre (a) Two boundary conditions are illustrated: when $\digamma$ = "0" where ($h-l=\theta) \wedge ([h]>[l]$); and when $\digamma$ = "1" where ($h-l=\theta) \wedge ([l]>[h])$. (b) Here we use $\digamma$ = "0" as an example to demonstrate the influence of flipped raw bits on their transformed 1-bit $\digamma$. The two $\ell 1$-Norms are firstly reshaped to a single row as shown above to backtrack to the same formulation strategy in S-Norm. The reshaped $\ell 1$-Norm is partitioned into two segments: A and B, the size of segment A is $n+\theta$, and the size of B is $n-\theta$ under the boundary condition $\digamma$ = "0". Consider three cases: \circled{1} any single raw bit flip in B will enhance the reliability of transformed $\digamma$ bit; otherwise, \circled{2} any single raw bit flip in segment A will degrade the reliability of ${\bf F}$. The case of $\digamma$ will fail/flip \circled{3} if there are $\theta$ or more raw bits flipped in segment A.}
	\label{fig:DHW_failure_rate}
\end{figure*}

Without losing generality, we focus on one case shown in \circled{1} in Fig.~\ref{fig:HW_failure_rate}. $\mystrut { \|\ff_i\|_1} = \ceil{\frac{n}{2}}+\theta$, the probability of having exact $x$ error bits in segment A can be expressed as $\mystrut {\rm Pr}_{|x| \in A}^{\rm flip} = \textsf{binopdf}(x,\frac{n}{2}+\theta,{\rm BER}_{\textbf{f}})$, given that each raw bit has a ${\rm BER}_{\textbf{f}}$ probability of flipping. Similarly, the probability of $y$ bits in segment B to be flipped is formulated as $\mystrut {\rm Pr}_{|y| \in  B}^{\rm flip} = \textsf{binopdf}(y,\frac{n}{2}-\theta,{\rm BER}_{\textbf{f}})$.

Although bit flip could occur in either segment A or B, consequential ${\rm BER}_{\bf F}$ of $\digamma$ bits are opposite: flipped bits in segment A reduction in the margin or potentially increases the ${\rm BER}_{\bf F}$ (shown as the dashed boundary line in Fig.~\ref{fig:HW_failure_rate} that moves toward the left). In contrast, flipped bits in segment B increase the margin or potentially decrease ${\rm BER}_{\bf F}$ (the boundary moves toward the right). If the boundary crosses the middle point of $\frac{n}{2}$, the $\|\ff_i\|_1$ falls below $\frac{n}{2}$, and consequently, the $\digamma$ bit flipped---exhibiting an error. 

Starting from the extreme but straightforward condition---there is no bit flip in segment B (i.e., $y = 0$). the maximum number of erroneous bits that can be tolerated is $\theta$ as discussed above. This can be expressed as $P_{\|\ff_i\|_1}^{\rm fail} = {\rm Pr}(x-y \geq \theta) = {\rm Pr}(x \geq \theta ~|~ y = 0) = {\rm Pr}(x \geq \theta) \times {\rm Pr}(y = 0)$, where the term ${\rm Pr}(x \geq \theta)$ can be expressed as $1 - {\rm Pr}(x < \theta) = 1 - \sum_{x = 0}^{\theta}\left({\rm Pr}_{|x| \in A}^{\rm flip} \right) = 1 - \textsf{binocdf}|(\theta,\ceil{\frac{n}{2}}+\theta,{\rm BER}_{\textbf{f}})$. By substituting ${\rm Pr}_{|y| \in B}^{\rm flip} = \textsf{binopdf}(0,\floor{\frac{n}{2}}-\theta,{\rm BER}_{\textbf{f}})$ into the $P_{\|\ff_i\|_1}^{\rm fail}$ equation, $P_{\|\ff_i\|_1}^{\rm fail}$ is expressed:
\begin{dmath*}
    P_{\|\ff_i\|_1}^{\rm fail} = \left(1 - \textsf{binocdf}(\theta,\ceil{\frac{n}{2}}+\theta,{\rm BER}_{\textbf{f}})\right) \times \textsf{binocdf}(0,\floor{\frac{n}{2}}-\theta,{\rm BER}_{\textbf{f}})
\end{dmath*}

However, there is more than one case that satisfies $x-y \geq \theta$ for $\{(x,y):|x| \in A, |y| \in B\}$. Since $A$ and $B$ are finite sets, the combination of $x$ and $y$ are numerable. Another property worth mentioning is $|A| > |B|$. Therefore the total number of combinations is up bounded by $|B| = \ceil{\frac{n}{2}}-\theta$ where ``$|~|$'' denotes the cardinality or the size of a set. If we enumerate and sum up all possible combinations, we obtain the complete form of equation~\eqref{eqn:HWBER}:

\begin{dmath*}
    {\rm BER}_{\bf F} = \sum_{i=0}^{\floor{\frac{n}{2}}-\theta} \biggl( \left(1 - \textsf{binocdf}(\theta + i,\ceil{\frac{n}{2}}+\theta,{\rm BER}_{\textbf{f}})\right) \times \textsf{binopdf}(i,\floor{\frac{n}{2}}-\theta,{\rm BER}_{\textbf{f}}) \biggr)
\end{dmath*}

\subsection{Unreliability Formalization of D-Norm Transformation}\label{app:DnormBER}
As discussed in Section~\ref{sec:noiseTolerantFingerprinting}. The transformed bit via D-Norm is determined as below:

\begin{dmath*}
    \digamma \leftarrow \mathcal{T}_{\rm DNorm}({\bf f}^{n\times m},\theta) = \begin{cases}
        1, & h -  l \geq \theta~\wedge~[h] < [l]\\
        0, &  h -  l \geq \theta~\wedge~[h] > [l]\\
        \bot, & h -  l < \theta
    \end{cases}
\end{dmath*}
where [] indicates the index.

To apply the same derivation strategy as the S-Norm, as illustrated in Fig.~\ref{fig:DHW_failure_rate} (b), the two $\ell 1$-Norms are reshaped into a single row, and four partitions are now rearranged as two segments: A and B. The length of segment A is eventually ${h}+(n-{ l})$ by considering the fact ${ h} = { l} + \theta$, whereas we can see that the length of A is $n+\theta$. The largest number of errors/flips within raw bits $ {\bf f}$ that still can not result in error or flip to the transformed $\digamma$ bit is $(n+\theta)-n = \theta$.

The rest of the steps are identical to those in S-Norm. Using  \circled{\small 1} as an example, on the one hand, Case \circled{1}, if one raw bit in segment B is flipped, it will increase the tolerance of the number of raw bits in segment A which allows being flipped {\it without influencing $\digamma$} bit. On the other hand, Case \circled{2} flipping one raw bit in segment A will potentially result in an error to the $\digamma$ bit. Further, for Case \circled{3}, supposing that segment B's raw bits remain unchanged, if more than $\theta$ raw bits flipped in segment A, the ${\bf F}$ will exhibit an error---flipping from ``0'' to ``1''. To be precise, the transformed ${\bf F}$ {\it will not exhibit} error unless more than $\theta+i$ raw bits in segment A flip. 

Now, an extreme condition is considered as a starting point: as shown in the third column in Fig.~\ref{fig:DHW_failure_rate}, we have two words, labeled with spatial index the ``first'' and the ``second''. We denote the raw bits as $\ff_{\rm first}$ and $\ff_{\rm second}$, respectively.

In the exemplified case, $\ff_{\rm first}$ has the lowest $\ell 1$-Norm while the $\ff_{\rm second}$ has the highest $\ell 1$-Norm in the $m$-word block.
(i.e., $\|\ff_{\rm first}\|_1 = l$, $\|\ff_{\rm second}\|_1 = h$). 
From the diagram, we can write the following equation:
\vspace{-1mm}
\begin{dmath*}
     \|\ff_{\rm second}\|_1 -  \|\ff_{\rm first}\|_1 = \theta 
\end{dmath*}
By substituting $\|\ff_{\rm second}\|_1 = h$ and $\|\ff_{\rm first}\|_1 = l$ into the equation above (the case \circled{1} in Fig.~\ref{fig:DHW_failure_rate} (a)), we obtain:
\vspace{-1mm}
\begin{dmath*}
    h -  l = \theta
\end{dmath*}
Add $n$ (number of bits in one word/group) to both sides of the equation. We obtain:
\vspace{-1mm}
\begin{dmath*}
    h + (n - l) = n + \theta
\end{dmath*}

If we reshuffle the four partitions in Fig.~\ref{fig:DHW_failure_rate} (b), the error rate of D-Norm can be formalized in a similar manner as the S-Norm (equation~\eqref{eqn:HWBER}). The margin (denoted as a dashed line) reduces and results in an unstable trend if any bit flips in the segment A. Once the margin crosses $n$ (marked as a solid line) from the right to the left, the transformed ${\bf F}$ is therefore erroneous. In contrast, bits flipped in segment B increase the margin and stabilize the ${\bf F}$.

The probability of $x$ error bits occurring in segment A can be expressed as ${\rm Pr}_{|x| \in A}^{\rm flip} = \textsf{binopdf}(x,n+\theta,{\rm BER}_{\textbf{f}})$. Similarly for $y$ bits in segment B to be flipped can be expressed as ${\rm Pr}_{|y| \in  B}^{\rm flip} = \textsf{binopdf}(y,n-\theta,{\rm BER}_{\textbf{f}})$.

Now consider the special case where there is no bit flip in segment B; then the highest number of bits allowed to be flipped in segment A is simply $\theta$. Otherwise, the ${\bf F}$ will exhibit errors. Consequently, the $P_{\rm DNorm}^{\rm fail}$ can be expressed as:

\begin{dmath*}
P_{\rm DNorm}^{\rm fail}(y=0) = \biggl(1 - 
\textsf{binocdf}(\theta-1,n+\theta,{\rm BER}_{\textbf{f}}) \biggr) 
\times \textsf{binopdf}(0,n-\theta,{\rm BER}_{\textbf{f}})
\end{dmath*}

If the number of flipped bits in segment B is non-zero, ${\rm Pr}_{|y| \in B}^{\rm flip} = \textsf{binopdf}(y,n-\theta,{\rm BER}_{\textbf{f}})$, where $y \in [0,|B|]$, $|B| = n-\theta$. In other words, flipped bits in segment B allows more tolerance of error bits in segment A, before ${\bf F}$ exhibiting error. Therefore, the D-Norm, ${\rm BER}_{\bf F}$, is the summation of $P_{\rm DNorm}^{\rm fail}(y)$ for all possible $y$, finally formulated as in equation:
\vspace{-1mm}
\begin{gather+}
{\rm BER}_{\bf F} = \sum_{y=0}^{n-\theta}\Biggl(\biggl(1 - 
\textsf{binocdf}(y + \theta-1,n+\theta,{\rm BER}_{\textbf{f}}) \biggr)\\ 
\times \textsf{binopdf}(y,n-\theta,{\rm BER}_{\textbf{f}}) \Biggr)
\end{gather+}

\subsection{Extraction Efficiency of S-Norm Transformation}\label{app:SnormEf}
For the S-Norm, if one group/word $\ff$ is selected, it must satisfy the selection criteria $\|\ff\|_1 \in [0,\floor{\frac{n}{2}}-\theta] \cup [\ceil{\frac{n}{2}}+\theta,n]$. Hence, the probability of a group being selected can be expressed as:
\vspace{-1mm}
\begin{align*}
P_{\rm SNorm}^{\rm select} &= {\rm Pr}(\|\ff\|_1 \leq \floor{\frac{n}{2}}-\theta) + {\rm Pr}(\|\ff\|_1 \geq \ceil{\frac{n}{2}}+\theta)\\
&= \sum_{i=0}^{\floor{\frac{n}{2}}-\theta} \Big( {\rm Pr}(\|\ff\|_1 = i)\Big) + \sum_{k=\ceil{\frac{n}{2}}+\theta}^{n} \Big( {\rm Pr}(\|\ff\|_1 = k)\Big)
\end{align*}

By substituting ${\rm Pr}(\|\ff\|_1 \leq i) = \textsf{binocdf}(i,n,0.5)$ and ${\rm Pr}(\|\ff\|_1 \geq k) = 1 - \textsf{binocdf}(k,n,0.5)$, we get:

\begin{dmath*}
    P_{\rm SNorm}^{\rm select} = \textsf{binocdf}(\ceil{\frac{n}{2}}-\theta-1,n,0.5) + \Big(1 - \textsf{binocdf}(\floor{\frac{n}{2}}+\theta,n,0.5)\Big)
\end{dmath*}
The $P_{\rm SNorm}^{\rm select}$ formulates the probability that one group is selected under S-Norm. The extraction efficiency ${\eta}_{\rm SNorm}$ can be directly expressed via $P_{\rm SNorm}^{\rm select}$:
\begin{dmath*}
    {\eta}_{\rm SNorm} = \frac{1}{n} \times P_{\rm SNorm}^{\rm select} \times (1024\times 8)
\end{dmath*}
Where $\frac{1}{n}$ means that a transformed $\digamma$ bit is from $n$ raw bits, The last term $1024\times 8$ is the conversion factor between bit and  KiByte (bit/KiB). By substituting $P_{\rm SNorm}^{\rm select}$ into ${\eta}_{\rm SNorm}$, we can finally obtain equation~\eqref{eqn:HW_SR}.

\begin{dmath*}
    {\eta}_{\rm SNorm} = \frac{1}{n} \times \Bigg( \textsf{binocdf}(\ceil{\frac{n}{2}}-\theta-1,n,0.5)\\ + \Big(1 - \textsf{binocdf}(\floor{\frac{n}{2}}+\theta,n,0.5)\Big) \Bigg) \times (1024\times 8)
\end{dmath*}

\begin{figure}[!ht]
	\centering
	\includegraphics[trim=0 0 0 0,clip,width=1.0\mylinewidth]{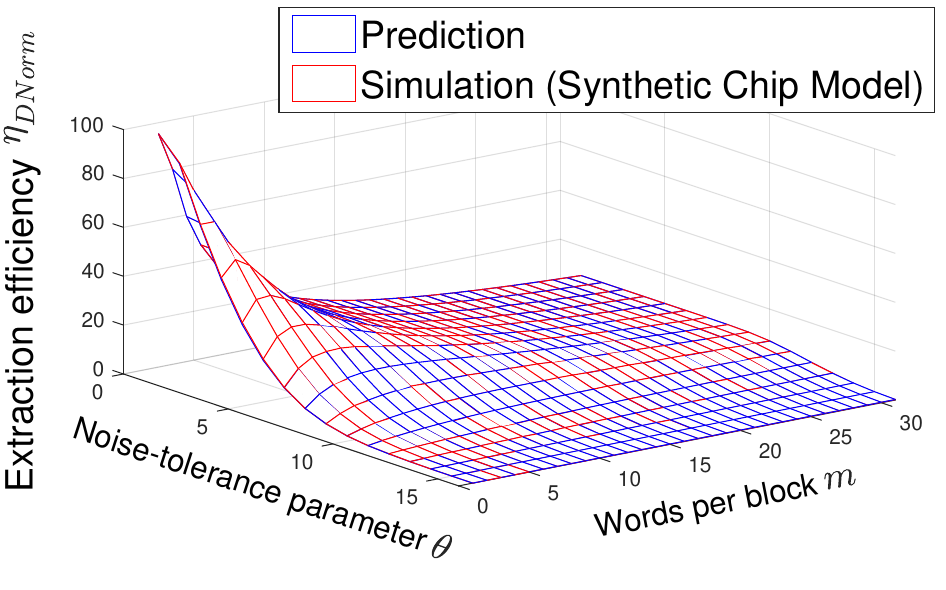}
	\caption{Validation on equation~\eqref{eqn:DHW_SR} (extraction efficiency of D-Norm) using a simulated chip (the \textbf{Simulation} test setting in Section~\ref{Sec:validation}). Here, $n=32$, while $m$ and the noise tolerance parameter $\theta$ are varied. 
	Overall,the simulation agrees well with the prediction, as two values overlaps.}
	\label{fig:DHWn32}
\end{figure}

\subsection{Extraction Efficiency of D-Norm Transformation}\label{app:DnormEf}
To estimate the extraction efficiency of D-Norm, what we need to do first is estimate the probability that among $m$ groups/words $\ff_1,\ff_2,\ldots,\ff_m)$, the minimum $\ell 1$-Norm $\|\ff_i\|_1$ is any given value $a$ from $0$ to $n$, and the maximum $\ell 1$ norm $\|\ff_i\|_1$ is another given value $z$ from $0$ to $n$:

\begin{gather+}[0.85]
    P(a,z) \triangleq {\rm Pr}\Big(l = a ~ \land ~ h = z\Big)
\end{gather+}
Recall that:
\begin{align*}
    h \triangleq \aggregate{arg~max}{\textbf{f}_i|i\in \{1,..,m\}} (\|\ff_i\|_1)\\
    l \triangleq \aggregate{arg~min}{\textbf{f}_i|i\in \{1,..,m\}} (\|\ff_i\|_1)
\end{align*}

Once we comply with the above principle, the $P_{\rm block}^{\rm select}$, that one block to be selected for noise-tolerant fingerprint extraction is simply the sum of all $P(a,z)$ over $z-a\ge \theta$.
\begin{equation*} \label{eqn:EtaGroup}
    P_{\rm block}^{\rm select}=\sum_{a=1}^{n-\theta}\sum_{z=a+\theta}^n P(a,z)
\end{equation*}

$P(a,z)$ is a non-trivial to estimate. Fortunately, we can solve an easier and related problem first:
\begin{gather+}[0.86]
Q(a,z)\triangleq{\rm Prob}\Big(l\ge a \land h\le z\Big)
= \Big( \sum_{i=a}^{z}\textsf{binopdf}(i,n,0.5) \Big)^m \qquad
\end{gather+}

Another angle to look at $Q(a,z)$ is: What is the probability that among $m$ words in a block with all $\ell 1$-Norm are at least $a$ and at most $z$? That question can be answered because it poses an independent question on each word $\ff_i$: is $a \le \|\ff_i\|_1 \le z$ or not? The answer must be "yes" for all $m$ words, and it is "yes" for a single word with probability $\sum_{i=a}^{z}\textsf{binopdf}(i,n,0.5)$ (the usual formula for the number of $\|\ff_i\|_1$ meet $\theta$ divided by the number of all $m$ words), and because those events are independent, the probabilities can be consequentially multiplied.

The question becomes: how do we get from $Q(a,z)$ to $P(a,z)$? 

Note that:
\begin{gather+}[0.70]
    \{(\ff_1,\ff_2,\ldots,\ff_m): (l=a  ~ \land ~ h=z)\}\\ 
    = \{(\ff_1,\ff_2,\ldots,\ff_m): (l\ge a  ~ \land ~ h=z)\} - \{(\ff_1,\ff_2,\ldots,\ff_m): (l\ge a+1  ~ \land ~ h=z)\}
\end{gather+}

because for the $l$ to be equal to $a$ it is equivalent to ask for the $l$ to be at least $a$ but not to be at least $a+1$. In addition, the set we are subtracting is actually a subset of the set we are subtracting from, so we obtain:
\begin{gather+}[0.70]
    P(a,z)= {\rm Prob}\{(\ff_1,\ff_2,\ldots,\ff_m): (l = a  ~ \land ~ h=z)\}\\
    = {\rm Pr}\{(\ff_1,\ff_2,\ldots,\ff_m): (l\ge a  ~ \land ~ h=z)\}
    - {\rm Pr}\{(\ff_1,\ff_2,\ldots,\ff_m): (l\ge a+1 ~ \land ~ h=z)\}
\end{gather+}

Our two operands are of the same type. We can do the same operation to reduce each probability to something expressible by some $Q(r,s)$:
\begin{gather+}[0.70]
    P(a,z)= \{(\ff_1,\ff_2,\ldots,\ff_m): (l\ge r \land h=z)\}\\ 
    = \{(\ff_1,\ff_2,\ldots,\ff_m): (l\ge r \land h \le z)\} 
    - \{(\ff_1,\ff_2,\ldots,\ff_m): (l\ge r \land h \le z-1)\} 
\end{gather+}

And we obtain:
\begin{gather+}[0.70]
    {\rm Pr}\{(\ff_1,\ff_2,\ldots,\ff_m): (l\ge r \land h=z)\} = \\
    {\rm Pr}\{(\ff_1,\ff_2,\ldots,\ff_m): (l\ge r \land h \le z)\}
    - {\rm Pr}\{(\ff_1,\ff_2,\ldots,\ff_m): (l\ge r \land h \le z-1)\} \\
    = Q(r,z) - Q(r,z-1) \qquad \qquad \qquad \qquad \qquad \qquad \qquad \qquad \qquad \qquad \qquad \qquad 
\end{gather+}

And finally, for $P(a,z)$, by substituting this in the above formula:
\begin{gather+}[0.82]
P(a,z)=(Q(a,z)-Q(a,z-1)) - (Q(a+1,z)- Q(a+1,z-1))
\end{gather+}

\begin{figure}[!ht]
    \centering
    \includegraphics[width=0.35\mylinewidth]{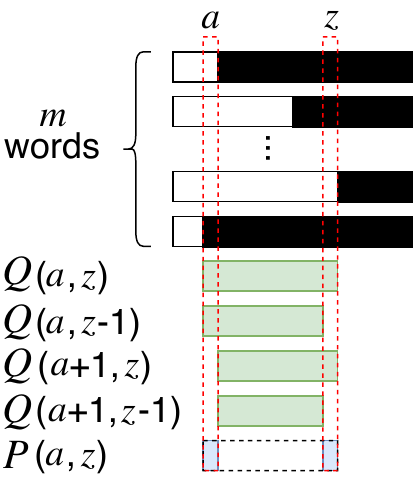}
    \caption{Showing the relationship between $P_{\rm block}$ and $Q$ terms}
    \label{fig:DHW_Q}
\end{figure}

The normalized D-Norm extraction efficiency ${\eta}_{\rm DNorm}$ is finally given:
\begin{gather+}
    {\eta}_{\rm DNorm} = \frac{1}{n \times m} \times P_{\rm block}^{\rm select} \times (1024 \times 8)
\end{gather+}
To be concise, we keep $P_{\rm block}^{\rm select}$, $P_{\rm block}$ and $Q(l,h)$ to be expressed separately.
The term of $\frac{1}{n\times m}$ stands for $n\times m$ raw bits producing a 1-bit noise-tolerant bit, per ($1024\times 8$) 1 KiB memory.
It tends be hard to follow when we substitute all terms and write a huge equation. To be concise, we keep express $P_{\rm DNorm}^{\rm select}$, $P_{\rm block}$ and $Q(l,h)$ separately.

\section{Remote Attestation}\label{app:remAtt}
The following description is based on the setting shown in Fig.~\ref{fig:Protocol_WORM} where the Prover device implements a secure WORM memory for storing the enrolled $\textbf{mask}$. During the one-time enrollment conducted by the trusted Verifier, we use a cabled JTAG interface and Segger J-link command-line tool to read out the start-up state (fingerprints) of Prover's (Nordic Semiconductor nRF52832) SRAM. Readout raw fingerprints are saved as binary files and then processed (using 
Matlab) for performing the D-Norm transform and selection (Section~\ref{Sec:method}). Such a process produces: i) a database entry containing Prover $\textbf{id}$ and selected reliable noise-tolerant ${ \digamma}$ bit and ii) a C language header file containing the $\textbf{mask}$ indicating the memory addresses of raw fingerprint bits to be employed for obtaining ${\digamma}$ bits to be compiled with the sensor node code. Next, we elaborate on an efficient means for organizing the memory addresses defined by the $\textbf{mask}$. 

\textbf{D-Norm Mask.~}The $\textbf{mask}$ first specify the starting address of the fingerprint zone, which is set to 0x4000, reserving the lower 16~KiB of SRAM space for system run-time operations. To reduce the storage footprint of the $\textbf{mask}$, only the relative offsets between selected memory addresses, rather than the 32-bit absolute addresses are recorded. Once the $\textbf{mask}$ is determined, the server computes a MAC $\textbf{tag}$ over the $\textbf{mask}$ with the derived key ${\bf F}$ for integrity checks. 

\textbf{Implemented System.~}During the attestation phase, we employ a command-line Verifier tool, \circled{1}  shown in Fig.~\ref{fig:Overview} (b), to randomly generate a challenge (nonce). We look up the \textbf{DB} according to the Prover's returned ${\bf id}$ and compute the expected response. To visualize the data exchange for {\it demonstration purposes}, we built our Gateway \circled{2} using an Android demo APP based on FastBLE library\footnote{FastBLE is available: \href{https://github.com/Jasonchenlijian/FastBle}{\underline{https://github.com/Jasonchenlijian/FastBle}}} and used the smartphone's built-in Bluetooth-LE interface to communicate with the Prover. In practice, the Gateway could be realized by any base station with a Bluetooth-LE transceiver. The Prover \circled{3} in this case study is a representative low-end sensor node equipped with an ARM-Cortex M4-based nRF52832 Bluetooth-LE SOC. The code to be attested on the Prover is statically allocated with a linker Preprocessor command.\footnote{For example \text{\_\_attribute\_\_((section(``.ARM.\_\_at_0x50000'')))} in Keil uVision specifies placing the function at memory starting from address 0x50000.} The noise-tolerant fingerprint regeneration function, the mask, and the immutable bootloader are placed in WORM memory using an ARM MPU.

\begin{figure}[!ht]
	\centering
	\includegraphics[trim=0 0 0 0,clip,width=0.3\textwidth]{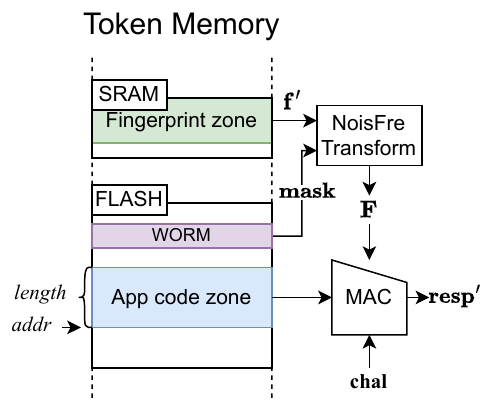}
	\caption{Memory management and data flow for the remote attestation at the Prover. Notably, the total SRAM memory size is 64~KiB. We only use 48~KiB for the fingerprint zone and reserve 16~KiB for system run-time operations.}
	\label{fig:dataFlow}
\end{figure}

\section{Memory Fingerprint Datasets}\label{app:memory_dataset}
Below we detail the collected NORDIC dataset since the details of the public SRAM datasets are in~\cite{guo2017scare,rahman2017systematic}.\Revsource{rev:flashEEPROM}{In addition to the SRAM datasets, we also employed a public FLASH dataset~\cite{guo2017ffd} and collected EEPROM datasets for generalization. The Winbond W29N02GV Flash used in ~\cite{guo2017ffd} is a single-level cell (SLC) Flash and its partial programming time is set to be 150 us~\cite{guo2017ffd}. The ${\rm BER}_{\textbf{f}}$ of Flash memory fingerprints are negligibly affected by changes in voltage and temperature but programming cycles impart chip aging. Hence, we only consider aging induced ${\rm BER}_{\textbf{f}}$ for Flash memory. Our method for collecting EEPROM data is similar to that employed for Flash~\cite{wang2012flash} based on leveraging partial programming. Different from Flash, EEPROM can only be programmed byte-by-byte. Consequently, partial programming consumes a longer period of time on EEPROM than on Flash. The partial programming latency depends on the temperature, at 25$\celsius$, it takes at least 40~ms to evaluate one byte.}

\vspace{0.1cm}
\noindent{\bf NORDIC.} We first collected 12 nRF52832 chips under each of the three operating temperatures (see one such chip in Fig.~\ref{fig:Overview}). The nRF52832 is a popular RF-enabled MCU, and supports various protocols including Bluetooth~5, Bluetooth mesh, ANT, and NFC. This chip has a 64~KiB SRAM memory. The NORDIC chip is representative of a typical low-cost IoT device MCU. Three temperature corners \{$-15\celsius$, $25\celsius$, $80\celsius$\} are evaluated to measure the reliability of raw bits using 100 repeated measurements taken under each operating corner. The worst-case ${\rm BER}_{\textbf{f}}$ of 6.09\% occurs under $80\celsius$ when the reference template is collected at $25\celsius$. \Revsource{rev:88chips}{We have collected a further 88 nRF52832 chips under room temperature conditions of $25\celsius$ to augment the dataset. This extensive dataset of 100 chips (at room temperature) is used to evaluate \textit{uniformity} and \textit{uniqueness} as these metrics benefit from a larger sample of chips but are normally insensitive to operating conditions. In our evaluations, unless otherwise stated, we use the 12-chip dataset (with evaluations at three operating corners) when referring to the NORDIC chip dataset.}

\end{document}